\documentclass[aps,prb,twocolumn,superscriptaddress,showpacs]{revtex4-1}

\usepackage{amsmath,amssymb}
\usepackage{bbm}
\usepackage{graphicx}
\usepackage{float}
\usepackage{color}
\usepackage{nicefrac}
\usepackage{longtable}
\usepackage{mathrsfs}
\usepackage{placeins}
\usepackage{calc}
\usepackage[linktocpage=true]{hyperref}

\renewcommand{\dag}{^\dagger}
\renewcommand{\vec}[1]{\boldsymbol #1}
\newcommand{\de}{\mathrm{d}}
\newcommand{\ii}{\mathrm{i}}

\newcommand{\h}{\hspace{1pt}}
\newcommand{\hh}{\hspace{0.5pt}}
\newcommand{\mh}{\hspace{-1pt}}
\newcommand{\pp}{'\hspace{-1pt}'}

\usepackage[T1]{fontenc}
\newcommand{\db}{\textnormal{\dj}}

\renewcommand{\k}{_{\vec k}}
\newcommand{\mk}{_{-\vec k}}
\newcommand{\T}{^{\rm T}}

\newcommand{\e}{\mathrm{e}}

\newcommand{\ket}[1]{|#1\rangle}
\newcommand{\bra}[1]{\langle #1|}

\newcommand{\Tr}{\mathrm{Tr}}

\DeclareMathOperator{\sgn}{sgn}
\DeclareMathOperator{\arsinh}{arsinh}

\begin{document}


\title{Functional renormalization and mean-field approach to multiband systems with \\[1pt] spin-orbit coupling: Application to the Rashba model with attractive interaction}


\author{G.\,A.\,H. Schober}
\author{K.-U. Giering}
\author{M.\,M. Scherer}
\affiliation{Institute for Theoretical Physics, Heidelberg University, Philosophenweg 19, D-69120 Heidelberg, Germany \vspace{1pt}}

\author{C. Honerkamp}
\affiliation{Institute for Theoretical Solid State Physics, RWTH Aachen University, D-52074 Aachen, Germany \vspace{2pt}}

\author{M. Salmhofer}
\affiliation{Institute for Theoretical Physics, Heidelberg University, Philosophenweg 19, D-69120 Heidelberg, Germany \vspace{1pt}}


\date{\today}

\begin{abstract}
The functional renormalization group (RG) in combination with Fermi surface patching is a well-established method for studying Fermi liquid instabilities of correlated electron systems. In this article, we further develop this method and combine it with mean-field theory to approach multiband systems with spin-orbit coupling, and we apply this  to a tight-binding Rashba model with an attractive, local interaction. The spin dependence of the interaction vertex is fully implemented in a RG flow without SU(2) symmetry, and its momentum dependence is approximated in a refined projection scheme. In particular, we discuss the necessity of including in the RG flow contributions from both bands of the model, even if they are not intersected by the Fermi level. As the leading instability of the Rashba model, we find a superconducting phase with a singlet-type 
interaction between electrons with opposite momenta. While the gap function has a singlet spin structure, the order parameter indicates an unconventional superconducting phase, with the ratio between singlet and triplet amplitudes being plus or minus one on the Fermi lines of the upper or lower band, respectively. We expect our combined functional RG and mean-field approach to be useful for an unbiased theoretical description of the low-temperature properties of spin-based materials.
\end{abstract}

\pacs{64.60.ae, 71.27.+a, 71.70.Ej, 74.20.Rp}

\maketitle

\tableofcontents

\section{Introduction\label{sec_Introduction}}

In the rapidly evolving field of spintronics, which aims at revolutionizing present-day computer logic and memory,\cite{Sinova} the Rashba model of a two-dimensional electron gas plays a paradigmatic r\^{o}le. By coupling the spin and orbital degrees of freedom, it provides a key to the manipulation of electron spins by means of electric fields.\cite{Barnes} Originally, the model was introduced in 1960 to describe the band splitting in non-centrosymmetric wurtzite semiconductors,\cite{Rashba60} but since then a vast number of materials have been found that are described by this model, ranging from semiconductor heterostructures\cite{Nitta97} to the Shockley surface states of noble metals\cite{LaShell96, Hoesch04} as well as noble-metal-based surface alloys incorporating heavy elements.\cite{Pacile, Ast07, Ast2} More recently, giant {\itshape bulk} Rashba spin splitting has been observed in the  semiconductor BiTeI,\cite{Ishizaka, Bahramy, BiTeImagn} at the Te-terminated surface of 
bismuth 
tellurohalides,\cite{Eremeev10, Eremeev13} and in other systems based on chemical elements with high atomic numbers.\cite{Rashba12} These materials are promising candidates for spintronics applications such as the Datta-Das spin transistor.\cite{Datta89} Indeed, much research is focused on the transport properties induced by the Rashba effect such as the intrinsic spin Hall effect,\cite{Sinova04} current-driven spin torques,\cite{Miron10} or spin precession phenomena.\cite{Wunderlich10, Jungwirth12}

From a many-body perspective, the equilibrium properties and in particular the low-temperature phase diagram of the Rashba model with  electron-electron interactions are also of great interest. Mean-field approaches have indicated the possibility of unconventional superconductivity in the Rashba model due to the absence of the SU(2) spin rotation symmetry.\cite{Edelstein89, GR01, Frigeri04, Min12} 
In combination with an $s$-wave pair potential and a sufficiently large Zeeman term, the Rashba spin-orbit coupling has been predicted to produce an effective chiral $p$-wave superconductor carrying Majorana bound states in its vortex cores.\cite{Fujimoto08, Sau10, Alicea10, Beenakker13} In a nanowire geometry, zero-energy states were then expected to appear at the two ends of the wire,\cite{Lutchyn10, Oreg10} a theoretical prediction which triggered the experimental hunt for Majorana fermions.\cite{Kouwenhoven12, GG12}

Despite the tremendous progress in theoretical understanding, much remains to be done to 
identify the symmetries of the superconducting phases of the Rashba model in an unambiguous way. Mean-field solutions of the gap equation yield the superconducting order parameter from a {\itshape given} gap function (pair potential). For the latter, one usually assumes the absence of interband pairing, which then allows for its parametrization in terms of spin-singlet and -triplet amplitudes.\cite{Min12} Their ratio in turn determines the topological properties of the superconducting phase.\cite{Neu10} It is therefore desirable to determine the gap function and the order parameter explicitly in a general setup such as the Rashba model with a local electron-electron interaction.

In this article, we address this problem for a tight-binding Rashba model on the hexagonal Bravais lattice with an attractive local interaction, using a combined functional RG and mean-field approach. The tight-bin\-{}ding model approximately describes the energy dispersion of the Rashba semiconductor BiTeI\cite{Ishizaka} near the high-symmetry point $A$ of the hexagonal Brillouin zone. The attractive {\itshape initial} interaction induces an instability towards a superconducting phase,\cite{BCS, Thouless} which can be detec\-{}ted by a divergence of the {\itshape effective} interaction vertex in the RG flow.\cite{SH00} In real materials, an initial attractive electron interaction may typically arise, for example, from an electron-phonon mechanism.\cite{BCS, Cooper} We remark that for a repulsive initial interaction, a superconducting  instability generally also occurs due to the Kohn-Luttinger effect.\cite{Kohn} However, in the absence of Fermi surface nesting, the critical scale below which the ordered phase occurs is much smaller and can therefore hardly be detected in our numerical RG approach. (For a detailed discussion of the Kohn-Luttinger effect in two-dimensional Fermi systems see Refs.~\onlinecite{Feldman, Sinclair}, and in the context of the functional RG see Ref.~\onlinecite{SH00}.) The superconducting instabilities in the Rashba model with a repulsive interaction have been investigated using RG methods by Vafek and Wang.\cite{Vafek11, Wang14}

Our aim is to characterize the superconducting phases of the Rashba model with an attractive, local initial interaction without a priori assumptions on the gap form factor. For this purpose, the functional RG in combination with a Fermi surface patching approximation has proven useful.\cite{SH00, Metzner} It allows to study the (possibly competing) Fermi liquid instabilities without making any potentially restrictive a priori assumptions (e.g., about the absence of interband pairing), and it yields the effective interaction for the electrons near the Fermi energy after integrating out the high-energy degrees of freedom. The RG procedure has already been successfully applied to models relevant for the high-temperature superconducting cuprates\cite{Ho2001a, Ho2001b, Husemann, Giering, Eberlein14, Eberlein14b} and
iron pnictides,\cite{DHLee, PlattNJP, Thomale09, Thomale, Platt11, Platt12, Lichtenstein} strontium ruthenate,\cite{QHWang} monolayer graphene,\cite{Ho2008, Kiesel, Wu, Janssen, ClassenPhon, Classen} bilayer\cite{Scherer1, Lang, Pena} and trilayer\cite{Scherer2} graphene, and many other correlated fermion systems (for recent reviews see Refs.~\onlinecite{Metzner} and \onlinecite{Platt}). The study of RG flow equations for non-SU(2)-invariant systems has recently become a topic of major interest.\cite{Maier, DScherer}

On the level of functional RG technique, the general set of RG equations was derived without the assumption of SU(2) symmetry in Ref.~\onlinecite{SH00}. Our work is novel in that we fully implement the RG flow in the case without SU(2) symmetry, taking into account the full spin dependence of the interaction vertex, and using a refined projection scheme for the momentum discretization of the interaction vertex, which admits projections on the representative momenta irrespectively of the band index. In particular, we find that contributions from both bands of the model have to be included in the RG flow in order to obtain the correct effective interaction at the critical scale. Our numerical solution of the RG equations within this refined projection scheme is confirmed by an analytical resummation of the particle-particle ladder.

Furthermore, we combine our functional RG approach with mean-field theory in order to provide a more detailed characterization of the superconducting phases. While the functional RG yields the superconducting interaction at the critical scale, mean-field theory allows to predict from this the superconducting gap function and the order parameter. 
In particular, we generalize the Bogoliubov transformation of Ref.~\onlinecite{Sig91} to the non-SU(2)-symmetric case,
and thereby obtain explicit expressions for the singlet and triplet amplitudes of the superconducting order parameter as a function of momentum and chemical potential. Finally, by analytically and numerically solving the scalar gap equation, we also predict the gap size as a function of the chemical potential and the interaction strength. A consistent fusion of functional RG methods and mean-field theory has already been discussed in Ref.~\onlinecite{Wang} (see also the earlier work Ref.~\onlinecite{Reiss}).

The article is organized as follows: In Sec.~\ref{sec_Rashba}, we define as our starting point a minimal tight-binding model on the hexagonal Bravais lattice which displays Rashba spin splitting near the center of the Brillouin zone. In Sec.~\ref{sec_RG}, we introduce the functional RG and describe the Fermi surface patching approximation used 
to solve the RG equations numerically. We discuss the advantages of our refined projection scheme, and after that, we present our results on the leading instabilities and corresponding effective interactions. In Sec.~\ref{sec_MF}, we describe the mean-field approach used to predict the gap function and the order parameter of the Rashba model. In Appendixes \ref{app_Rashba} and \ref{app_not}, we fix our conventions for the tight-binding description of electronic states on the hexagonal Bravais lattice and for the corresponding temperature Green functions. Furthermore, Appendix~\ref{app_Rashba} contains an elementary derivation of our minimal tight-binding model from symmetry conditions only, and Appendix \ref{app_not} a brief review of the most important definitions and properties of temperature Green functions. These two appendixes are essentially self-contained and can be read independently as pedagogical reviews. Finally, in Appendix \ref{app_proj} we derive the projected RG equations which we have implemented numerically, and we show how the mean-field interaction is deduced from the resulting projected interaction vertex. \\

\section{Rashba model with local interaction} \label{sec_Rashba}

\begin{figure}
\hspace{-0.5cm}\includegraphics[width=0.85\columnwidth]{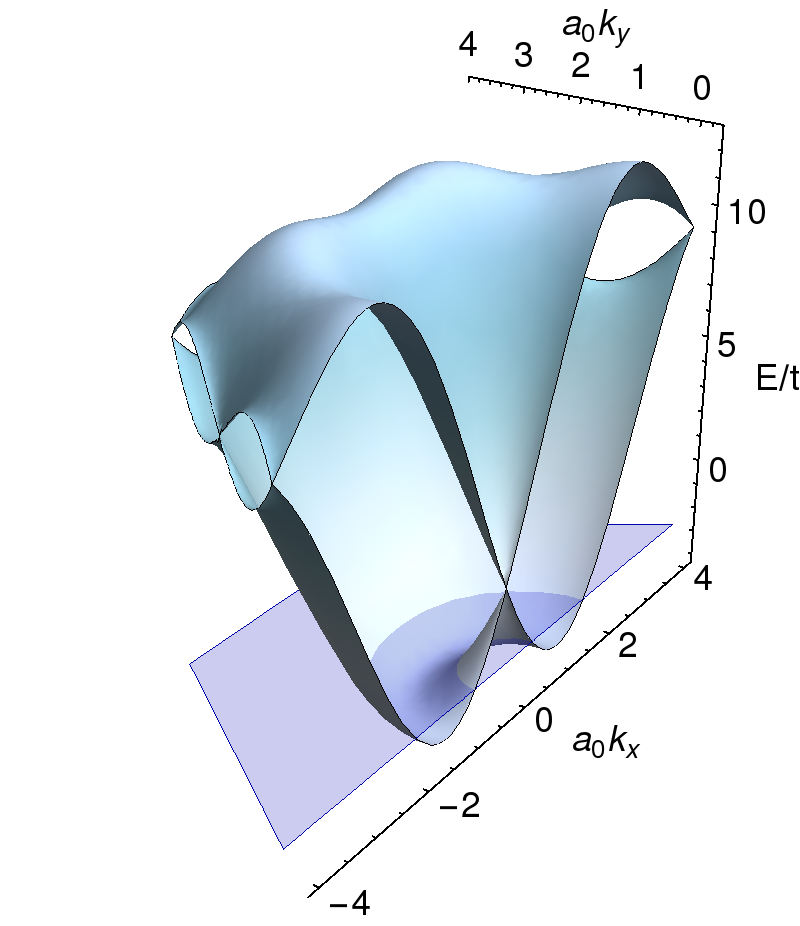}
\caption{Energy bands of the minimal tight-binding model~for $\alpha/t=2$, with the chemical potential $\mu/t = -2$ below the band~crossing at $\vec k = \vec 0$. To show more clearly the energy dis\-{}persion near the band crossing, only half the Brillouin zone ($k_y > 0$) is shown.\label{fig_disp}}
\end{figure}

We study a minimal tight-binding model on the hexagonal Bravais lattice in two dimensions, which displays Rashba spin 
splitting near the center of the Brillouin zone. The free Hamiltonian is given by
\begin{equation} \label{start}
 \hat H^0 = \int \! \db^2 \vec k \, \sum_{s, \h s'} H^0_{ss'}(\vec k) \, \hat a_s\dag(\vec k) \h \hat a_{s'}(\vec k) \,.
\end{equation}
Here, the Bloch momentum $\vec k$ ranges over the (first) Brillouin zone $\mathcal B$, and we use a normalized measure
\begin{equation} \label{nmeasure}
 \int \! \db^2 \vec k = \frac{1}{|\mathcal B|} \h \int_{\mathcal B} \de^2 \vec k \,,
\end{equation}
where $|\mathcal B|$ is the area of the Brillouin zone. Furthermore, $s, s' \in \{\uparrow, \downarrow\}$ are {\itshape spin indices.} For a more detailed exposition of our conventions, see Appendix \ref{subsec_basis}. We can expand the Hamiltonian matrix $H^0(\vec k) \equiv H^0_{s s'}(\vec k)$ as
\begin{equation}
 H^0(\vec k) = f(\vec k) \h \mathbbm 1 + \vec g(\vec k) \cdot \vec \sigma \,, \label{eq_Ham} 
\end{equation}
where $\mathbbm 1$ denotes the $(2 \times 2)$ identity matrix and $\vec \sigma = (\sigma_x, \sigma_y, \sigma_z)^{\rm T}$ the vector of the Pauli matrices. In terms of the dimensionless quantity
\begin{equation} \label{dimensionless_kappa}
 \vec \kappa = a_0 \hh \vec k \h \equiv (\kappa_x, \h \kappa_y)^{\rm T} \,,
\end{equation}
where $a_0$ denotes the lattice constant, the coefficient functions of our minimal tight-binding model are given explicitly by
\begin{align}
 & \hspace{-0.87cm} f(\vec \kappa) = 6t -2 t \, \big\{ \cos(\kappa_x) + 2 \cos\!\big(\mbox{$\frac 1 2$} \kappa_x\big) \cos\!\big(\mbox{$\frac{\sqrt{3}}{2}$} \kappa_y\big) \big\} \,, \label{eq_f} \\[3pt]
 g_x(\vec \kappa) & = -2 \alpha \, \sqrt{3} \cos\!\big(\mbox{$\frac 1 2$} \kappa_x\big) \sin\!\big(\mbox{$\frac{\sqrt 3}{2} \kappa_y$}\big) \,, \\[3pt]
 g_y(\vec \kappa) & = 2 \alpha \, \big\{ \sin(\kappa_x) + \sin\!\big(\mbox{$\frac 1 2$}\kappa_x\big) \cos\!\big(\mbox{$\frac{\sqrt 3}{2}$}\kappa_y \big) \big\} \,, \\[6pt]
 g_z(\vec \kappa) & = 0 \,, \label{eq_gz}
\end{align}
where $t$ and $\alpha$ are constants. In Appendixes~\ref{sec_sym}--\ref{sec_min}, we show that this model corresponds in direct space to nearest-neighbor hopping with the amplitudes $t$ and $\alpha$, and that this model can be derived from symmetry considerations only. The constant term $6t$ has been added to the Hamiltonian such that $f(\vec 0) = 0$ (and consequently both eigenvalues at $\vec k = \vec 0$ are zero). Since $|\vec g| \not = 0$, the model Hamiltonian \eqref{eq_Ham} is not invariant under general SU(2) spin rotations. For small momenta, we can Taylor expand the above functions to quadratic order in the momentum as
\begin{align}
 f(\vec \kappa) & = \frac {3t} 2 (\kappa_x^2 + \kappa_y^2) \,, \\[3pt]
 g_x(\vec \kappa) & = -3 \alpha \h \kappa_y \,, \\[6pt]
 g_y(\vec \kappa) & = 3 \alpha \h \kappa_x \,, \\[6pt]
 g_z(\vec \kappa) & = 0 \,.
\end{align}
This shows that near the center of the Brillouin zone, the model is described by the {\itshape Rashba Hamiltonian}
\begin{equation} 
H_{\rm R}(\vec k) = \frac{3 t}{2} \h a_0^2 \h (k_x^2 + k_y^2) \h \mathbbm 1 + 3 \alpha \h a_0 \h ( k_x \sigma_y - k_y \sigma_x ) \,.
\end{equation}
This can be written equivalently as
\begin{equation} \label{eq_rashba}
 H_{\rm R}(\vec k) = E_{\rm R} \h \bigg( \frac{k_x^2 + k_y^2}{\raisebox{-1pt}{$k_{\rm R}^2$}} \, \mathbbm 1 \h + \h 2 \, \frac{k_x \sigma_y - k_y \sigma_x}{k_{\rm R}} \bigg) \,,
\end{equation}
where the {\itshape Rashba energy} $E_{\rm R}$ and the {\itshape Rashba wave vector} $k_{\rm R}$ are related to the parameters of the tight-binding model by
\begin{equation}
 E_{\rm R} = \frac{3 \hh \alpha^2}{2 \hh t} \,, \qquad k_{\rm R} = \frac{\alpha}{a_0 \hh t} \,.
\end{equation}
The characteristics of the Rashba model dispersion are (i) the band crossing at \h$\vec k = \vec 0$, (ii) the approximately linear dispersion for small wave vectors, and (iii) the band minimum which is attained on the circle $|\vec k| = k_{\rm R}$ around \linebreak the axis of symmetry.\cite{Boiko60} The Rashba energy $E_{\rm R}$, which equals the energy difference between the band crossing and the band minimum, is often used to quantify the Rashba spin splitting of the energy bands in real materials.\cite{Ishizaka} In the following, we define all quantities with the dimension of an energy by specifying their ratio to the hopping parameter $t$. In particular, we choose the model parameter~$\alpha$ such that
\begin{equation}
 \alpha / t = 2 \,.
\end{equation}
Now, we come to the diagonalization of the Hermitian matrix \eqref{eq_Ham}. Generally, we have
\begin{equation} \label{defU}
\sum_{s, \h s'} U_{ns}^\dagger(\vec k) \, H^0_{ss'}(\vec k) \, U_{s'n'}(\vec k) = \delta_{nn'} E_{n'}(\vec k) \,,
\end{equation}
where $n, n' \in \{-, +\}$ are {\itshape band indices,} and the eigenvalues are given by
\begin{equation} \label{ev}
 E_{\mp}(\vec k) = f(\vec k) \mp |\vec g(\vec k)| \,.
\end{equation}
We refer to $E_-(\vec k)$ and $E_+(\vec k)$ as the {\itshape lower} and the {\itshape upper energy band,} respectively. 
Their dispersion is shown in Fig.~\ref{fig_disp}. 
The unitary matrix $U_{sn}(\vec k)$ mediates between the {\itshape spin basis} and the {\itshape band basis.} It contains the normalized eigenvectors as column vectors and is given by
\begin{align}
 & U(\vec k) \equiv \left( \begin{array}{cc} U_{\uparrow -}(\vec k) & U_{\uparrow +}(\vec k) \\[5pt] U_{\downarrow -}(\vec k) & U_{\downarrow +}(\vec k) \end{array} \right) \label{expU1} \\[5pt]
 & = \frac{1}{\sqrt{2 \h |\vec g|}} \left( \begin{array}{cc} \sqrt{|\vec g| - g_z} & \sqrt{|\vec g| + g_z} \\[5pt] -\sqrt{|\vec g| + g_z} \,\h \e^{\ii \varphi} & \sqrt{|\vec g| - g_z} \,\h \e^{\ii \varphi} \end{array} \right) ,\label{expU}
\end{align}
 \ \\[-5pt]
where we have suppressed the $\vec k$ dependencies on the right-hand side of the equation. Furthermore,
\begin{equation}
 |\vec g| = \sqrt{g_x^2 + g_y^2 + g_z^2} \,,
\end{equation}
and $\varphi \equiv \varphi(\vec k) \in [0, 2\pi)$ is defined by
\begin{align}
g_x & = \sqrt{g_x^2 + g_y^2} \, \cos \varphi \,, \\[2pt]
g_y & = \sqrt{g_x^2 + g_y^2} \, \sin \varphi \,.
\end{align}
Equivalently, we can write this as
\begin{equation} \label{phi}
 \e^{\ii\varphi} = \frac{g_x + \ii g_y}{\sqrt{g_x^2 + g_y^2}} = \frac{g_x + \ii g_y}{\sqrt{(|\vec g|-g_z) \h (|\vec g|+g_z)}} \,.
\end{equation}
\begin{figure}
\begin{center}
\includegraphics[width=\columnwidth]{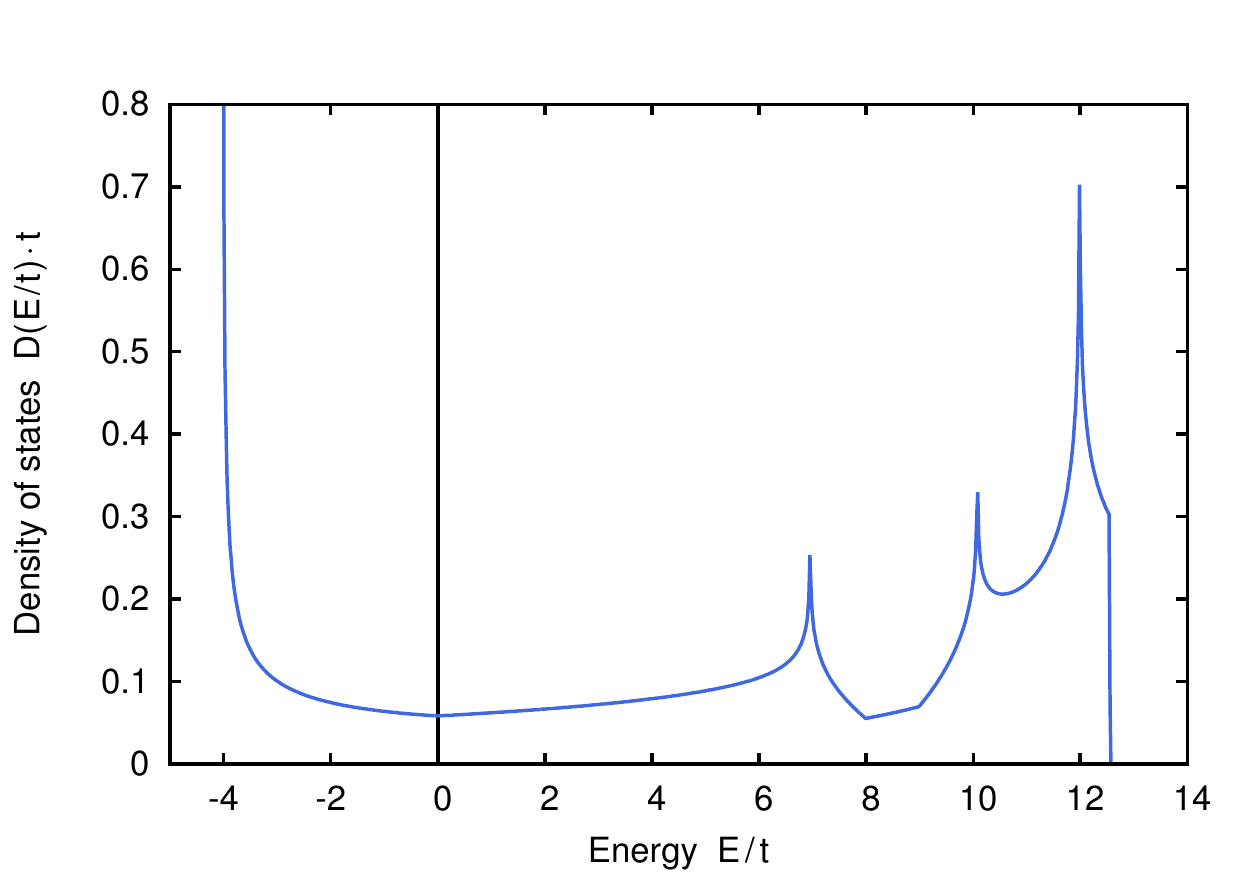}
\caption{Density of states of the minimal tight-binding model with the same parameters as in Fig.~\ref{fig_disp}. The vertical line marks the position of the band crossing at the center of the Brillouin zone. \label{fig_dos}}
\end{center}
\end{figure} \ \\[-5pt]
Note that $\varphi(\vec k)$ is {\itshape not} the polar angle of the vector $\vec k$, but of the vector
$(g_x(\vec k), \, g_y(\vec k))^{\rm T}$\h. Since $g_z(\vec k) \equiv 0$ in our model, the unitary matrix \eqref{expU} simplifies to
\begin{equation} \label{expU_simpl}
 U(\vec k) = \frac{1}{\sqrt{2}} \left( \begin{array}{cc} 1 & 1 \\[5pt] -\e^{\ii \varphi(\vec k)} & \e^{\ii \varphi(\vec k)} \end{array} \right) .
\end{equation}
The density of states (DOS) is defined separately for each energy band as
\begin{equation} \label{dos_def}
 D_{\mp}(E) = \int \! \db^2 \vec k \ \delta( E - E_{\mp}(\vec k) ) \,,
\end{equation}
while the total DOS is defined by their respective sum,
\begin{equation} \label{total_dos}
 D(E) = D_-(E) + D_+(E) \,.
\end{equation}
The total DOS of the minimal tight-binding model is shown in Fig.~\ref{fig_dos}. In particular, we read off the bandwidth (i.e.~the difference between the band maximum and the band minimum) of the model,
\begin{equation} \label{bandwidth}
 (\Delta E)_{\rm max} \h / \hh t \approx 16.5 \,.
\end{equation}
For the {\itshape ideal} Rashba Hamiltonian~\eqref{eq_rashba}, the DOS can be calculated analytically and is given by (see e.g.~Ref.~\onlinecite{Winkler}, Eqs.~(6.20a)--(6.20b))
\begin{equation} \label{dos_explicit}
\begin{aligned}
 & \mh D_{\rm R}(E) = \\[5pt]
 & \left\{ \begin{array}{ll} \displaystyle \frac{\sqrt 3}{4\pi} \, \frac{(a_0 k_{\rm R})^2}{E_{\rm R}} \, \frac{1}{\sqrt{1 + E / E_{\rm R}}} \,, & \textnormal{\itshape if} \ \, E \leq 0 \,, \\[12pt]
 \displaystyle \frac{\sqrt 3}{4\pi} \, \frac{(a_0 k_{\rm R})^2}{E_{\rm R}} \,, & \textnormal{\itshape if} \ \, E \geq 0 \,. \end{array} \right.
\end{aligned}
\end{equation}
This function has the following properties: (i) it is constant for $E \geq 0$\h, (ii) it is continuous but has a kink at the band crossing (where $E = 0$), and (iii) it diverges at the minimum of the lower band, $E = -E_{\rm R}$.

In addition to the quadratic Hamiltonian \eqref{start}, we consider a quartic interaction term of the form
\begin{align} \label{def_coeff_m}
 \hat V^0 & = -\frac 1 2 \, \int \! \db^2 \vec k_1 \int \! \db^2 \vec k_2 \int \! \db^2 \vec k_3 \nonumber \\[2pt]
 & \quad \, \times \sum_{s_1, \ldots, s_4} V^0_{s_1 \ldots s_4}(\vec k_1, \vec k_2, \vec k_3) \\[2pt] 
 & \quad \, \times \hat a\dag_{s_1}(\vec k_1) \h \hat a\dag_{s_2}(\vec k_2) \h \hat a_{s_3}(\vec k_3) \h \hat a_{s_4}(\vec k_4) \,, \nonumber
\end{align}
which is a shorthand notation for Eq.~\eqref{def_coeff}. In particular, $\vec k_4$ is fixed by Bloch momentum conservation,
\begin{equation} \label{mom_cons}
 \vec k_4 = \vec K + \vec k_1 + \vec k_2 - \vec k_3 \,,
\end{equation}
where the reciprocal lattice vector $\vec K$ ensures that $\vec k_4$ lies in the first Brillouin zone.
We choose a momentum-independent interaction kernel,
\begin{equation} \label{eq_onsite}
 V^0_{s_1 \ldots s_4}(\vec k_1, \vec k_2, \vec k_3) = \frac{U}{2} \h ( \delta_{s_1 s_3} \h \delta_{s_2 s_4} - \delta_{s_1 s_4} \h \delta_{s_2 s_3} ) \,.
\end{equation}
The corresponding operator $\hat V^0$ then coincides with the normal-ordered operator
\begin{align} \label{int_local}
 \hat V^0 & = U \h \sum_{\vec R} : \hat n_{\uparrow}(\vec R) \h \hat n_{\downarrow}(\vec R) : \\[5pt]
 & = U \h \sum_{\vec R} \hat a\dag_{\uparrow}(\vec R) \h \hat a\dag_{\downarrow}(\vec R) \h \hat a_{\downarrow}(\vec R) \h \hat a_{\uparrow}(\vec R) \,,
\end{align}
where the spin-resolved density operator is defined as
\begin{equation}
 \hat n_s(\vec R) = \hat a_s\dag(\vec R) \h \hat a_s(\vec R) \,.
\end{equation}
An interaction of the form \eqref{int_local} is called {\itshape local}, because it contains only products of electronic density operators at the same lattice site. We choose the parameter $U$ as
\begin{equation} \label{u}
 U / t = -2 \,.
\end{equation}
In particular, the negative sign means that we consider an {\itshape attractive} interaction between electrons. The interaction \eqref{eq_onsite} is SU(2) invariant. The interaction kernel in the band basis is defined in terms of its counterpart \eqref{eq_onsite} in the spin basis by
\begin{equation} \label{spintoband}
\begin{aligned}
 & V^0_{n_1 \ldots n_4}(\vec k_1, \vec k_2, \vec k_3) = \sum_{s_1, \ldots, s_4} U_{n_1 s_1}^\dagger(\vec k_1) \, U_{n_2 s_2}^\dagger(\vec k_2) \\[2pt]
 & \times V^0_{s_1 \ldots s_4}(\vec k_1, \vec k_2, \vec k_3) \, U_{s_3 n_3}(\vec k_3) \, U_{s_4 n_4}(\vec k_4) \,,
\end{aligned}
\end{equation}
with $U(\vec k)$ given by Eq.~\eqref{expU_simpl}. For momentum combinations with
$\vec k_1 = -\vec k_2$, and hence by Eq.~\eqref{mom_cons} also $\vec k_4 = -\vec k_3$, we obtain the following explicit expression \linebreak

\pagebreak \noindent
for the interaction kernel in the band basis:
\begin{equation} \label{this}
\begin{aligned}
 & V^0_{n_1 n_2 n_3 n_4}(-\vec k_2, \vec k_2, \vec k_3) \\[5pt]
 & = -\frac{U}{2} \, \delta_{n_1 n_2} \, \delta_{n_3 n_4} \, n_2 \h n_3 \, \e^{\ii \varphi(\vec k_3) - \ii \varphi(\vec k_2)} \,.
\end{aligned}
\end{equation}
We recall that $n_i \in \{- , +\}$, so
\begin{equation}
 n_2 \h n_3 = \left\{ \begin{array}{ll} 1 & \textnormal{if } n_2 = n_3 \,, \\[5pt]
 -1 & \textnormal{if } n_2 \not = n_3 \,, \end{array} \right.
\end{equation}
The result \eqref{this} is derived from Eq.~\eqref{eq_onsite} using
\begin{equation} \label{id_u}
 \sum_{s} U^\dagger_{ns}(\vec k) \h U_{sn'}(\vec k') = \frac 1 2 \left(1 + n \hh n' \h \e^{\ii \varphi(\vec k') - \ii \varphi(\vec k)} \right),
\end{equation}
which follows from the explicit form \eqref{expU_simpl} of the unitary matrix.

\section{Functional renormalization group} \label{sec_RG}

\subsection{RGE without spin rotation invariance} \label{subsec_RGE}

To obtain the phase diagram of the Rashba model, we use the functional RG approach which describes the evolution from the initial interaction at the ultraviolet scale to an effective interaction at low energies.\cite{Metzner} Importantly, this method accounts for the interplay between different ordering tendencies in an unbiased way. Explicitly, we employ the fermionic RG for the one-line irreducible Green functions with a momentum cut-off as introduced in Ref.~\onlinecite{SH00}. The generating functional $\Gamma$ for the one-line irreducible Green functions is the Legendre transform of the generating functional $\mathcal{W}$ for the connected Green functions (see Appendix \ref{app_not}). To induce the RG flow, we modify the free two-point Green function of the model by introducing an infrared regulator suppressing modes with energies less than a scale $\Lambda$. This leads to a scale dependence of the 
generating 
functional, $\Gamma \rightarrow \Gamma_\Lambda$, and a hierarchy of {\itshape renormalization group equations} (RGE) for the one-line irreducible Green functions. The resulting RG flow smoothly interpolates between the initial interaction defined at the ultraviolet scale $\Lambda_0$ and the low-energy effective interaction for $\Lambda\rightarrow 0$.

In the truncation of the hierarchy used here, typical RG flows then show the following behavior: As the energy scale is lowered, the effective two-particle interaction $V_\Lambda$ diverges already at a nonzero {\itshape critical scale} $\Lambda_{\rm c}>0$ for particular combinations of momenta and further indices such as spins or bands. This is interpreted as a signal for ``an instability leading to an ordered phase via spontaneous symmetry breaking'' (Ref.~\onlinecite{Reiss}; see also Ref.~\onlinecite{Kohn}, footnote~2). The divergence of the effective interaction is due to the truncation, which in particular restricts to the symmetric phase. It has been shown\cite{SHML, Gersch} that the flow can be continued into the symmetry-broken phase and down to $\Lambda = 0$ if the symmetry-breaking terms \linebreak indicated by the effective interaction above $\Lambda_{\rm c}$ are included. The level-two truncation is therefore not used down to $\Lambda = \Lambda_{\rm c}$\h, but the flow is stopped at a scale $\Lambda_* > \Lambda_{\rm c}$ where the coupling reaches a certain threshold (but is still finite). This can be justified in a certain scale range depending on the Fermi surface curvature.\cite{SH00} From the momentum structure of the near-critical two-particle interaction at $\Lambda_*$ one can then construct an effective low-energy Hamiltonian and determine the leading Fermi liquid instability. The effective interaction just above $\Lambda_{\rm c}$ is often referred to as ``the interaction at $\Lambda_{\rm c}$\hh.'' We will also follow this slightly loose convention below, but note here that more strictly, this is to mean ``the interaction \linebreak at $\Lambda_*$\hh.''

In the following, we will give explicit expressions for the RG equations in our model. We use the conventions described in Appendix \ref{app_not}. The free two-point Green function (or covariance) in the spin basis, which is defined as the Fourier transform of Eq.~\eqref{eq_formal_prod_equiv} with respect to the spatial variables, obeys the equation of motion
\begin{equation} \label{eom}
\begin{aligned}
 & \sum_{s\pp } \left( \delta_{ss\pp } \frac{\partial}{\partial \tau} + (H^0_{ss\pp }(\vec k) -\mu \h \delta_{ss\pp })/\hbar \right) \\
 & \qquad \hspace{0.2cm} \times C_{s\pp  s'}(\vec k, \tau - \tau') = \delta_{s s'} \h \delta(\tau - \tau') \,,
\end{aligned}
\end{equation}
where $H_{ss'}^0(\vec k)$ has been defined by Eqs.~\eqref{eq_Ham}--\eqref{eq_gz}. This can be shown directly from the imaginary-time analog of 
the Heisenberg equation of motion for the field operators,
\begin{equation}
 \hbar \h \frac{\partial}{\partial \tau} \h \hat a(\vec k, \tau) = \big[ \hat H^0 - \mu \hat N, \h \hat a(\vec k, \tau) \big] \,,
\end{equation}
where $\hat N$ denotes the particle-number operator. In the frequency domain, Eq.~\eqref{eom} is equivalent to (see Eqs. \eqref{Fourier_Tr_1}--\eqref{Fourier_Tr_2})
\begin{equation}
\begin{aligned}
 & \sum_{s\pp } \big({-\ii\omega} \h \delta_{ss\pp } + (H_{ss\pp }^0(\vec k) - \mu \h \delta_{ss\pp }) / \hbar \h \big) \\[-1pt]
 & \qquad \hspace{0.2cm} \times C_{s\pp s'}(\vec k, \omega) = (\hbar\beta)^{-1} \h \delta_{ss'} \,.
\end{aligned}
\end{equation}
The covariance in the band basis is defined by means of the unitary matrix $U_{sn}(\vec k)$ (see Eq.~\eqref{defU}) as
\begin{equation}
\sum_{s, \h s'} U_{ns}^\dagger(\vec k) \, C_{ss'}(\vec k) \, U_{s'n'}(\vec k) = C_{nn'}(\vec k) \,.
\end{equation}
It obeys the equation of motion
\begin{equation}
 \big( {-\ii \omega} + e_n(\vec k)/\hbar \h \big) \, C_{n n'}(\vec k, \omega) = (\hbar\beta)^{-1}\h\delta_{n n'} \,,
\end{equation}
and is therefore given explicitly by
\begin{equation}
 C_{n n'}(\vec k, \omega) = \delta_{n n'} \, \frac{1}{\beta} \,\frac{1}{-\ii\hbar \omega + e_n(\vec k)} \,.
\end{equation}
Here and in the following, we denote by
\begin{equation} \label{def_e}
 e_n(\vec k) \equiv E_n(\vec k) - \mu
\end{equation}
the eigenvalues of the single-particle Hamiltonian measured from the chemical potential.

The {\itshape scale-dependent covariance} is now defined in the band basis by
\begin{equation} \label{scdep}
 (C_\Lambda)_{n n'}(\vec k, \omega) = \delta_{n n'} \, \frac{1}{\beta} \, \frac{\chi_\Lambda(e_n(\vec k))}{-\ii\hbar \omega + e_n(\vec k)} \,,
\end{equation}
where $\chi_\Lambda$ denotes the {\itshape regulator function}. The latter can be chosen either as a {\itshape strict cut-off function,} which is a smooth function with the properties that
\begin{equation} \label{strict}
 \chi_\Lambda(e) = \left\{ \begin{array}{ll} 0 \,, & \textnormal{if} \ \, |e| < 0.5 \h \Lambda \,, \\[6pt]
 1 \,, & \textnormal{if} \ \, |e| > 1.5 \h \Lambda \,. \end{array} \right.
\end{equation}
For this choice, the numerator of Eq.~\eqref{scdep} vanishes if
\begin{equation}
 |e_n(\vec k)| \equiv |E_n(\vec k) - \mu | < 0.5 \h \Lambda \,,
\end{equation}
which means that all momenta inside a shell of thickness~$\Lambda$ around the Fermi lines are {\itshape cut off.} 
For the concrete implementation of the RG equations, we will use instead the regulator function
\begin{equation} \label{eq_regulator}
\chi_\Lambda(e) = \Big( 10^{(\Lambda - |e|)/(0.05\Lambda)} + 1 \h \Big)^{\!-1} .
\end{equation}
This is always greater than zero and smaller than one, hence Eq.~\eqref{strict} holds only up to terms of the order $10^{-10}$. Correspondingly, in Eq.~\eqref{scdep} all momenta inside a shell of thickness~$\Lambda$ around the Fermi lines are {\itshape suppressed} (but not cut off). We will, however, do our calculations at a tiny positive temperature (such that $\beta \hh t = 10^{10}$), where $\chi_\Lambda$ can be used down to scales $\Lambda \approx 10^{-10} \h t$. 

The meaning of the regulator function can most easily be explained by referring to the strict cut-off function: The scale-dependent covariance \eqref{scdep} approaches the original free two-point Green function in the {\itshape infrared limit},
\begin{equation} \label{ir}
 \lim_{\Lambda \to 0} C_\Lambda = C \,.
\end{equation}
Furthermore, by defining the {\itshape ultraviolet scale} (or {\itshape initial scale}) $\Lambda_0$ much larger than the bandwidth of the model, such that
\begin{equation} \label{eq_cond}
 |E_n(\vec k) - \mu| < 0.5 \h \Lambda_0 \quad \textnormal{for all } n \textnormal{ and } \vec k \,,
\end{equation}
the covariance vanishes at this scale, i.e.,
\begin{equation} \label{uv}
 C_{\Lambda_0} = 0 \,.
\end{equation}
Now, the scale dependence of the covariance induces by means of the Feynman graph expansion a scale dependence of all interacting temperature Green functions. In particular, the one-line irreducible Green functions (see Appendix \ref{subsec_oli}) become scale dependent,
\begin{equation}
 \varGamma^{2n} = \varGamma_{\Lambda}^{2n} \,.
\end{equation}
Again, the infrared limit $\Lambda \to 0$ simply yields back the original one-line irredu\-{}cible Green functions: for $n \geq 1$,
\begin{equation}
\lim_{\Lambda \to 0} \varGamma_\Lambda^{2n}(x_1, \ldots, x_{2n}) = \varGamma^{2n}(x_1, \ldots, x_{2n}) \,.
\end{equation}
Here and in the following,
\begin{equation} \label{multiindexx}
 x = (\vec R, s, \tau)
\end{equation}
denotes a multi-index composed of the lattice site $\vec R$, the spin variable $s$ and the imaginary-time variable $\tau$. On the other hand, at the ultraviolet scale $\Lambda = \Lambda_0$, the one-line irreducible two-point function $\varGamma_\Lambda^2$ is not well defined. This is because $\varGamma_\Lambda^2$ is the inverse of the (full) two-point Green function $G_{\Lambda}^2$ (see Eqs.~\eqref{n1a} and \eqref{n1b}), but the covariance $C_\Lambda$ and hence also $G_\Lambda^2$ vanish identically at the ultraviolet scale. For~\mbox{$n \geq 2$,} the one-line irreducible Green functions $\varGamma^{2n}_\Lambda$ can still be defined at the ultraviolet scale, because the inverse of $C_\Lambda$ does not appear in their Feynman graph expansions (see Appendix \ref{subsec_oli}). In particular, for $n = 2$ we obtain \footnote{Note the prefactor $(-2\beta)$ in Eq.~\eqref{incond}: The factor 2 comes from the fact that there are indeed two first-order Feynman graphs contributing to $\varGamma^4$ (see Table \ref{tab_feyn}),
while $(-\beta)^k$ appears as a prefactor of every \mbox{$k$-th} order Feynman graph. In particular, both sides of Eq.~\eqref{incond} have the same dimension, because by our conventions the interaction kernel $V$ has the dimension of an energy, whereas all Green functions are dimensionless (see Appendix \ref{app_not}).}
\begin{align} \label{incond}
 \varGamma_{\Lambda_0}^4(x_1, \ldots, x_4) = -2 \hh \beta \h V^0(x_1, \ldots x_4) \,,
\end{align}
where $V^0(x_1, \ldots x_4)$ is the interaction kernel given by
\begin{align}
 & V^0(x_1, x_2, x_3, x_4) = V^0_{s_1 \ldots s_4}(\vec R_1, \ldots \vec R_4) \label{is_int_ker} \\[5pt] \nonumber
 & \times (\hbar\beta)^3 \, \delta(\tau_2 - \tau_3) \h \delta(\tau_1 - \tau_3) \h \delta(\tau_2 -\tau_4) \,,
\end{align}
(see Appendix \ref{subsec_Grassmann}), and where
\begin{align}
 & V^0_{s_1 \ldots s_4}(\vec R_1, \ldots \vec R_4) \\[5pt]
 & = \frac{U}{2} \h ( \delta_{s_1 s_3} \h \delta_{s_2 s_4} - \delta_{s_1 s_4} \h \delta_{s_2 s_3} ) \, \delta_{\vec R_1, \vec R_2} \, \delta_{\vec R_2, \vec R_3} \, \delta_{\vec R_3, \vec R_4} \nonumber
\end{align}
is the inverse Fourier transform of Eq.~\eqref{eq_onsite}. Furthermore, for $n \geq 3$ we have
\begin{equation}
 \varGamma_{\Lambda_0}^{2n}(x_1, \ldots, x_{2n}) = 0 \,,
\end{equation}
because all Feynman graphs contributing to $\varGamma^{2n}_\Lambda$ contain at least one internal line carrying the covariance $C_\Lambda$.

\begin{table*}[ht]
\normalsize
\begin{tabular}{c}
\hline\hline \\[0.1cm]
\hspace{0.2cm} \includegraphics{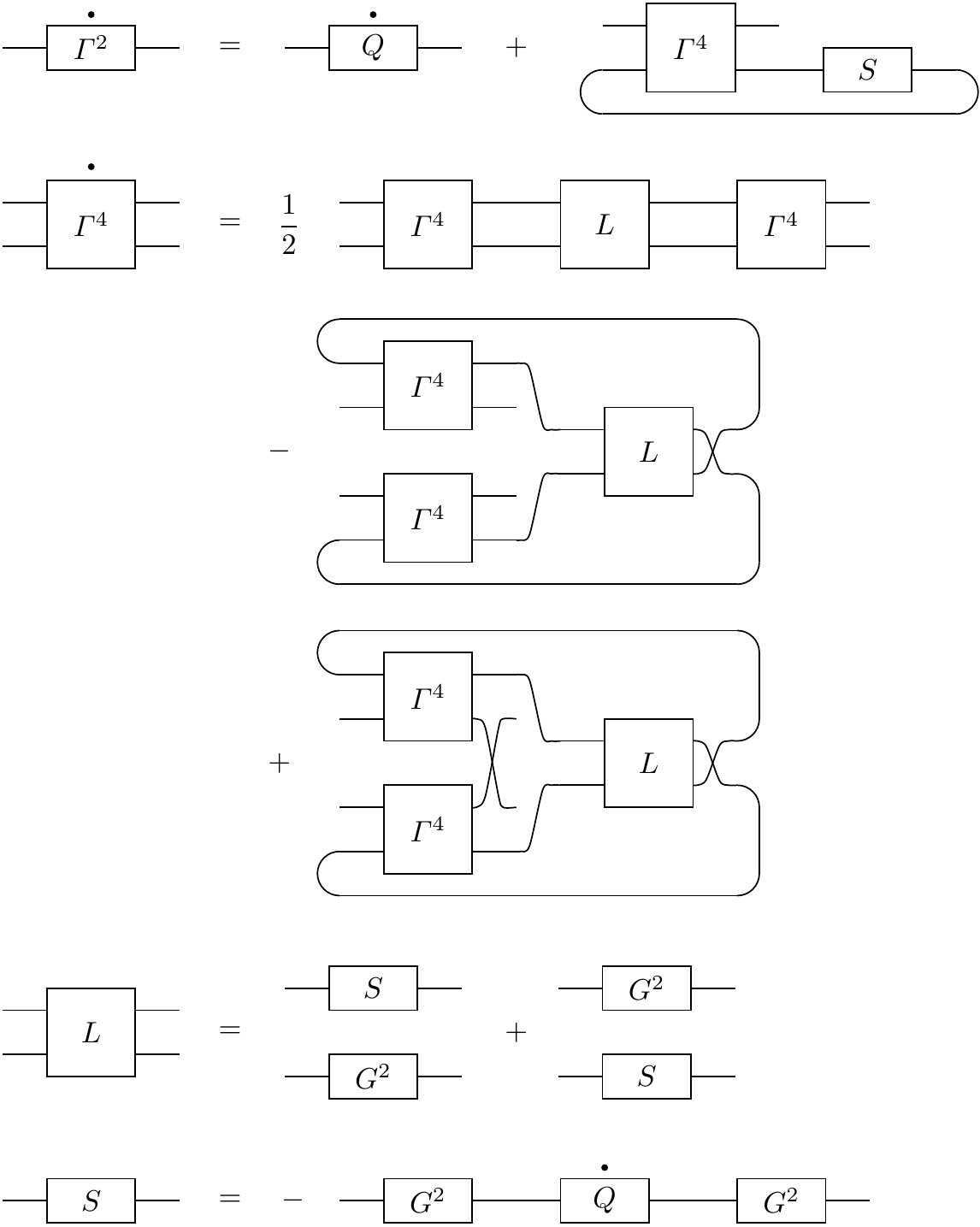} \hspace{0.2cm} \\[0.5cm]
\hline\hline
\end{tabular}

\vspace{5pt}
\caption{RG equations for the one-line irreducible Green functions in the level-two truncation. The three terms on the right-hand side of the second equation are the particle-particle term, the crossed particle-hole term, and the direct particle-hole term (in this order).} \label{tab_rge_oli}

\end{table*}

The {\itshape RG equations in the one-line irreducible scheme} constitute an infinite hierarchy of coupled differential equations, which is exactly solved by the scale-dependent one-line irreducible Green functions $\varGamma_\Lambda^{2n}$ (see Ref.~\onlinecite{SH00}).
In practice, this hierarchy has to be {\itshape truncated} in order to allow for approximate solutions. Our above-mentioned standard truncation is the {\itshape level-two truncation}, where one keeps only the two- and the four-point function in the flow and sets
\begin{equation}
 \varGamma^{2n} = 0 \qquad \textnormal{for} \ n \geq 3 \,. 
\end{equation}
The RG equations in the level-two truncation read explicitly as follows: For $n = 1$,
\begin{equation} \label{eq_rge_oli_2}
\begin{aligned}
 & \dot \varGamma^2_\Lambda(x_1, x_2) = \dot Q_\Lambda(x_1, x_2) \\[5pt]
 & + \int \! \de y_1 \int \! \de y_2 \,\h \varGamma^4_\Lambda(x_1, y_1, x_2, y_2) \, S_\Lambda(y_2, y_1) \,.
\end{aligned}
\end{equation}
Here, the dot denotes the derivative with respect to the scale parameter $\Lambda$. Furthermore, \vspace{-1pt}
\begin{equation}
 Q_\Lambda = (C_\Lambda)^{-1} \medskip
\end{equation}
is the inverse of the scale-dependent covariance, and $S_\Lambda$ is
the \emph{single-scale Green function}. The latter is defined in terms of $Q_\Lambda$ and the scale-dependent (full) two-point Green function $G_\Lambda^2$ as the operator product
\begin{equation} \label{eq_def_ssgreen}
 S_\Lambda = -G^2_{\Lambda} \, \dot Q_\Lambda \h G^2_{\Lambda} \,.
\end{equation}
In terms of integral kernels, this means
\begin{equation}
\begin{aligned}
 & S_\Lambda(x, x') \\[2pt]
 & = -\int \! \de y \int \! \de y' \, G^2_\Lambda(x, y) \, \dot Q_\Lambda(y, y') \, G^2_\Lambda(y', x') \,,
\end{aligned}
\end{equation}
where the integrations over multi-indices (see Eq.~\eqref{multiindexx}) are defined as
\begin{equation} \label{int_multi_index}
 \int \! \de x = \sum_{\vec R} \sum_s \frac{1}{\hbar\beta} \int_0^{\hbar\beta} \! \de \tau \,.
\end{equation}
For $n = 2$, the RG equation in the level-two truncation reads as \vspace{-2pt}
\begin{equation} \label{eq_rge_oli_4}
\begin{aligned}
 & \dot \varGamma^4_\Lambda(x_1, \ldots, x_4) \\[5pt]
 & = \big[ \varPhi^{\rm pp}_\Lambda + \varPhi^{\rm ph, c}_\Lambda + \varPhi^{\rm ph, d}_\Lambda \h \big](x_1, \ldots, x_4) \,,
\end{aligned}
\end{equation}
where the three terms on the right-hand side are called the \emph{particle-particle term}, the {\itshape crossed particle-hole term} and the \emph{direct particle-hole term}. These three terms are given explicitly by \vspace{-0.3cm}
\begin{widetext}
\begin{align}
 \varPhi^{\rm pp}_\Lambda(x_1, x_2, x_3, x_4) & = \frac{1}{2} \h \int \! \de y_1 \ldots \int \! \de y_4 \,\h L_\Lambda(y_1, y_2, y_3, y_4) \, \varGamma^4_\Lambda(x_1, x_2, y_1, y_2) \, \varGamma^4_\Lambda(y_3, y_4, x_3, x_4) \,, \label{eq_pp} \\[8pt]
 \varPhi_\Lambda^{\rm ph, c}(x_1, x_2, x_3, x_4) & = -\mh\int \! \de y_1 \ldots \int \! \de y_4 \,\h L_\Lambda(y_3, y_4, y_2, y_1) \, \varGamma^4_\Lambda(y_1, x_1, y_3, x_3) \, \varGamma^4_\Lambda(x_2, y_2, x_4, y_4) \,, \label{eq_ph} \\[10pt]
 \varPhi_\Lambda^{\rm ph, d}(x_1, x_2, x_3, x_4) & = -\h\varPhi_\Lambda^{\rm ph, c}(x_1, x_2, x_4, x_3) \,.
\end{align}

\vspace{0.4cm} \noindent
Here, the \emph{loop function} $L_\Lambda$ is defined in terms of the full two-point Green function $G_{\Lambda}^2$ and the single-scale Green \end{widetext}
function $S_\Lambda$ by
\begin{equation} \label{eq_def_lgreen}
\begin{aligned}
 L_\Lambda(y_1, y_2, y_3, y_4)&  = S_\Lambda(y_1, y_3) \, G^2_{\Lambda}(y_2, y_4) \\[5pt]
 & \quad \, + G^2_{\Lambda}(y_1, y_3) \, S_\Lambda(y_2, y_4) \,. \vspace{0.5cm}
\end{aligned}
\end{equation}
Graphically, the RG equations \eqref{eq_rge_oli_2} and \eqref{eq_rge_oli_4} are represented by means of {\itshape universal Feynman graphs} in Table~\ref{tab_rge_oli} (see Ref.~\onlinecite{Ronald}, and Table \ref{tab_feyn} for our conventions).

In order to simplify the RG equations, we first switch to the Fourier domain, where the above equations \eqref{eq_rge_oli_2}--\eqref{eq_def_lgreen} hold in precisely the same form but with the multi-indices~$x$ replaced by
\begin{equation}
 k = (\vec k, s, \omega) \,.
\end{equation}
Here, $\vec k$ ranges in the first Brillouin zone $\mathcal B$, and $\omega$ is a
fermionic Matsubara frequency (see Appendix \ref{sec_temperature_green_functions}). Then, the summations over $\vec R$ get replaced by integrations over $\vec k$, and the $\tau$-integrals get replaced by summations over the Matsubara frequencies, i.e.,
\begin{equation}
 \int \! \de k = \frac{1}{|\mathcal B|} \h \int_{\mathcal B} \de^2 \vec k \h \sum_s \sum_{\omega} \,.
\end{equation}
We further employ translation invariance, which effectively reduces the number of arguments of all Green functions by one (see Appendix \ref{sec_temperature_green_functions}). Next, we switch from the spin basis to the band basis by means of the unitary matrix $U_{sn}(\vec k)$ from Sec.~\ref{sec_Rashba}. For example, the four-point function in the band basis is defined in terms of its counterpart in the spin basis by (cf.~Eq.~\eqref{spintoband})
\begin{equation}
\begin{aligned}
 & (\varGamma_\Lambda^4)_{n_1 \ldots n_4}(\vec k_1, \omega_1; \vec k_2, \omega_2; \vec k_3, \omega_3) \\[5pt]
 & = \sum_{s_1, \ldots, s_4} (\varGamma_\Lambda^4)_{s_1 \ldots s_4}(\vec k_1, \omega_1; \vec k_2, \omega_2; \vec k_3, \omega_3) \\[2pt]
 & \quad \, \times U_{n_1 s_1}^\dagger(\vec k_1) \, U_{n_2 s_2}^\dagger(\vec k_2) \, U_{s_3 n_3}(\vec k_3) \, U_{s_4 n_4}(\vec k_4) \,.
\end{aligned}
\end{equation}
Furthermore, we employ the following approximations to the RG equations (besides the level-two truncation), 
which have already been established in many works before:\cite{Metzner, Platt} We neglect the self-energy $\varSigma_\Lambda$, which is 
defined by the equation (cf.~\eqref{def_self_energy})
\begin{equation}
 (G_\Lambda^2)^{-1} = (C_\Lambda)^{-1} - \varSigma_\Lambda \,.
\end{equation}
This means, we replace the full two-point Green function $G_\Lambda^2$ by the covariance $C_\Lambda$. Thus, the RG equation for the four-point function $\varGamma_\Lambda^4$ becomes 
closed and decouples from the equation for $\varGamma^2_\Lambda$\h. The single-scale Green function simplifies to
\begin{equation}
 S_\Lambda = -C_\Lambda \h \dot Q_\Lambda \h C_\Lambda = \dot C_\Lambda \,,
\end{equation}
and the loop function becomes (in a symbolic notation)
\begin{equation}
 L_\Lambda = S_\Lambda \otimes C_\Lambda + C_\Lambda \otimes S_\Lambda \,.
\end{equation}
Moreover, we neglect the frequency dependencies of the four-point function, i.e., we replace
\begin{equation}
\begin{aligned}
 & (\varGamma_\Lambda^4)_{n_1 \ldots n_4}(\vec k_1, \omega_1\h ;\h \vec k_2, \omega_2\h ;\h \vec k_3, \omega_3) \\[5pt]
 & \mapsto \, (\varGamma_\Lambda^4)_{n_1 \ldots n_4}(\vec k_1, \vec k_2, \vec k_3) \,. 
\end{aligned}
\end{equation}
This approximation in combination with the choice of a momentum regulator (see Eq.~\eqref{scdep}) allows us to perform the remaining frequency sums on the right-hand side of the RG equations analytically and thereby to obtain explicit expressions for the particle-particle and particle-hole loops (see Eq.~\eqref{def_lmp} below).
Finally, we reformulate the RG equation in terms of the scale-dependent {\itshape effective interaction} (or {\itshape interaction vertex}) $V_\Lambda$, which is related to the one-line irreducible 
four-point Green function by
\begin{equation} \label{fund_rel}
V_\Lambda \equiv -\frac{1}{2\beta} \h \varGamma^4_\Lambda \,.
\end{equation}
From this definition and from Eq.~\eqref{incond}, we see that the initial condition for $V_\Lambda$ at the ultraviolet scale is precisely given by the original interaction kernel,
\begin{equation} \label{in1}
\begin{aligned}
 (V_{\Lambda_0})_{n_1 \ldots n_4}(\vec k_1, \vec k_2, \vec k_3) \equiv V^0_{n_1 \ldots n_4}(\vec k_1, \vec k_2, \vec k_3) \,. \\ \phantom{a}
\end{aligned}
\end{equation}
The latter is given in the spin basis by Eq.~\eqref{eq_onsite} and in the band basis by Eq.~\eqref{spintoband}. With these simplifications, the RG equations for the interaction vertex read as follows:
\begin{equation} \label{rg_eq}
\begin{aligned}
 & \frac{\de}{\de \Lambda} (V_\Lambda)_{n_1 n_2 n_3 n_4}(\vec p_1, \vec p_2, \vec p_3) \\[6pt]
 & = \big[ \varPhi_\Lambda^{\rm pp} + \h \varPhi_\Lambda^{\rm ph, c} + \h \varPhi_\Lambda^{\rm ph, d} \h \big]_{n_1 n_2 n_3 n_4} (\vec p_1, \vec p_2, \vec p_3) \,,
\end{aligned}
\end{equation}
where the particle-particle term, the crossed particle-hole term, and the direct particle-hole term are given by\cite{SH00}
\begin{widetext}
\begin{align}
 (\varPhi_\Lambda^{\rm pp})_{n_1 n_2 n_3 n_4}(\vec p_1, \vec p_2, \vec p_3) & = -{\sum_{\ell_1, \h \ell_2}} \frac{1}{|\mathcal B|} \int_{\mathcal B} \! \de^2 \vec k_1 \int_{\mathcal B} \! \de^2 \vec k_2 \, \sum_{\vec K} \delta^2(\vec K + \vec p_1 + \vec p_2 - \vec k_1, \vec k_2) \label{eq_phipp} \\[5pt] \nonumber
 & \qquad \quad \, \times (L_\Lambda^-)_{\ell_1 \ell_2}(\vec k_1, \vec k_2) \,\h (V_\Lambda)_{n_1 n_2 \ell_1 \ell_2}(\vec p_1, \vec p_2, \vec k_1) \,\h (V_\Lambda)_{\ell_1 \ell_2 n_3 n_4}(\vec k_1, \vec k_2, \vec p_3) \,, \\[15pt]
 (\varPhi_\Lambda^{\rm ph, c})_{n_1 n_2 n_3 n_4}(\vec p_1, \vec p_2, \vec p_3) \label{eq_phiphc} & = -2 \sum_{\ell_1, \h \ell_2} \frac{1}{|\mathcal B|} \int_{\mathcal B} \! \de^2 \vec k_1 \int_{\mathcal B} \! \de^2 \vec k_2 \, \sum_{\vec K} \delta^2(\vec K + \vec p_1 - \vec p_3 + \vec k_1, \vec k_2) \\[5pt] \nonumber
 & \qquad \quad \, \times (L_\Lambda^+)_{\ell_2 \ell_1}(\vec k_2, \vec k_1) \,\h (V_\Lambda)_{\ell_1 n_1 \ell_2 n_3}(\vec k_1, \vec p_1, \vec k_2) \,\h (V_\Lambda)_{n_2 \ell_2 \ell_1 n_4}(\vec p_2, \vec k_2, \vec k_1) \,, \\[15pt]
 (\varPhi_\Lambda^{\rm ph, d})_{n_1 n_2 n_3 n_4}(\vec p_1, \vec p_2, \vec p_3)& = -(\varPhi_\Lambda^{\rm ph, c})_{n_2 n_1 n_3 n_4}(\vec p_2, \vec p_1, \vec p_3) \,. \label{eq_phiphd}
\end{align}

\vspace{10pt} \noindent
In these equations, the {\itshape particle-particle loop} $L_\Lambda^-$, and the {\itshape particle-hole loop} $L_\Lambda^{+}$ are given by
\begin{equation} \label{def_lmp}
 (L_\Lambda^\mp)_{\ell_1 \ell_2}(\vec k_1, \vec k_2)= \frac{\de}{\de\Lambda} \h \Big( \chi_\Lambda(e_{\ell_1}\mh(\vec k_1) ) \,\h \chi_\Lambda(e_{\ell_2}\mh(\vec k_2) ) \Big) \, (F_\Lambda^\mp)_{\ell_1 \ell_2}(\vec k_1, \vec k_2) \,, \vspace{0.5cm}
\end{equation}
\end{widetext}
with the functions $F_\Lambda^{\mp}$ defined by
\begin{equation} \label{disc_2a}
 F^-_{\ell_1 \ell_2}(\vec k_1, \vec k_2) = \frac{1 - f(e_{\ell_1}\mh(\vec k_1) ) - f(e_{\ell_2}\mh(\vec k_2) )}{e_{\ell_1}\mh(\vec k_1) + e_{\ell_2}\mh(\vec k_2)} \,,
\end{equation}
and respectively
\begin{equation} \label{disc_2b}
 F^+_{\ell_1 \ell_2}(\vec k_1, \vec k_2) = \frac{f(e_{\ell_1}\mh(\vec k_1) ) - f(e_{\ell_2}\mh(\vec k_2) )}{e_{\ell_1}\mh(\vec k_1) - e_{\ell_2}\mh(\vec k_2)} \,.
\end{equation}
Recall that $e_\ell(\vec k) = E_\ell(\vec k) - \mu$ are the eigenvalues of the single-particle Hamiltonian measured relative to the chemical potential. Furthermore,
\begin{equation}
 f(e) = \frac{1}{\e^{\beta e} + 1}
\end{equation}
denotes the Fermi distribution function, which in the zero-temperature limit $(\beta \to \infty)$ reduces to
\begin{equation}
 f(e) = \varTheta(- e) \quad \ (\beta \to \infty)
\end{equation}
(except at $e = 0$), where $\varTheta$ denotes the Heaviside step function. Note that in Eqs.~\eqref{eq_phipp}--\eqref{eq_phiphd}, the reciprocal lattice vector $\vec K$ is fixed in each term by the condition that all external momenta $\vec p_1, \vec p_2, \vec p_3$ and all internal momenta $\vec k_1, \vec k_2$ lie in the Brillouin zone $\mathcal B$. The vector $\vec k_2$ is then fixed by the Dirac delta distribution, and hence the right-hand side of the RG equation effectively requires only a summation over two band indices $\ell_1, \ell_2$ and an integration over one Bloch momentum $\vec k_1$.

The {\itshape initial value problem} for the scale-dependent interaction vertex is defined by the RG equation \eqref{rg_eq} and the initial condition \eqref{in1}. This initial value problem has a unique solution, which, however, may typically not be continued down to $\Lambda = 0$. In fact, as mentioned above, the interaction vertex $V_\Lambda$ diverges already at a nonzero scale $\Lambda_{\rm c}$, signaling an instability towards an ordered phase. We define the stopping scale $\Lambda_{\rm *}$ (which is greater, but close to $\Lambda_{\rm c}$) as the scale where the supremum of $V_\Lambda$ exceeds a certain threshold value $S$. More precisely, with the scale-dependent {\itshape vertex supremum}
\begin{equation}
 V^{\rm sup}(\Lambda) := \sup \h \{ \h | \h (V_{\Lambda})_{n_1 \ldots n_4}(\vec k_1, \vec k_2, \vec k_3) \h |\h \} \,,
\end{equation}
we define the stopping condition for the RG flow as
\begin{equation} \label{stopcond}
 V^{\rm sup}(\Lambda_*) \h = \h S \,.
\end{equation}
The effective interaction at the stopping scale $\Lambda_*$ can then be transformed back into the spin basis by using the unitary matrix $U_{sn}(\vec k)$ from Sec.~\ref{sec_Rashba}, i.e.,
\begin{align}
 & (V_\Lambda)_{s_1 \ldots s_4}(\vec k_1, \vec k_2, \vec k_3) = \sum_{n_1, \ldots, n_4} U_{s_1 n_1}(\vec k_1) \, U_{s_2 n_2}(\vec k_2) \nonumber \\[5pt]
 & \times (V_\Lambda)_{n_1 \ldots n_4}(\vec k_1, \vec k_2, \vec k_3) \, U_{n_3 s_3}^\dagger(\vec k_3) \, U_{n_4 s_4}^\dagger(\vec k_4) \,.
\end{align}

\pagebreak \noindent
In our implementation, we use a threshold of
\begin{equation} \label{val_th}
 S / t = 40 \,,
\end{equation}
which is more than an order of magnitude larger than the initial interaction, $|U/2| \hh /\hh t = 1$ (see Eq.~\eqref{u}, and note that by Eq.~\eqref{eq_onsite} the supremum of the initial interaction kernel is indeed $|U/2|$\h).

\subsection{Fermi surface patching approximation} \label{sec_FSpatching}

In order to solve the RG equations numerically, we discretize the momentum dependence of the interaction vertex $V_\Lambda$. A standard approximation is {\itshape Fermi surface patching,} where each momentum is projected on a representative momentum on the Fermi surface (or {\itshape Fermi line} in two dimensions).\cite{SH00, Metzner} By projecting the three momentum arguments of the interaction vertex to the Fermi line, the RG equations can be reformulated in terms of finitely many parameters. For one-band systems, this fixes the projection unambiguously, whereas for multiband systems, one still has to specify the dependence on the band indices. In this work, we use a {\itshape refined projection scheme} that makes no assumption on the relative importance of the contributions from different bands.

\begin{figure}[t]
\smallskip
\includegraphics[width=0.75\columnwidth]{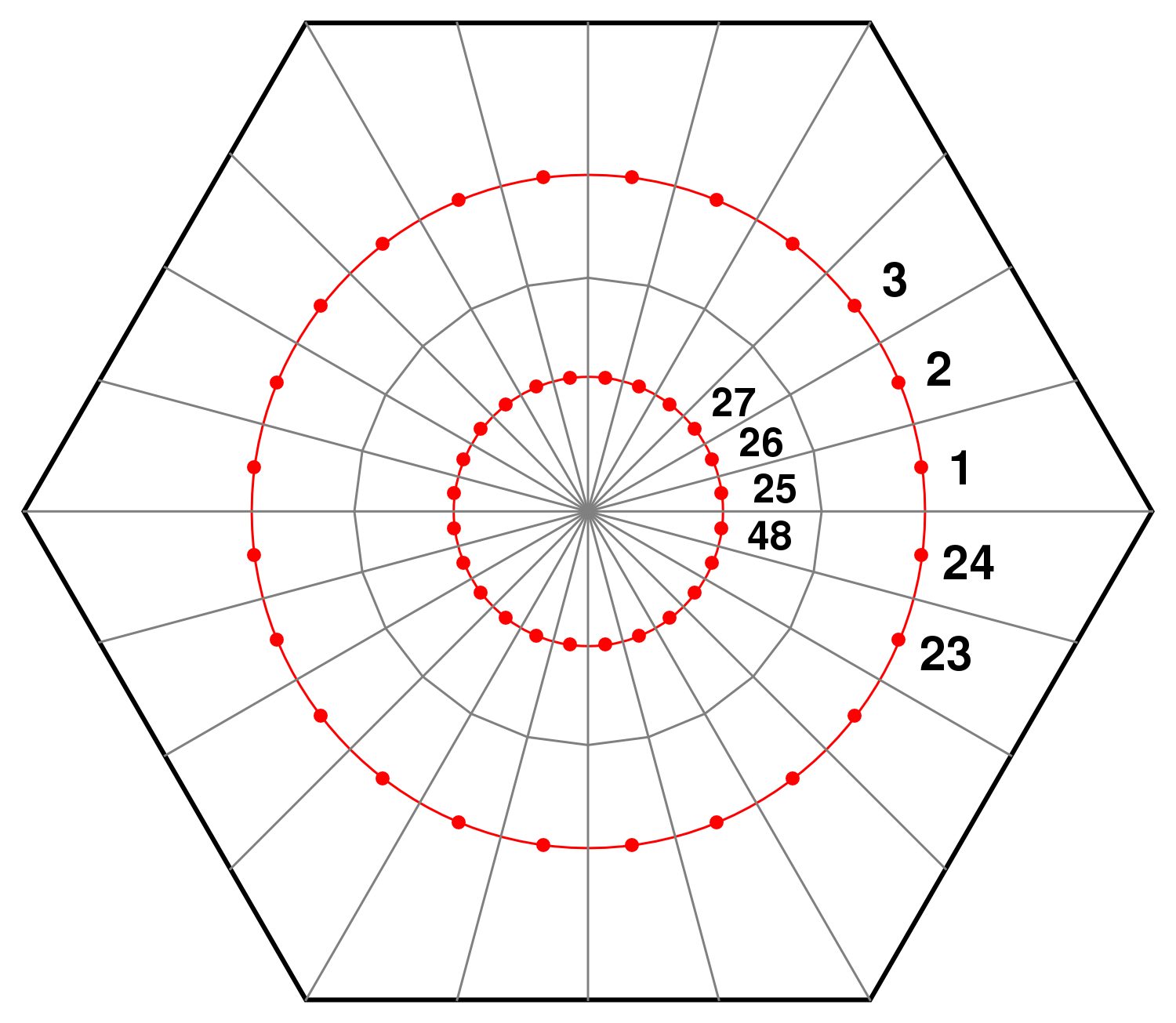}
\caption{Division of the Brillouin zone into $N = 48$ patches and representative momenta on the two Fermi lines. The latter are only schematically represented here as perfect circles, which is indeed a good approximation for small Fermi energies (near the band crossing, see~Fig.~\ref{fig_disp}). The patches are labeled counterclockwise, with patches on the outer Fermi line having smaller indices than those on the inner Fermi line. \label{fig_patches}}
\end{figure}

 To explain this in more detail, we choose for each band~$n$ intersected by the Fermi level a number $N_n$ of {\itshape representative momenta} $\vec \pi_{\alpha}^{n}$, $\alpha \in \{1, \ldots, N_n\}$, which lie on the respective Fermi line of the  band $n$. Furthermore, we denote by
\begin{equation}
 N = \sum_{n} N_n
\end{equation}
the total number of representative momenta, and let
\begin{equation}
 i \in \{1, \ldots, N_1, N_1 + 1, \ldots, N \}
\end{equation}
be an index which labels {\itshape all} representative momenta $\vec \pi_i$ (on all Fermi lines). This means, we identify
\begin{equation}
\begin{aligned}
 & \vec \pi_{1} \equiv \vec \pi_1^1 \,, \ \ldots \,, \ \vec \pi_{N_1} \equiv \vec \pi_{N_1}^1 \,, \\[5pt]
 & \vec \pi_{N_1 + 1} \equiv \vec \pi_{1}^2 \,, \ \ldots \,, \ \vec \pi_{N_1 + N_2} \equiv \vec \pi_{N_2}^2 \,, \\[5pt]
 & \ldots \\[5pt]
 & \vec \pi_{N_1 + \ldots + N_{n-1} + 1} \equiv \vec \pi_{1}^n \,, \ \ldots \,, \ \vec \pi_N \equiv \vec \pi_{N_n}^n \,.
\end{aligned}
\end{equation}
In our refined projection scheme, we divide the Brillouin zone~$\mathcal B$ into $N$ disjoint 
patches,
\begin{equation}
 \mathcal B = \bigcup_{i=1}^N \mathcal B_i \,,
\end{equation}
where the patch $\mathcal B_i$ is defined as the set of all momenta in $\mathcal B$ which lie closer to the representative momentum $\vec \pi_i$ than to any other representative momentum (see Fig.~\ref{fig_patches}). We then make the following ansatz, which assumes the interaction vertex in the band basis to be constant on each patch:
\begin{equation} \label{ansatz}
\begin{aligned}
 & (V_\Lambda)_{n_1\ldots n_4}(\vec k_1, \vec k_2, \vec k_3) \\[3pt]
 & = \sum_{i_1 = 1}^N \h \sum_{i_2 = 1}^N \h \sum_{i_3 = 1}^N \h (V_\Lambda)_{n_1\ldots n_4}(i_1, i_2, i_3) \\[5pt]
 & \quad \, \times \mathbbm 1(\vec k_1 \in \mathcal B_{i_1}) \, \mathbbm 1(\vec k_2 \in \mathcal B_{i_2}) \, \mathbbm 1(\vec k_3 \in \mathcal B_{i_3}) \,,
\end{aligned}
\end{equation}
where $\mathbbm 1$ is the characteristic function defined by
\begin{equation}
\mathbbm 1(\vec k \in \mathcal B_i) = \left\{ \begin{array}{ll} 1 \,, & \textnormal{if} \ \, \vec k \in \mathcal B_i \,, \\[3pt]
 0 \,, & \textnormal{otherwise} \,, \end{array} \right.
\end{equation}
and where $V_\Lambda$ is approximated patch-wise by its value at the representative momenta, i.e.,
\begin{equation} \label{thiseq}
 (V_\Lambda)_{n_1\ldots n_4}(i_1, i_2, i_3) \equiv (V_\Lambda)_{n_1\ldots n_4}(\vec \pi_{i_1}, \vec \pi_{i_2}, \vec \pi_{i_3}) \,.
\end{equation}
Thus, we are left with a finite set of parameters for the interaction vertex labeled by four band indices $n_1, \ldots, n_4$ and three patch indices $i_1, \ldots, i_3$. Now, for each fixed combination $(n_1, \ldots, n_4)$ of band indices, the projected interaction vertex \eqref{ansatz} gets contributions from \linebreak $N \times N \times N$ combinations $(\vec \pi_{i_1}, \vec \pi_{i_2}, \vec \pi_{i_3})$ of representative momenta. For example, the representative momentum $\vec \pi_{i_1}$ may lie on the Fermi line of {\itshape any} band $m_1$, i.e., not necessarily $m_1 = n_1$. Therefore, if $L$ denotes the number of bands (in our case $L = 2$), there are in total $N^3 \times L^4$ complex numbers---given by Eq.~\eqref{thiseq}---which parametrize the interaction vertex.

For a clearer comparison, let us contrast our projection scheme with the projection scheme described, e.g., in Ref.~\onlinecite{Platt}: There, one divides the Brillouin zone~$\mathcal B$ for each band $n$ separately into $N_n$ disjoint patches,
\begin{equation}
 \mathcal B = \bigcup_{\alpha = 1}^{N_n} \mathcal B_\alpha^n \qquad (\textnormal{for each} \ n) \,.
\end{equation}
Each representative momentum $\vec \pi_{\alpha}^{n}$ lies on the Fermi line of the band $n$ within the patch $\mathcal B_{\alpha}^{n}$. One then parametrizes the interaction vertex as follows:
\begin{equation} \label{ansatz_w}
\begin{aligned}
 & (V_\Lambda)_{n_1\ldots n_4}(\vec k_1, \vec k_2, \vec k_3) \\[3pt]
 & = \sum_{\alpha_1 = 1}^{N_{n_1}} \h \sum_{\alpha_2 = 1}^{N_{n_2}} \h \sum_{\alpha_3 = 1}^{N_{n_3}} \h (V_\Lambda)_{n_1\ldots n_4}(\alpha_1, \alpha_2, \alpha_3) \\[5pt]
 & \quad \, \times \mathbbm 1(\vec k_1 \in \mathcal B_{\alpha_1}^{n_1}) \, \mathbbm 1(\vec k_2 \in \mathcal B_{\alpha_2}^{n_2}) \, \mathbbm 1(\vec k_3 \in \mathcal B_{\alpha_3}^{n_3}) \,,
\end{aligned}
\end{equation}
where
\begin{equation} \label{thateq}
\begin{aligned}
 & (V_\Lambda)_{n_1\ldots n_4}(\alpha_1, \alpha_2, \alpha_3) \\[5pt]
 & \equiv (V_\Lambda)_{n_1\ldots n_4}(\vec \pi_{\alpha_1}^{n_1}, \vec \pi_{\alpha_2}^{n_2}, \vec \pi_{\alpha_3}^{n_3}) \,.
\end{aligned} \smallskip
\end{equation}
Now, for each fixed combination $(n_1, \ldots, n_4)$ of band indices, the projected interaction vertex \eqref{ansatz_w} gets contributions from only $N_{n_1} \times N_{n_2} \times N_{n_3}$ combinations of representative momenta. For example, $\vec \pi_{\alpha_1}^{n_1}$ in Eq.~\eqref{thateq} necessarily lies on the Fermi line of the band $n_1$. If the fourth band index $n_4$ is not further specified (and hence allowed to take values in each band of the model), then this projection scheme leaves $N^3 \times L$ complex numbers parametrizing the interaction vertex. If $n_4$ is fixed by another condition (e.g.~by requiring that $\vec \pi_{\alpha_4}^{n_4}$ is the representative momentum with the smallest distance to $\vec k_4 \equiv \vec k_1 + \vec k_2 - \vec k_3$), then even only $N^3$ parameters of the interaction vertex remain. This is by a factor of $L^4$ smaller than the number of parameters in our refined projection scheme. 

For the Rashba model considered in this article, however, it turns out that all $N^3 \times L^4$ parameters of the discretized inter\-{}action vertex need to be considered in the RG flow. This means, only the refined projection ansatz \eqref{ansatz} yields a meaningful approximation to the exact solution of the initial value problem described in the previous subsection (the RG equation \eqref{rg_eq} together with the initial condition \eqref{in1}). By contrast, the ansatz \eqref{ansatz_w} with a reduced number of parameters leads to a qualita\-{}tively different result for the effective interaction (see the discussion in the following subsections \ref{sec_scint}--\ref{sec_pp}).

Finally, we derive the explicit RG equations for the discretized interaction vertex in the refined projection scheme, which can be directly implemented numerically (see Appendix \ref{sec_num}). By putting our ansatz \eqref{ansatz} into Eqs. \eqref{rg_eq}--\eqref{eq_phiphd}, we obtain the following approximate equations 
for the finitely many parameters of the interaction vertex:
\begin{equation} \label{disc_1}
\begin{aligned}
 & \frac{\de}{\de\Lambda} \, (V_\Lambda)_{n_1 n_2 n_3 n_4}(i_1, i_2, i_3) \\[5pt]
 & = \big[ \varPhi_\Lambda^{\rm pp} + \h \varPhi_\Lambda^{\rm ph, c} + \h \varPhi_\Lambda^{\rm ph, d} \h \big]_{n_1 n_2 n_3 n_4} (i_1, i_2, i_3)\,, 
\end{aligned} \smallskip
\end{equation}
where the three terms on the right-hand side are, respectively, given by
\begin{widetext}
\begin{align}
(\varPhi_\Lambda^{\rm pp})_{n_1 n_2 n_3 n_4}(i_1, i_2, i_3) \label{eq_18}
 & = -{\sum_{\ell_1, \h \ell_2}} \ \sum_{j_1 = 1}^N \,\h \sum_{j_2 = 1}^N \,\h \sum_{\vec K} \mathbbm 1(\vec K + \vec \pi_{i_1} + \vec \pi_{i_2} - \vec \pi_{j_1} \in \mathcal B_{j_2}) \ (L^-_\Lambda)_{\ell_1 \ell_2}(i_1, i_2, j_1) \\[5pt] \nonumber
 & \qquad \quad \, \times \Big[ \h (V_\Lambda)_{n_1 n_2 \ell_1 \ell_2}(i_1, i_2, j_1) \ (V_\Lambda)_{\ell_1 \ell_2 n_3 n_4}(j_1, j_2, i_3) \, + \, (j_1, \ell_1) \h \leftrightarrow \h (j_2, \ell_2) \, \Big] \,, \\[10pt]
 (\varPhi_\Lambda^{\rm ph, c})_{n_1 n_2 n_3 n_4}(i_1, i_2, i_3) \label{eq_20}
 & = {-\h2} \h\sum_{\ell_1, \h \ell_2} \,\h \sum_{j_1 = 1}^N \,\h \sum_{j_2 = 1}^N \,\h \sum_{\vec K} \mathbbm 1(\vec K + \vec \pi_{i_1} - \vec \pi_{i_3} + \vec \pi_{j_1} \in \mathcal B_{j_2}) \\[5pt] \nonumber
 & \qquad \quad \, \times (L^+_\Lambda)_{\ell_1 \ell_2}(i_1, i_3, j_1) \ (V_\Lambda)_{\ell_1 n_1 \ell_2 n_3}(j_1, i_1, j_2) \ (V_\Lambda)_{n_2 \ell_2 \ell_1 n_4}(i_2, j_2, j_1) \\[10pt] \nonumber
 & \quad {-\h2} \h\sum_{\ell_1, \h \ell_2} \,\h \sum_{j_1 = 1}^N \,\h \sum_{j_2 = 1}^N \,\h \sum_{\vec K} \mathbbm 1(\vec K + \vec \pi_{i_3} - \vec \pi_{i_1} + \vec \pi_{j_1} \in \mathcal B_{j_2}) \\[5pt] \nonumber
 & \qquad \quad \, \times (L_\Lambda^+)_{\ell_1 \ell_2}(i_3, i_1, j_1) \ (V_\Lambda)_{\ell_2 n_1 \ell_1 n_3}(j_2, i_1, j_1) \ (V_\Lambda)_{n_2 \ell_1 \ell_2 n_4}(i_2, j_1, j_2) \,, \\[10pt]
 (\varPhi_\Lambda^{\rm ph, d})_{n_1 n_2 n_3 n_4}(i_1, i_2, i_3) & = -(\varPhi_\Lambda^{\rm ph, c})_{n_2 n_1 n_3 n_4}(i_2, i_1, i_3) \,. \label{eq_21}
\end{align}
\end{widetext}
Here, we have defined
\begin{equation} \label{eq_22}
\begin{aligned}
 & (L^\mp_\Lambda)_{\ell_1 \ell_2}(i_1, i_2, j_1) = \frac{1}{|\mathcal B|} \int_{\mathcal B_{j_1}} \!\!\! \de^2 \vec k \\[3pt]
 & \times \dot \chi_\Lambda(e_{\ell_1}\mh(\vec k)) \,\h \chi_\Lambda(e_{\ell_2}\mh(\vec K + \vec \pi_{i_1} \pm \vec \pi_{i_2} \mp \vec k)) \\[7pt]
 & \times F^\mp_{\ell_1 \ell_2}(\vec k, \vec K + \vec \pi_{i_1} \pm \vec \pi_{i_2} \mp \vec k) \,,
\end{aligned}
\end{equation}
with the functions $F^{\mp}_{\ell_1 \ell_2}$ defined by Eqs.~\eqref{disc_2a}--\eqref{disc_2b}. 
Note that in the particle-particle term \eqref{eq_18}, the reciprocal lattice vector $\vec K$ is fixed by the condition
\begin{equation}
 \vec K + \vec \pi_{i_1} + \vec \pi_{i_2} - \vec \pi_{j_1} \in \mathcal B \,,
\end{equation}
and therefore depends on only three patch indices $i_1$, $i_2$, and $j_1$. The stricter condition
\begin{equation}
 \vec K + \vec \pi_{i_1} + \vec \pi_{i_2} - \vec \pi_{j_1} \in \mathcal B_{j_2}
\end{equation}
then also fixes the patch index $j_2$. Hence, the right-hand side of Eq.~\eqref{eq_18} effectively requires only the summation over two band indices $\ell_1, \ell_2$ and one patch index $j_1$. The same applies to the particle-hole terms.

In our numerical implementation, we have directly solved the RG equations \eqref{disc_1}--\eqref{eq_21} for the discretized interaction vertex. The solution $V_\Lambda$ with the given initial interaction $V_{\Lambda_0}$ can formally be written as
\begin{align}
 V_\Lambda & = V_{\Lambda_0} + \int_{\Lambda_0}^{\Lambda} \de \Lambda \, \frac{\de}{\de \Lambda} \h V_\Lambda \\[5pt]
 & = V_{\Lambda_0} + \int_{\Lambda_0}^{\Lambda} \de \Lambda \, \big[ \varPhi_\Lambda^{\rm pp} + \h \varPhi_\Lambda^{\rm ph, c} + \h \varPhi_\Lambda^{\rm ph, d} \h \big] \,.
\end{align}
The scale integral has been performed numerically by starting at the initial scale $\Lambda_0$ and stepwise determining $V_{\Lambda+ \de \Lambda}$ from the previously calculated $V_\Lambda$. The integration steps $\de \Lambda$ have been adjusted in each step depending on how fast the interaction vertex changes in the flow. In this way, the divergence at the critical scale could be approached numerically by gradually decreasing the step size. We remark that our implementation for the Rashba model uses $48$ patch momenta, $24$ on each of the two Fermi lines as shown schematically in Fig.~\ref{fig_patches}. We have checked that our results do not change significantly by further increasing the number of patches, i.e., by taking 72 instead of 48 patches.

\subsection{Effective superconducting interaction} \label{sec_scint}

\begin{figure}[t]
\includegraphics[width=\columnwidth]{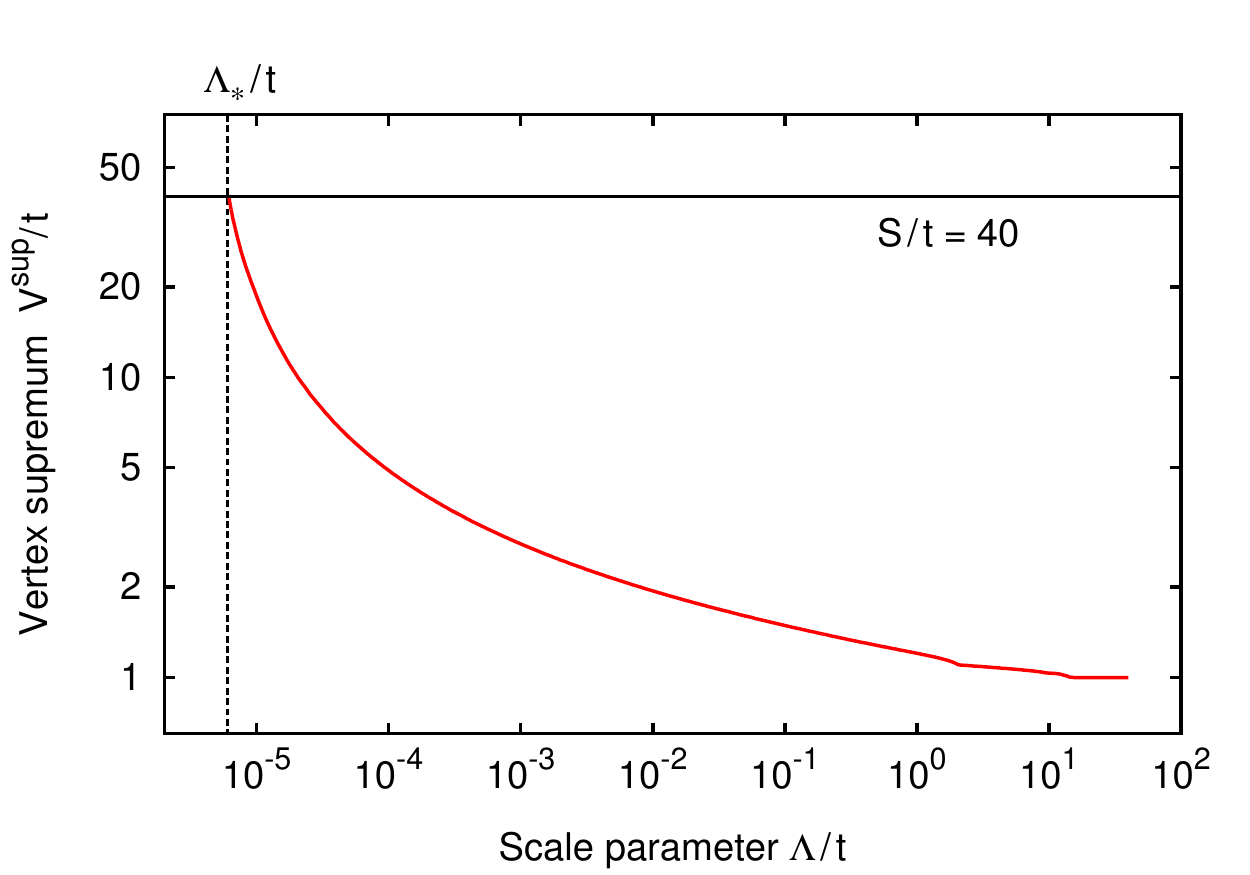}
\caption{Double-logarithmic plot of the scale-dependent vertex supremum $V^{\rm sup}(\Lambda)$, for the chemical potential $\mu / t = -2$. The~RG flow is stopped at the scale $\Lambda_*$ where $V^{\rm sup}$ exceeds the threshold parameter $S$. \label{fig_vs}}
\end{figure}

\begin{figure}
\includegraphics[width=0.45\columnwidth]{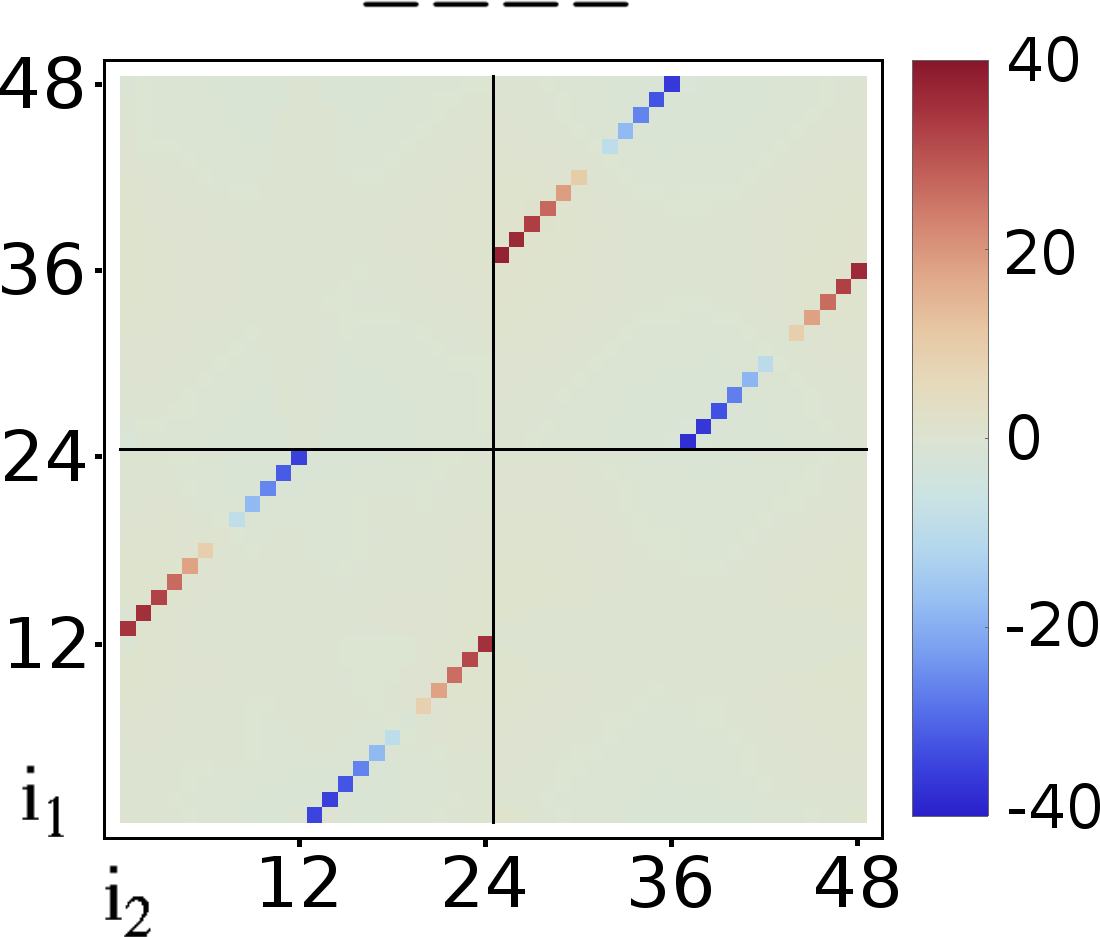}\hspace{0.08\columnwidth}\includegraphics[width=0.45\columnwidth]{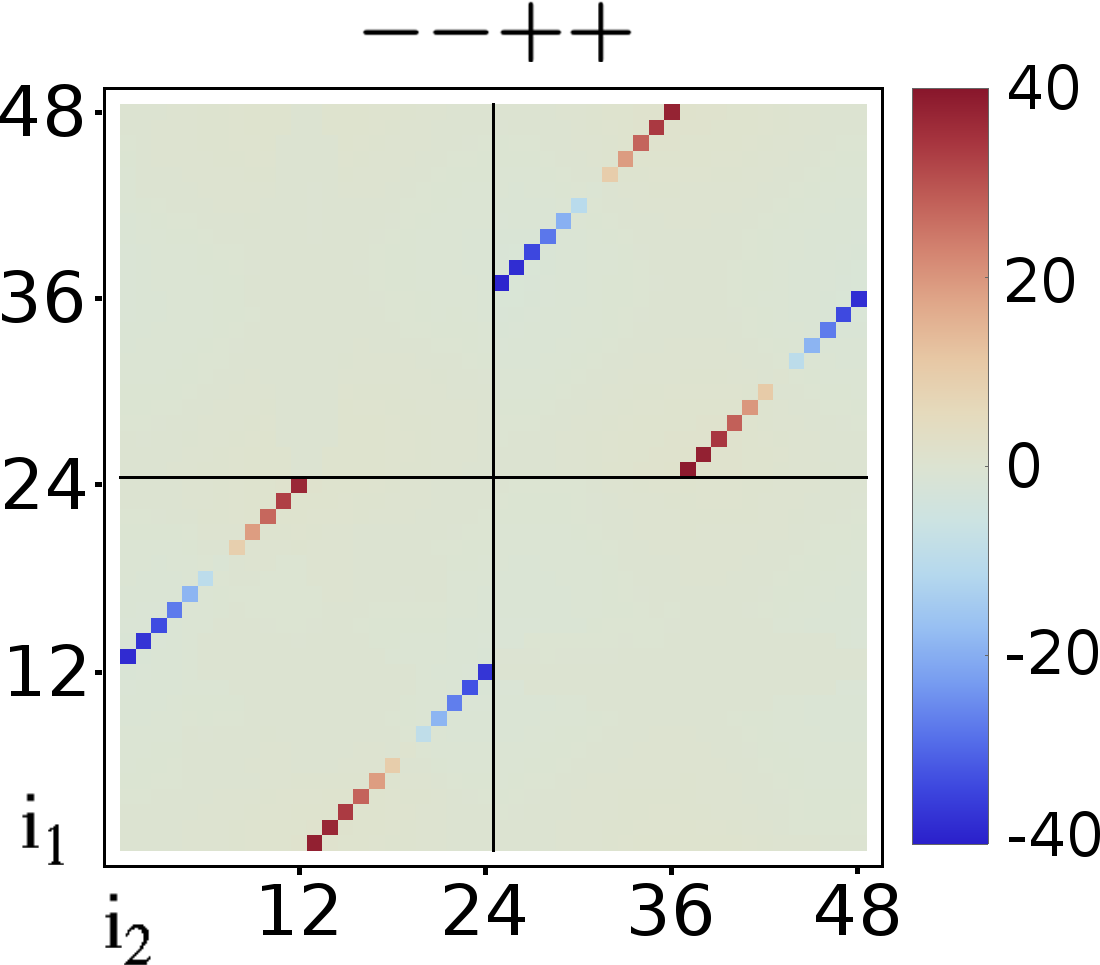}

\bigskip
\smallskip
\includegraphics[width=0.45\columnwidth]{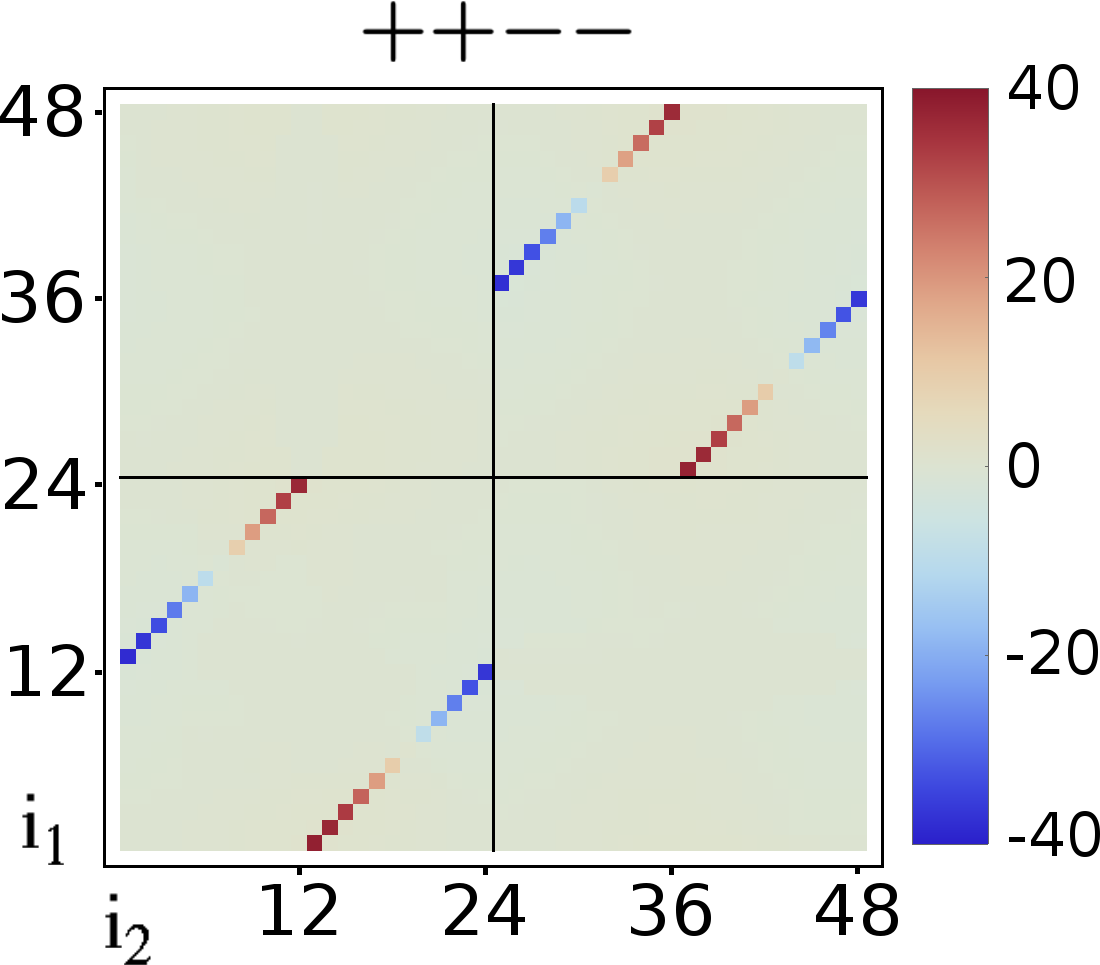}\hspace{0.08\columnwidth}\includegraphics[width=0.45\columnwidth]{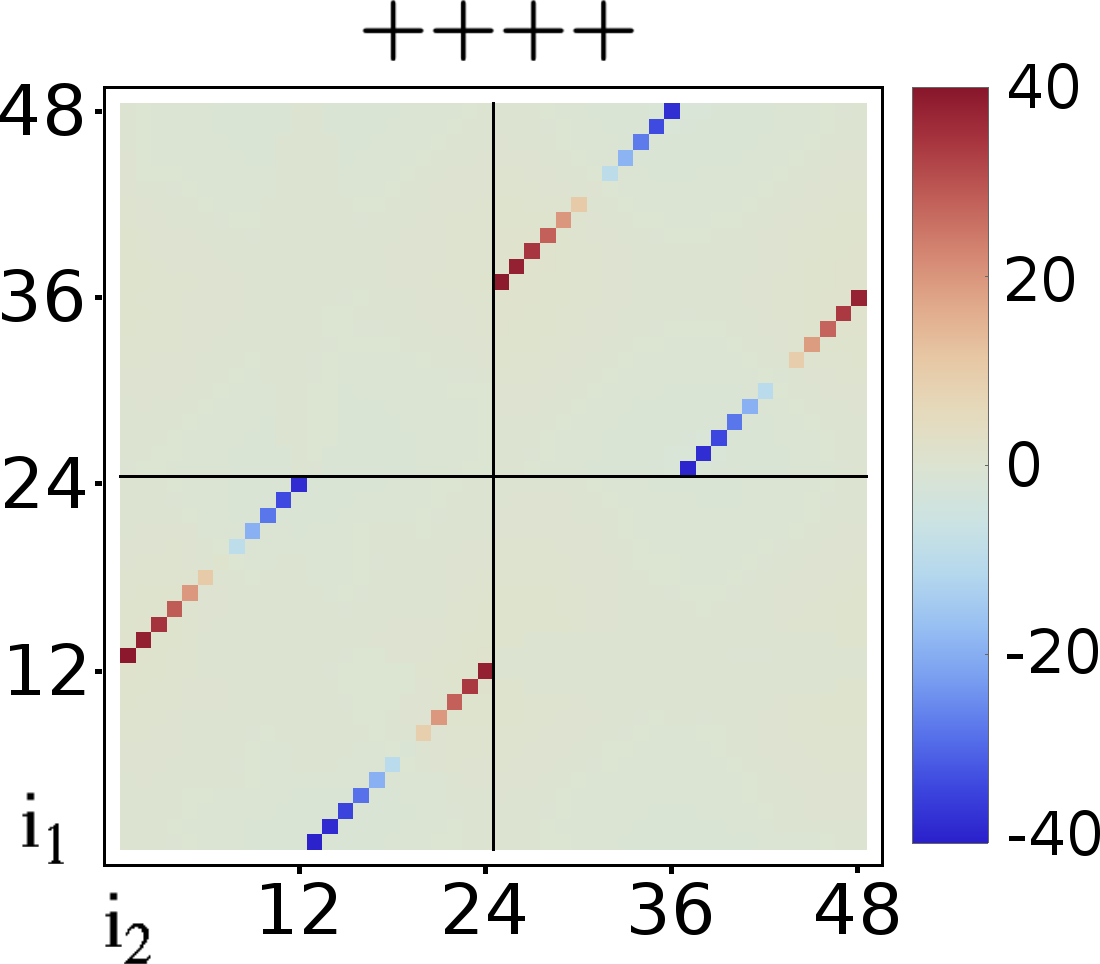}

\medskip
\caption{Real part of the interaction vertex in the band basis, $\mathrm{Re} \, (V_\Lambda)_{n_1 n_2 n_3 n_4}(i_1, i_2, i_3) / t$, after following the RG flow down to the stopping scale $\Lambda = \Lambda_{*}$ (for $\mu/t = -2$). Shown are the four non-vanishing contributions with band indices $n_1 \hh n_2 \hh n_3 \hh n_4$ and the dependence on two patch indices $i_1$ and $i_2$ (while the third patch index is fixed as $i_3 = 1$). The patches are labeled as shown schematically in Fig.~\ref{fig_patches}. \label{fig_veff_pos_band}}
\end{figure}

\begin{figure} \vspace{-2pt}
\includegraphics[width=0.45\columnwidth]{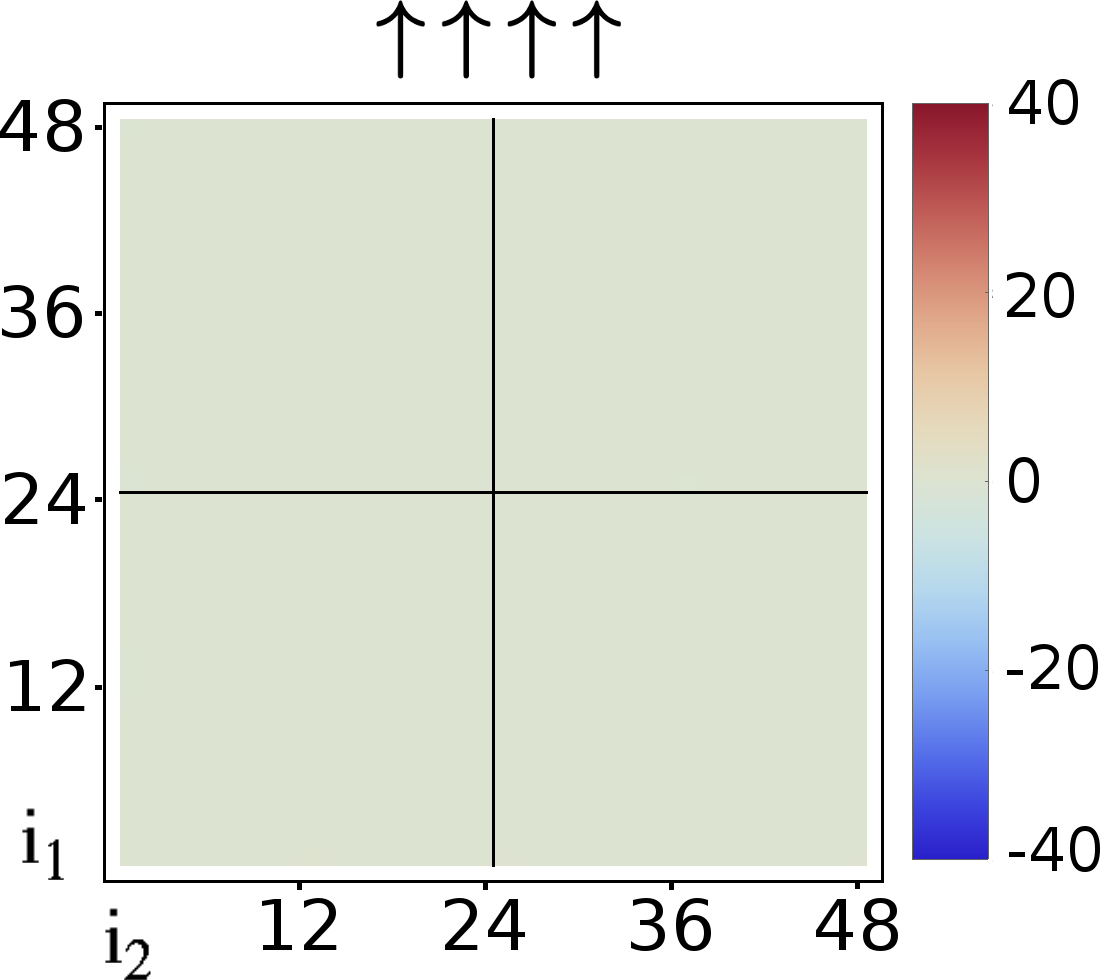}\hspace{0.08\columnwidth}\includegraphics[width=0.45\columnwidth]{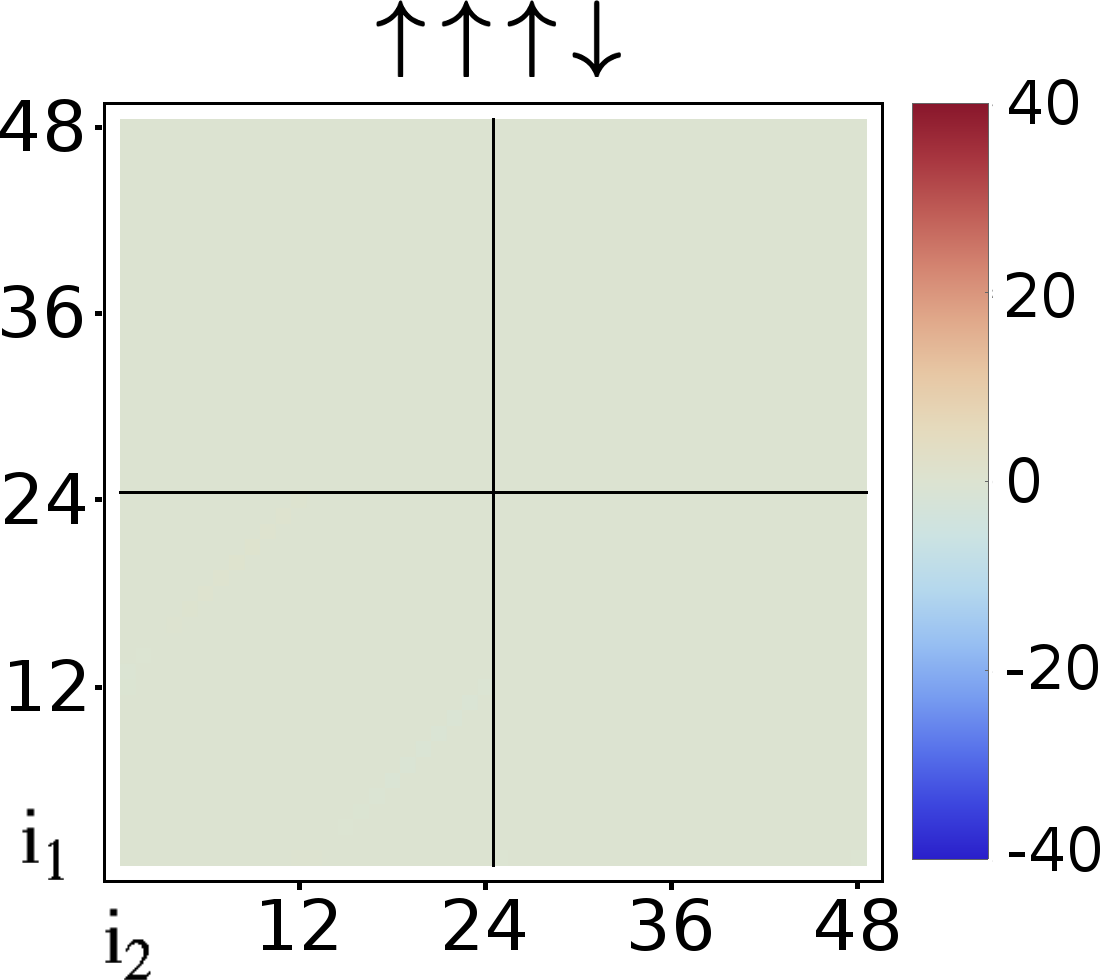}

\bigskip
\smallskip
\includegraphics[width=0.45\columnwidth]{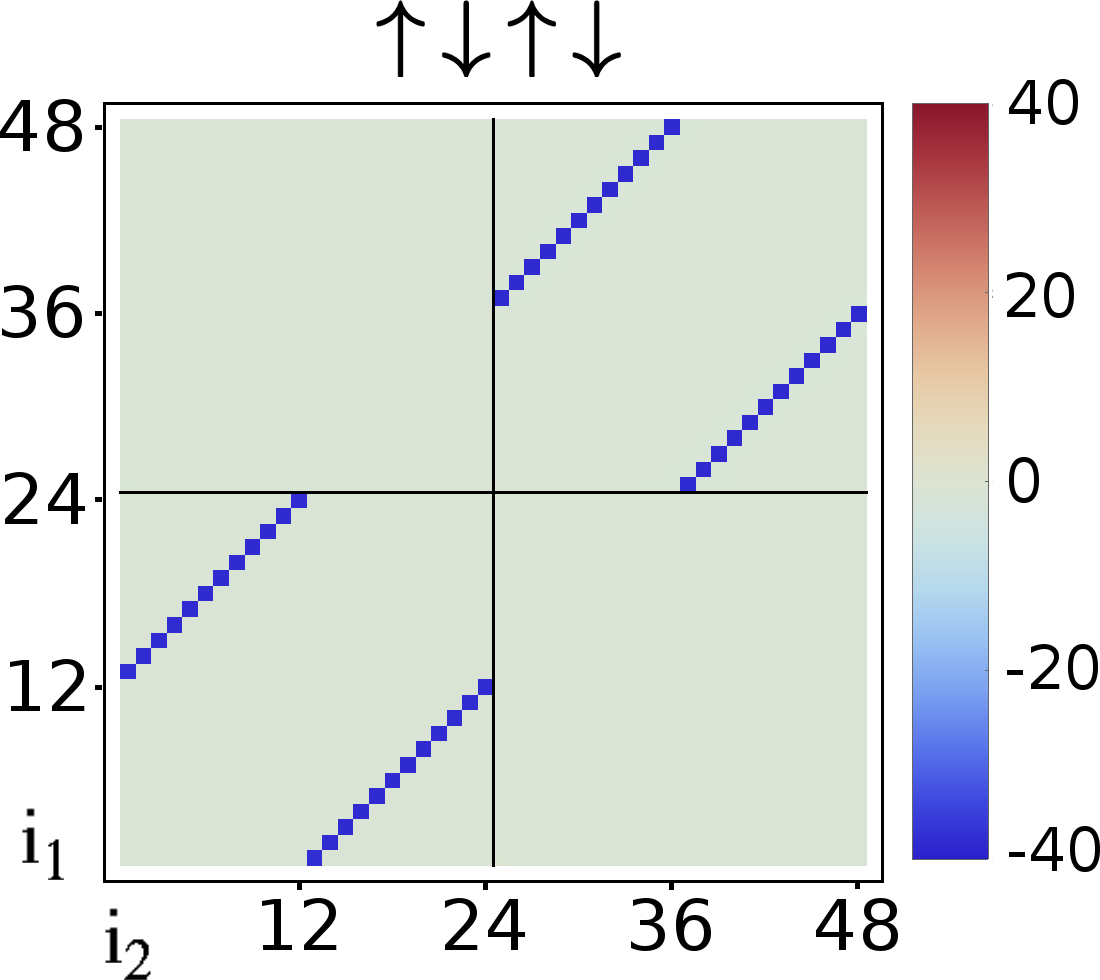}\hspace{0.08\columnwidth}\includegraphics[width=0.45\columnwidth]{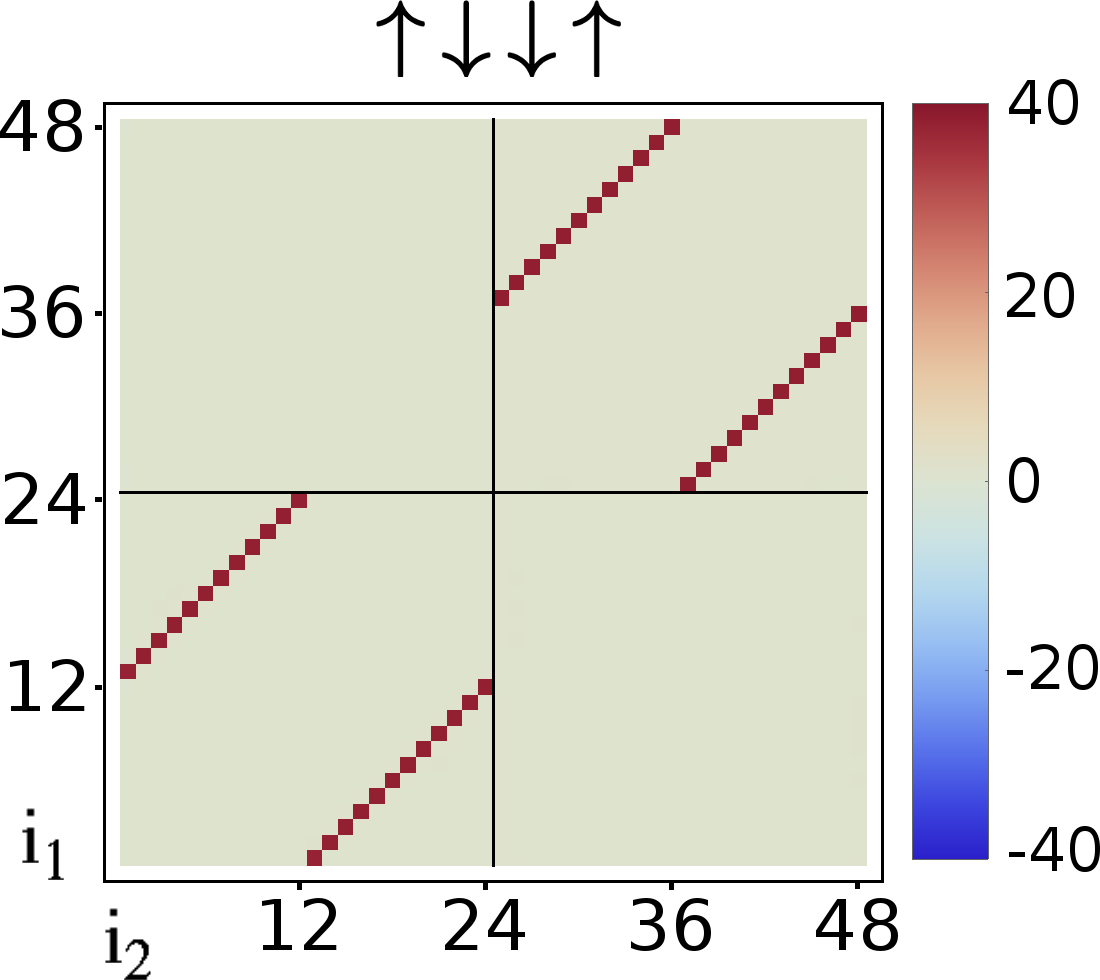}

\medskip
\caption{Interaction vertex $(V_\Lambda)_{s_1 s_2 s_3 s_4}(i_1, i_2, i_3) / t$ \,in the spin basis at the stopping scale $\Lambda = \Lambda_{\rm *}$ (for $\mu/t = -2$). Shown are four representative spin configurations $s_1 s_2 s_3 s_4$ and the dependence on two patch indices $i_1$ and $i_2$ (while $i_3 = 1$). The patches are labeled again as shown in Fig.~\ref{fig_patches}. \label{fig_veff_pos}}
\end{figure}

As described above, we have numerically solved the discretized RG equations \eqref{disc_1}--\eqref{eq_21} with an attractive onsite interaction at the initial scale $\Lambda_0$. The latter was chosen much larger than the bandwidth of the model (given by Eq.~\eqref{bandwidth}), i.e.,
\begin{equation}
 \Lambda_0 / t = 40 \,.
\end{equation}
Thus, the condition \eqref{eq_cond} is fulfilled, and our results do not change by further increasing $\Lambda_0$. We have stopped the RG flow at the stopping scale $\Lambda_*$ defined by Eq.~\eqref{stopcond}, which is close to the critical scale $\Lambda_{\rm c}$ where the interaction vertex diverges. (As mentioned above, in the following, we do not distinguish explicitly between these two scales.) Our numerical result for the vertex supremum $V^{\rm sup}(\Lambda)$ as a function of the scale parameter $\Lambda$ is shown in Fig.~\ref{fig_vs}. One clearly sees that the interaction vertex grows with decreasing $\Lambda$ and eventually approaches a divergence at the critical scale. The numerical result for the effective interaction at the stopping scale is shown in the band basis in Fig.~\ref{fig_veff_pos_band} and in the spin basis in Fig.~\ref{fig_veff_pos}. We have fixed the third patch index $i_3 = 1$ and analyzed the dependence of the effective interaction on $i_1$ and $i_2$
for all possible band and spin combinations (of which four representative ones are shown in Figs.~\ref{fig_veff_pos_band} and \ref{fig_veff_pos}, respectively). The result clearly signals a superconducting instability, where pairing occurs between opposite momenta on the same Fermi line. 
The discretized effective interaction at the stopping scale is well represented in the band basis~by
\begin{equation} \label{num_res_band}
\begin{aligned}
 & (V_{\Lambda_{*}})_{n_1 \ldots n_4}(i_1, i_2, i_3) \h = \h \mathbbm 1(\vec \pi_{i_1} \mh = \mh -\vec \pi_{i_2}) \\[5pt]
 & \quad \times S \h \delta_{n_1 n_2} \h \delta_{n_3 n_4} \h n_2 \h n_3 \,\h \e^{\ii \varphi(\vec \pi_{i_3}) - \ii \varphi(\vec \pi_{i_2})} \,,
\end{aligned}
\end{equation}
and in the spin basis by
\begin{equation} \label{eq_eff}
\begin{aligned}
 & (V_{\Lambda_{*}})_{s_1\ldots s_4}(i_1, i_2, i_3) \h = \h \mathbbm 1(\vec \pi_{i_1} \mh = \mh -\vec \pi_{i_2}) \\[6pt]
 & \quad \times (-S) \h ( \delta_{s_1 s_3} \delta_{s_2 s_4} - \delta_{s_1 s_4} \delta_{s_2 s_3} ) \,,
\end{aligned}
\end{equation}
where $S$ is the threshold parameter (see Eq.~\eqref{val_th}). The corresponding interaction operator (which is  obtained by inserting Eq.~\eqref{eq_eff} into the projection ansatz \eqref{ansatz}) is then approximately given by
\begin{equation} \label{to_derive}
\begin{aligned}
 \hat V_{\Lambda_*} & = \frac {S}{2N} \int \! \db^2 \vec k_2 \int \! \db^2 \vec k_3 \\[5pt]
 & \quad \, \times \sum_{s_1, \ldots, s_4} ( \delta_{s_1 s_3} \delta_{s_2 s_4} - \delta_{s_1 s_4} \delta_{s_2 s_3} ) \\[5pt]
 & \quad \, \times \hat a\dag_{s_1}(-\vec k_2) \h\hh \hat a\dag_{s_2}(\vec k_2) \h\hh \hat a_{s_3}(\vec k_3) \h\hh \hat a_{s_4}(- \vec k_3) \,,
\end{aligned}
\end{equation}
where $N$ is the number of patches. The factor $1/N$ corresponds to the area of a single $\vec k$-space patch, which arises because our effective interaction \eqref{eq_eff} turns out to have~a \h$\vec k_1 = -\vec k_2$ restriction on the level of patches (see the derivation in Appendix \ref{app_sc_int}). By explicitly performing the spin sums and using the canonical anticommutation relations of the creation and 
annihilation operators, we further obtain the equivalent expression
\begin{equation} \label{eq_eff_coeff}
\begin{aligned}
 \hat V_{\Lambda_*} & = -g \int \! \db^2 \vec k \int \! \db^2 \vec k'\\[2pt]
 & \quad \, \times \hat a_{\uparrow}\dag(-\vec k) \h\hh \hat a_{\downarrow}\dag(\vec k) \h\hh \hat a_{\downarrow}(\vec k') \h\hh \hat a_{\uparrow}(-\vec k') \,,
\end{aligned}
\end{equation}
where we have defined the coupling constant
\begin{equation}
 g := \frac{2S}{N} > 0 \,.
\end{equation}
The interaction \eqref{eq_eff_coeff} is a {\itshape singlet} superconducting interaction. We have obtained this result for the effective interaction at the critical scale independently of the chemical potential~$\mu$, whether it is above ($\mu > 0$) or below ($\mu < 0$) the band crossing of the Rashba dispersion. 

We stress that the form of the effective interaction crucially depends on the projection scheme used to discretize the scale-dependent interaction vertex (see Sec.~\ref{sec_FSpatching}). Our result given by Eqs.~\eqref{num_res_band}--\eqref{eq_eff} has been obtained by using the refined projection scheme, while a qualitatively different result would be obtained in the projection scheme of Ref.~\onlinecite{Platt}.
The reason for the difference between the two projection schemes can in fact already be understood by considering the discretized {\itshape initial} interaction. The latter is given in the spin basis by (see Eq.~\eqref{eq_onsite})
\begin{equation} \label{eq_onsite_discrete}
 V^0_{s_1 s_2 s_3 s_4}(i_1, i_2, i_3) = \frac U 2 \, ( \delta_{s_1 s_3} \h \delta_{s_2 s_4} - \delta_{s_1 s_4} \h \delta_{s_2 s_3} ) \,,
\end{equation}
and in the band basis by
\begin{equation}
\begin{aligned} \label{eq_onsite_discrete_band}
 & V^0_{n_1 \ldots n_4}(i_1, i_2, i_3) = \sum_{s_1, \ldots, s_4} U_{n_1 s_1}^\dagger(\vec \pi_{i_1}) \, U_{n_2 s_2}^\dagger(\vec \pi_{i_2}) \\[2pt]
 & \times V^0_{s_1 \ldots s_4}(i_1, i_2, i_3) \, U_{s_3 n_3}(\vec \pi_{i_3}) \, U_{s_4 n_4}(\vec \pi_{i_4}) \,,
\end{aligned}
\end{equation}
where $\vec \pi_{i_4}$ is defined by the condition
\begin{equation}
 \vec K + \vec \pi_{i_1} + \vec \pi_{i_2} - \vec \pi_{i_3} \in \mathcal B_{i_4} \,,
\end{equation}
with some reciprocal lattice vector $\vec K$. For momentum combinations where $\vec \pi_{i_1} = -\vec \pi_{i_2}$, and consequently also $\vec \pi_{i_4} = -\vec \pi_{i_3}$, we obtain the following explicit expression (analo\-{}gous to Eq. \eqref{this}):
\begin{equation} \label{prop_1}
\begin{aligned}
 & V^0_{n_1 n_2 n_3 n_4}(i_1, i_2, i_3) \\[5pt]
 & = -\frac{U}{2} \, \delta_{n_1 n_2} \h \delta_{n_3 n_4} \h n_2 \h n_3 \,\h \e^{\ii \varphi(\vec \pi_{i_3}) - \ii \varphi(\vec \pi_{i_2})} \,.
\end{aligned}
\end{equation}
Now, let $\vec \pi_{i_2}$ be a representative momentum, say, on the Fermi line of the band $m_2$. Furthermore, let $\vec \pi_{i_1} = -\vec \pi_{i_2}$ be the opposite momentum on the same Fermi line. Then, Eq.~\eqref{prop_1} implies  that
\begin{equation} \label{thisone}
 | V^0_{n_1 n_2 n_3 n_4}(i_1, i_2, i_3) | = \frac {|U|} 2 \, \delta_{n_1 n_2} \h \delta_{n_3 n_4} \,.
\end{equation}

 \ \\[-20pt]
Hence, the two components of the discretized interaction vertex with \,$n_1 = n_2 = -$ \,or \,$n_1 = n_2 = +$ are equal in magnitude. For each representative momentum $\vec \pi_{i_2}$ on the Fermi line of the band $m_2$, one therefore has to take into account both components \,$n_2 = -$ \,and $n_2 = +$ \,of the discretized interaction (not only those with $n_2 = m_2$). 
Furthermore, if we transform Eq.~\eqref{prop_1} back to the spin basis by
\begin{align}
 & V^0_{s_1 \ldots s_4}(i_1, i_2, i_3) = \sum_{n_1, \ldots, n_4} U_{s_1 n_1}(\vec \pi_{i_1}) \, U_{s_2 n_2}(\vec \pi_{i_2}) \nonumber \\[2pt]
 & \times V^0_{n_1 \ldots n_4}(i_1, i_2, i_3) \, U_{n_3 s_3}^\dagger(\vec \pi_{i_3}) \, U_{n_4 s_4}^\dagger(\vec \pi_{i_4}) \,,
\end{align}
we recover again the initial interaction \eqref{eq_onsite_discrete}. Thus, the transformation of the discretized interaction between the band basis and the spin basis is exact in the refined projection scheme. On the other hand, if we neglected all components in Eq.~\eqref{prop_1} with $n_2 \not = m_2$ (where $m_2$ is the band index of the representative momentum $\vec \pi_{i_2}$), then we would lose information on the interaction kernel, and by 
transforming the result back to the spin basis we would obtain a qualitatively different interaction.

Our numerical implementation of the RG equations further shows that the property \eqref{prop_1} remains unchanged in the flow, i.e., the scale-dependent interaction vertex $V_\Lambda$ has this property for every~$\Lambda$ (and in fact near the critical scale, all other contributions with $\vec \pi_{i_1} \not = -\vec \pi_{i_2}$ become negligible). This is most clearly seen in Fig.~\ref{fig_veff_pos_band}, which shows the four contributions
\begin{equation}
\begin{aligned}
 & (V_\Lambda)_{----} \,, \quad \ (V_\Lambda)_{--++} \,, \\[3pt]
 & (V_\Lambda)_{++--} \,, \quad \ (V_\Lambda)_{++++} \,,
\end{aligned}
\end{equation}
of the interaction vertex in the band basis at the stopping scale $\Lambda = \Lambda_{*}$. The four contributions are of equal magnitude, and the momentum dependence is well described 
by Eq.~\eqref{num_res_band}. We emphasize that this comes out even in the case where the chemical potential is in the lower band ($\mu/t = -2$ in Fig.~\ref{fig_veff_pos_band}), which implies that the upper band is completely empty (at $T = 0$). The unexpected result that even in this case contributions to the interaction vertex with an upper band index cannot be neglected in RG flow will be explained further by means of an analytical resummation of the particle-particle ladder in Sec.~\ref{sec_pp}.

\subsection{Critical scale and phase diagram}

\begin{figure}[t]
\includegraphics[width=\columnwidth]{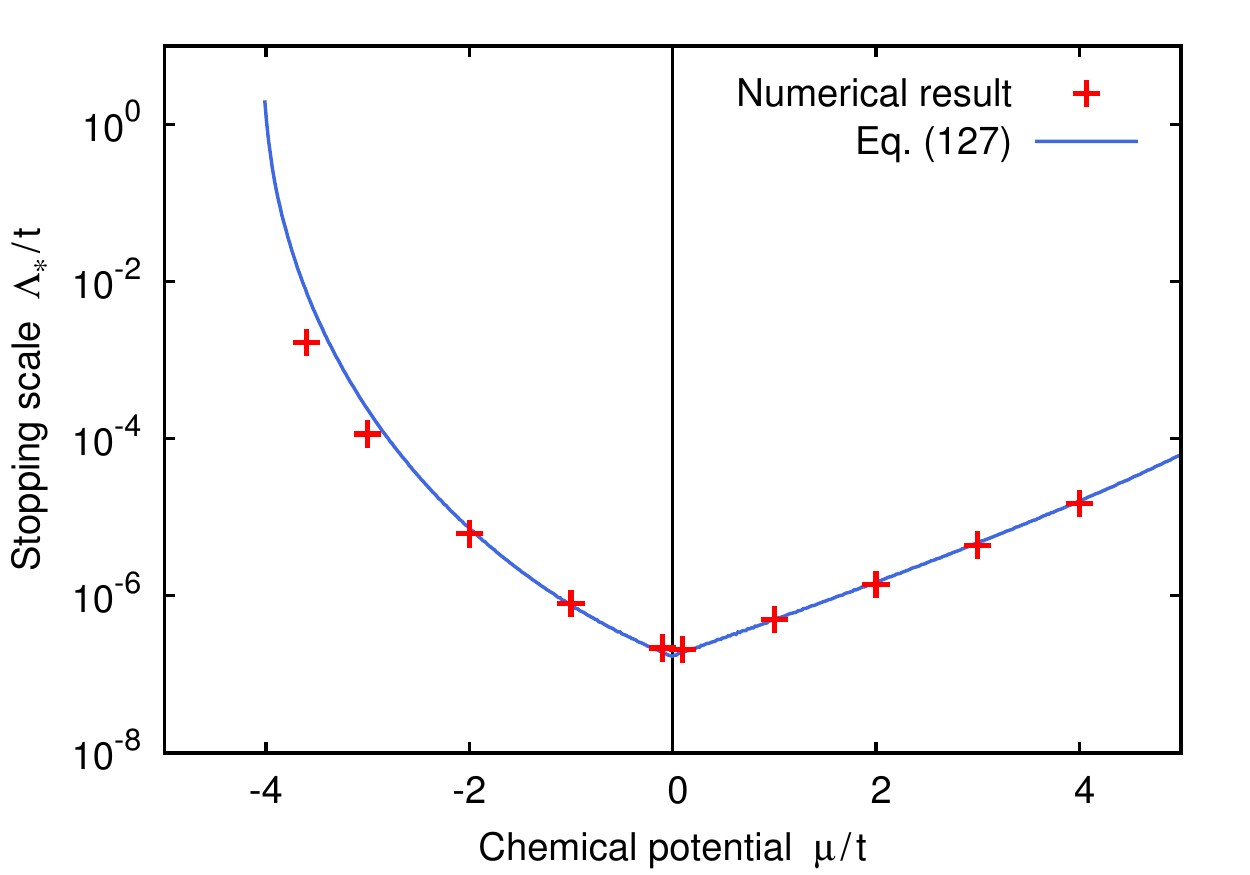}
\caption{Logarithmic plot of the stopping scale $\Lambda_{*}$ as a function of the chemical potential $\mu$, for an initial interaction of $U/t = -2$. The vertical line (where $\mu = 0$) marks the position of the band crossing of the Rashba dispersion. The red points show the stopping scales obtained from the numerical implementation of the RG flow. The blue curve corresponds to Eq.~\eqref{fit}, which can be motivated by an analytical resummation of the particle-particle ladder. \label{fig_phase}}
\end{figure}

The RG flow is stopped at the scale $\Lambda_{*}$ where the vertex exceeds a threshold value and hence a divergence is approached, which signals the breakdown of the Fermi liquid description. Figure~\ref{fig_phase} shows $\Lambda_{*}$ as a function of the chemical potential $\mu$.
The numerical data turns out to be well represented by the following formula:
\begin{equation} \label{fit}
 \Lambda_{*} / t = 5.0 \, \exp \! \left({-\frac{2}{|U| D(\mu)}} \right) , 
\end{equation}
where $U$ is the initial interaction strength (given by Eq. \eqref{u}), and $D(\mu)$ is the density of states of the minimal tight-binding model as shown in Fig.~\ref{fig_dos}. The exponent in this formula can in fact be motivated by an analytical resummation of the particle-particle ladder as performed in the next subsection (see the text following Eq.~\eqref{expon}). In particular, the sharp increase of $\Lambda_{*}$ for small $\mu$ reflects the diverging density of states at the band minimum of the Rashba dispersion, and the kink at $\mu = 0$ corresponds to the kink in the density of states at the band crossing (cf.~Eq.~\eqref{dos_explicit} for the ideal Rashba model).

We stress again that the choice of the momentum projection scheme has a strong impact on the simulational result for the form of the effective interaction near the critical scale, and even for the critical scale itself. In fact, the characteristic dependence of the critical scale on the density of states as shown in Fig.~\ref{fig_phase}---and described by Eq.~\eqref{fit}, which is consistent with the particle-particle ladder resummation---could only be reproduced in our refined projection scheme. We conclude that it is imperative to include all $N^3 \times L^4$ parameters of the interaction vertex in the RG flow, i.e., to use the refined projection scheme, in order to obtain the correct results for the effective interaction and the critical scale in our model. It would be interesting to compare the two projection schemes (described in Sec.~\ref{sec_FSpatching}) also for other models with several bands, in order to see whether their difference is 
a special feature of the Rashba model or a general issue to be considered in the Fermi surface patching approximation for multiband systems.

We conclude this subsection with a remark about the interpretation of Fig.~\eqref{fig_phase} as a ``phase diagram.'' In two dimensions, the critical scale $\Lambda_{\rm c}$ can be regarded as an estimate for a true temperature of a phase transition, in which the ordering sets in, in the following sense: In cases where the symmetry that gets broken is discrete, the breaking is strictly allowed. When, as in our case, a continuous symmetry is involved, no long-range order can exist in an infinite two-dimensional system due to the Mermin--Wagner theorem. However, for a finite (large) system, a sufficiently slow decay of correlations becomes indistinguishable from long-range order.
Moreover, in the case of three-dimensional materials with a layered structure such as BiTeI, correlated fermion models have a much smaller hopping amplitude in the direction perpendicular to the layers than in the layers, but a slow decay of correlations in a single layer means that the order-parameter field is almost constant in large domains of the layer. The typical area of these domains can then scale up even small interlayer couplings between the order-parameter fields, hence at some low temperature make the dynamics three-dimensional so that ordering can set in. Finally, we note that mean-field theory yields a quantitative relation between the critical scale $\Lambda_{\rm c}$ and the critical temperature $T_{\rm c}$, which is obtained from the gap equation by assuming that the superconducting gap vanishes as the temperature $T$ approaches $T_{\rm c}$ (see Eq.~\eqref{est_temp} in Sec.~\ref{sec_gapeq}).

\subsection{Particle-particle ladder resummation} \label{sec_pp}

In our numerical evaluation of the RG equations for an initial attractive, local interaction, the particle-hole terms remain small, and the full RG flow is close to the result of a particle-particle ladder resummation (see Ref. \onlinecite{Salmhofer}, Sec.~4.5.4). Therefore, we restrict ourselves in the following to the particle-particle flow. We first give a heuristic argument which shows that when a superconducting instability is approached, the scale-dependent interaction vertex has relevant contributions from both bands of the model even if the Fermi level intersects only the lower band (such that the upper band is empty at $T = 0$). After that, we provide a general analytical solution of the particle-particle flow in the spin basis, which applies to the case where the single-particle Hamiltonian is not SU(2) invariant. This solution is consistent with our numerical results as presented in the previous subsections.

The particle-particle flow for the discretized interaction vertex is defined by the RG equation \eqref{disc_1} where only the particle-particle term \eqref{eq_18} is kept on the right-hand side. Thus, we consider the equation
\begin{widetext}
\begin{equation} \label{pp}
\begin{aligned}
 \frac{\de}{\de\Lambda} \, (V_\Lambda)_{n_1 n_2 n_3 n_4}(i_1, i_2, i_3) & = {-\sum_{\ell_1, \h \ell_2}} \ \sum_{j_1 = 1}^N \,\h \sum_{j_2 = 1}^N \,\h \sum_{\vec K} \mathbbm 1(\vec K + \vec \pi_{i_1} + \vec \pi_{i_2} - \vec \pi_{j_1} \in \mathcal B_{j_2}) \ (L^-_\Lambda)_{\ell_1 \ell_2}(i_1, i_2, j_1) \\[5pt]
 & \quad \, \times \Big[ \h (V_\Lambda)_{n_1 n_2 \ell_1 \ell_2}(i_1, i_2, j_1) \ (V_\Lambda)_{\ell_1 \ell_2 n_3 n_4}(j_1, j_2, i_3) \, + \, (j_1, \ell_1) \h \leftrightarrow \h (j_2, \ell_2) \, \Big] \,,
\end{aligned}
\end{equation}
\end{widetext}
with $L_\Lambda^-$ given by Eq.~\eqref{eq_22}. A special solution of this equation is a superconducting interaction of the form 
\begin{equation}
\begin{aligned}
 & (V_{\Lambda})_{s_1 \ldots s_4}(i_1, i_2, i_3) \h = \h -\frac{g_\Lambda}{2} \h \mathbbm 1(\vec \pi_{i_1} \mh = \mh -\vec \pi_{i_2}) \\[5pt]
 & \quad \times (\delta_{s_1 s_3} \delta_{s_2 s_4} - \delta_{s_1 s_4} \delta_{s_2 s_3})
\end{aligned}
\end{equation}
in the spin basis, or
\begin{equation} \label{scint}
\begin{aligned}
 & (V_{\Lambda})_{n_1 \ldots n_4}(i_1, i_2, i_3) \h = \h \frac{g_\Lambda}{2} \h \mathbbm 1(\vec \pi_{i_1} \mh = \mh -\vec \pi_{i_2}) \\[5pt]
 & \quad \times \delta_{n_1 n_2} \h \delta_{n_3 n_4} \, n_2 \h n_3 \,\h \e^{\ii \varphi(\vec \pi_{i_3}) - \ii \varphi(\vec \pi_{i_2})}
\end{aligned}
\end{equation}
in the band basis (cf.~Eqs.~\eqref{num_res_band}--\eqref{eq_eff}). Here, $g_\Lambda > 0$ \h is a scale-dependent coupling parameter. By putting this ansatz into Eq.~\eqref{pp}, we obtain the differential equation
\begin{equation} \label{tosolve}
 \dot g_\Lambda = (B_-(\Lambda) + B_+(\Lambda)) \, g_\Lambda^2 \,,
\end{equation}
where we have defined for $\ell \in \{-, +\}$,
\begin{equation} \label{expr}
\begin{aligned}
 B_\ell(\Lambda) & = \int \! \db^2 \vec k \,\h \dot \chi_\Lambda(e_{\ell}(\vec k) ) \,\chi_\Lambda(e_{\ell}(-\vec k) ) \\
 & \quad \, \times \frac{1 - f(e_{\ell}(\vec k) ) - f(e_{\ell}(-\vec k))}{e_{\ell}(\vec k) + e_{\ell}(-\vec k) } \,,
\end{aligned}
\end{equation}
with $e_\ell(\vec k) = E_\ell(\vec k) - \mu$. Before coming to the solution of Eq.~\eqref{tosolve}, we simplify the above expression. We use that by time-reversal symmetry,
\begin{equation}
 e_\ell(\vec k) = e_\ell(-\vec k)
\end{equation}
and, furthermore,
\begin{equation}
 1 - 2 \h f(e) = \tanh \mh \left(\frac{\beta e}{2}\right).
\end{equation}
Thus, we obtain
\begin{equation}
\begin{aligned}
 B_\ell(\Lambda) & = \int \! \db^2 \vec k \,\h \dot \chi_\Lambda(e_{\ell}(\vec k) ) \, \chi_\Lambda(e_{\ell}(\vec k)) \\
 & \quad \, \times \frac{1}{2 \h e_\ell(\vec k)} \h \tanh \! \left(\frac{\beta e_\ell(\vec k)}{2}\right).
\end{aligned}
\end{equation}
In the zero-temperature limit, $\beta \to \infty$, this yields
\begin{equation} \label{20}
 B_\ell(\Lambda) = \int \! \db^2 \vec k \,\h \dot \chi_\Lambda(e_{\ell}(\vec k) ) \, \chi_\Lambda(e_{\ell}(\vec k)) \, \frac{1}{2 \h |e_\ell(\vec k)|} \,.
\end{equation}
This expression can be written as a scale derivative,
\begin{equation} \label{scale_derivative}
 B_\ell(\Lambda) = \frac{\de}{\de \Lambda} \h \beta_\ell(\Lambda) \,,
\end{equation}
of the function
\begin{equation} \label{of_the_function}
 \beta_\ell(\Lambda) = \frac 1 2 \int \! \db^2 \vec k \,\h [ \h \chi_\Lambda(e_{\ell}(\vec k))]^2 \, \frac{1}{2 \h |e_\ell(\vec k)|} \,.
\end{equation}
Further defining
\begin{equation}
 \beta(\Lambda) = \beta_-(\Lambda) + \beta_+(\Lambda) \,,
\end{equation}
we can write Eq.~\eqref{tosolve} as
\begin{equation}
 \dot g_\Lambda = \dot \beta(\Lambda) \, g_\Lambda^2 \,.
\end{equation}
The unique solution of this differential equation with the initial condition
\begin{equation}
 g_{(\Lambda = \Lambda_0)} = g_0 \,,
\end{equation}
is now given by
\begin{equation} \label{expl_sol}
 g_\Lambda = \frac{g_0}{1 - g_0 \h ( \beta(\Lambda) - \beta(\Lambda_0) )} \,.
\end{equation}
To obtain an even more concrete expression, we replace the regulator function by a {\itshape sharp cut-off function,} i.e.,
\begin{equation}
 \chi_\Lambda(e) = \varTheta(|e| - \Lambda) \,.
\end{equation}
This has the property $\chi_\Lambda^2 = \chi_\Lambda$, and its scale derivative is given by
\begin{equation}
 \dot \chi_\Lambda(e) = -\delta(|e| - \Lambda) \,.
\end{equation}
From Eqs.~\eqref{scale_derivative}--\eqref{of_the_function}, we therefore obtain
\begin{equation} \label{bell}
 B_\ell(\Lambda) = -\frac 1 {4 \Lambda} \int \! \db^2 \vec k \,\h \delta(|e_\ell(\vec k)| - \Lambda ) \,.
\end{equation}
(For the general treatment of the sharp cut-off limit, see the Appendix of Ref.~\onlinecite{sharp}.)
In terms of the densities of states of each band (defined by Eq.~\eqref{dos_def}), we can further write Eq.~\eqref{bell} as
\begin{equation}
 B_\ell(\Lambda) = -\frac{D_\ell(\mu + \Lambda) + D_\ell(\mu - \Lambda)}{4 \Lambda} \,.
\end{equation}
This expression approaches
\begin{equation} \label{B}
 B_\ell(\Lambda) = -\frac{D_\ell(\mu)}{2 \Lambda} + \mathcal O(1)
\end{equation}
as $\Lambda \to 0$, if $D_\ell$ is regular at $\mu$. The singular term can be integrated explicitly to yield
\begin{equation}
 \beta_\ell(\Lambda) - \beta_\ell(\Lambda_0) = -\frac{D_\ell(\mu)}{2} \h \ln \mh \left(\mh \frac{\Lambda}{\Lambda_0}\right).
\end{equation}
Thus, the solution \eqref{expl_sol} turns into
\begin{equation}
 g_\Lambda = g_0 \left( 1 + \frac{g_0 \hh D(\mu)}{2} \h \ln \mh \left(\mh \frac{\Lambda}{\Lambda_0}\right) \right)^{\!\!-1} , \label{gmm1}
\end{equation}
where $D(\mu) = D_-(\mu) + D_+(\mu)$ is the total density of states. In particular, $g_\Lambda$ diverges at a critical scale $\Lambda_{\rm c}$, which is determined through the equation
\begin{equation}
 1 = -\frac{g_0 \hh D(\mu)}{2} \h \ln \mh \left(\mh \frac{\Lambda_{\rm c}}{\Lambda_0}\right),
\end{equation}
and given explicitly by
\begin{equation} \label{expon}
 \Lambda_{\rm c} = \Lambda_0 \h \exp \mh \left({-\frac{2}{g_0 \hh D(\mu)}} \right) .
\end{equation}
This result can already be compared with our numerical result for the stopping scale shown in Fig.~\ref{fig_phase}. However, since Eq.~\eqref{expon} has been derived by solving the particle-particle flow in the limit $\Lambda \to 0$, we cannot expect an exact agreement with our numerical solution, which had been obtained by starting the RG flow at an initial scale $\Lambda_0$ much larger than the bandwidth of the model. Nevertheless, it turns out that the formula \eqref{fit} agrees well with our numerical data. This is obtained from Eq.~\eqref{expon} by identifying $g_0$ with the strength of the initial onsite interaction $|U|$ and replacing the prefactor $\Lambda_0/t$ by a numerical factor $5.0$ (which in the logarithmic plot corresponds to a constant shift of the whole curve).

It is now instructive to consider an ansatz for the solution of Eq.~\eqref{pp} which is more general than Eq.~\eqref{scint}, namely
\begin{equation} \label{genint}
\begin{aligned}
 & (V_{\Lambda})_{n_1 \ldots n_4}(i_1, i_2, i_3) \h = \h \frac{1}{2} \h \mathbbm 1(\vec \pi_{i_1} \mh = \mh -\vec \pi_{i_2}) \\[6pt]
 & \times \delta_{n_1 n_2} \h \delta_{n_3 n_4} \h g_{n_2 n_3}(\Lambda) \h \e^{\ii \varphi(\vec \pi_{i_3}) - \ii \varphi(\vec \pi_{i_2})} \,,
\end{aligned} \smallskip \vspace{2pt}
\end{equation}
with a matrix $g_{n_2 n_3}(\Lambda)$ of generalized coupling parameters. In the following, we will suppress the $\Lambda$ dependencies in the notation. Hermiticity requires that
\begin{equation}
 g_{n \hh n'} = g_{n' n}^* \,,
\end{equation}
such that $g_{++}, \,\h g_{--}$ are real and $g_{+-} = g_{-+}^*$\h. For the particular choice
\begin{align}
 g_{++} = g_{--} & = g \,, \label{choice_1} \\[3pt]
 g_{+-} = g_{-+} & = -g \,, \label{choice_2}
\end{align}
we recover again the superconducting interaction \eqref{scint}. We focus on the case where the chemical potential $\mu$ is in the lower band, such that
\begin{equation}
 D_+(\mu) = 0 \,,
\end{equation}
and the total density of states is determined only by the lower band,
\begin{equation}
 D(\mu) = D_-(\mu) \,.
\end{equation}
Na\"ively, one could expect that in this case only the coupling \h$g_{--}$ \h is important in the flow, while all couplings with at least one upper band index do not play any r\^{o}le. We will now show, however, that this is not true.

By putting the generalized ansatz \eqref{genint} into the RG equation \eqref{pp}, we obtain after a straightforward calculation the following coupled differential equations for the generalized coupling constants:
\begin{equation} \label{sys}
 \dot g_{nn'} = \sum_\ell B_\ell \, g_{n \ell} \, g_{\ell n'} \,,
\end{equation}
or equivalently,
\begin{align}
 \dot g_{++} & = B_{+} \, g_{++}^2 \h + \h B_- \, |g_{+-}|^2 \,, \label{1} \\[5pt]
 \dot g_{--} & = B_- \, g_{--}^2 \h +\h B_+ \, |g_{+-}|^2 \,, \label{2} \\[5pt]
 \dot g_{+-} & = B_+ \,g_{++} \, g_{+-} \h+\h B_- \, g_{--} \,g_{+-} \,, \label{3}
\end{align}
with the coefficient functions $B_\ell$, $\ell \in \{+, \h -\}$ given by Eq.~\eqref{expr}.
By our result \eqref{B}, the vanishing of $D_+(\mu)$ also implies
\begin{equation}
 B_+ = 0 \,,
\end{equation}
and hence the above system simplifies to
\begin{align}
 \dot g_{++} & = B_- \, |g_{+-}|^2 \,, \label{first} \\[5pt]
 \dot g_{--} & = B_- \, g_{--}^2 \label{second_diff} \, \\[5pt]
 \dot g_{+-} & = B_- \, g_{--} \, g_{+-} \label{third} \,.
\end{align}
Let us assume that at some initial scale $\Lambda_0$ we have the equalities
\begin{equation} \label{in2}
 g_{++}(\Lambda_0) \h = \h g_{--}(\Lambda_0) \h = \h -g_{+-}(\Lambda_0) \h\hh \equiv \h\hh g_0 \,.
\end{equation}
The second equation \eqref{second_diff} is closed and can be solved readily: with the above initial condition, we find
\begin{equation}
 g_{--}(\Lambda) = g_0 \left( 1 + \frac{g_0 \hh D_-(\mu)}{2} \h \ln \mh \left(\mh \frac{\Lambda}{\Lambda_0}\right) \right)^{\!\!-1} \,, \label{gmm2}
\end{equation}
precisely analogous to Eq.~\eqref{gmm1}. Next, consider 
the equation \eqref{third} for the coupling $g_{+-}$\h. The point is now that \h$g_{--}$ \h also appears on the right-hand side of this equation, and thereby drives the flow of $g_{+-}$\h. In particular, as $\Lambda \to \Lambda_{\rm c}$\h, the growing of $g_{--}$ also leads to a divergence of $g_{+-}$\h. Concretely, the solution of Eq. \eqref{third} with the initial condition \eqref{in2} is simply
\begin{equation}
 g_{+-}(\Lambda) = -g_{--}(\Lambda).
\end{equation}
Similarly, we see from Eq.~\eqref{first} that $g_{+-}$ drives the flow of $g_{++}$, and in the end we obtain the solution
\begin{equation}
 g_{++}(\Lambda) = g_{--}(\Lambda) = -g_{+-}(\Lambda) \,.
\end{equation}
Thus, we have shown that all couplings $g_{nn'}(\Lambda)$ remain of equal magnitude in the flow and together approach a divergence as $\Lambda \to \Lambda_{\rm c}$\h, even though the upper band is completely empty. In other words, if the condition \eqref{in2} is satisfied at some initial scale $\Lambda_0$, then this property of the effective interaction remains invariant in the particle-particle flow.

Finally, we generalize the above heuristic argument in order to provide an analytical solution of the particle-particle flow for a general class of SU(2)-symmetric initial interactions. For this purpose, we switch to the spin basis, where the particle-particle flow equation reads as (cf.~Eqs.~\eqref{rg_eq}--\eqref{eq_phipp} in the band basis)
\begin{widetext}
\begin{equation} \label{pp_spin}
\begin{aligned}
 \frac{\de}{\de \Lambda} (V_\Lambda)_{s_1 s_2 s_3 s_4}(\vec p_1, \vec p_2, \vec p_3)
 & = {-\sum_{t_1, \ldots, t_4}} \h \frac{1}{|\mathcal B|} \int_{\mathcal B} \! \de^2 \vec k_1 \int_{\mathcal B} \! \de^2 \vec k_2 \, \sum_{\vec K} \delta^2(\vec K + \vec p_1 + \vec p_2 - \vec k_1, \vec k_2) \\[5pt]
 & \quad \, \times (L_\Lambda^-)_{t_1 t_2 t_3 t_4}(\vec k_1, \vec k_2) \,\h (V_\Lambda)_{s_1 s_2 t_1 t_2}(\vec p_1, \vec p_2, \vec k_1) \,\h (V_\Lambda)_{t_3 t_4 s_3 s_4}(\vec k_1, \vec k_2, \vec p_3) \,.
\end{aligned}
\end{equation}
\end{widetext}
Note that the particle-particle loop in the spin basis depends on {\itshape four} spin indices and is given in terms of its counterpart in the band basis, Eq.~\eqref{def_lmp}, by
\begin{align}
 & (L_\Lambda^-)_{t_1 t_2 t_3 t_4}(\vec k_1, \vec k_2) = \sum_{\ell_1 \mh, \, \ell_2} U_{t_1 \ell_1}(\vec k_1) \, U_{t_2 \ell_2}(\vec k_2) \nonumber \\[3pt]
 & \times (L_\Lambda^-)_{\ell_1 \ell_2}(\vec k_1, \vec k_2) \, U_{\ell_1 t_3}^\dagger(\vec k_1) \, U_{\ell_2 t_4}^\dagger(\vec k_2) \,. \label{asdf}
\end{align}
We assume that the initial interaction is SU(2) invariant and of the following form:
\begin{equation} \label{form}
\begin{aligned}
 & V_{s_1 \ldots s_4}(\vec k_1, \vec k_2, \vec k_3) \\[2pt]
 & = -\frac 1 2 \, g(\vec k_1 + \vec k_2) \, (\delta_{s_1 s_3} \h \delta_{s_2 s_4} - \delta_{s_1 s_4} \h \delta_{s_2 s_3} ) \,,
\end{aligned}
\end{equation}
or more precisely,
\begin{equation}
\begin{aligned}
 & V_{s_1 \ldots s_4}(\vec k_1, \vec k_2, \vec k_3) \\[4pt]
 & = -\frac 1 2 \h \int_{\mathcal B} \de^2 \vec k \,\h \sum_{\vec K} \delta^2(\vec K + \vec k_1 + \vec k_2, \h \vec k) \\[2pt]
 & \quad \, \times g(\vec k) \, (\delta_{s_1 s_3} \h \delta_{s_2 s_4} - \delta_{s_1 s_4} \h \delta_{s_2 s_3} ) \,,
\end{aligned}
\end{equation}
with an arbitrary function $g(\vec k)$. For example, an onsite attractive interaction corresponds to
\begin{equation}
 g(\vec k) = -U
\end{equation}
with $U < 0$, while a superconducting interaction is represented by
\begin{equation} \label{show}
 g(\vec k) = g \, |\mathcal B| \, \delta^2(\vec k, \h \vec 0) \smallskip
\end{equation}
with $g > 0$. Now, given any initial interaction of the form 
\eqref{form}, one can show that the effective interaction retains this form in the particle-particle flow. In particular, this means that the SU(2) invariance of the effective interaction is preserved in the particle-particle flow even if the single-particle Hamiltonian does not possess this symmetry. This can be proven easily by putting the ansatz \eqref{form} into the RG equation \eqref{pp_spin} and using the identity 
\begin{equation}
 \delta_{s_1 s_3} \h \delta_{s_2 s_4} - \delta_{s_1 s_4} \h \delta_{s_2 s_3} = [\ii\sigma_y]_{s_1 s_2} \h [\ii\sigma_y]_{s_3 s_4} \,,
\end{equation}
(see Ref.~\onlinecite{Edelstein89}, Eq.~(10)). The RG equation for the interaction vertex then reduces to a decoupled system of differential equations, one for each component $g_\Lambda(\vec p)$ with $\vec p \in \mathcal B$. Explicitly, this reads as
\begin{equation} \label{diffeq}
 \dot g_\Lambda(\vec p) = B_\Lambda(\vec p) \, g_\Lambda(\vec p)^2 \,,
\end{equation}
where the coefficient functions are given by
\begin{equation}
\begin{aligned}
 B_\Lambda(\vec p) & = \frac 1 2 \h \int \! \db^2 \vec k \sum_{t_1, \ldots, t_4} (\delta_{t_1 t_3} \h \delta_{t_2 t_4} - \delta_{t_1 t_4} \h \delta_{t_2 t_3} ) \\[2pt]
 & \quad \, \times (L_\Lambda^-)_{t_1 \ldots t_4}(\vec k, \vec p - \vec k) \,.
\end{aligned}
\end{equation}
A straightforward calculation using Eq.~\eqref{asdf} and the property \eqref{id_u} of the unitary matrix $U(\vec k)$ further yields
\begin{equation} \label{blambdap}
\begin{aligned}
 B_\Lambda(\vec p) & = \frac 1 2 \h \sum_{\ell_1, \h \ell_2} \int \! \db^2 \vec k \, (L_\Lambda^-)_{\ell_1 \ell_2} (\vec k, \vec p - \vec k) \\[2pt]
 & \quad \, \times \frac 1 2 \h \big( 1 - \ell_1 \ell_2 \h \cos(\varphi(\vec k) - \varphi(\vec p - \vec k))\big) \,.
\end{aligned}
\end{equation}
In particular, for $\vec p = \vec 0$, we have
\begin{equation}
 \varphi(-\vec k) = \varphi(\vec k) + \pi \,,
\end{equation}
and hence,
\begin{align}
 B_\Lambda(\vec 0) & = \frac 1 2 \h \sum_\ell \int \! \db^2 \vec k \, (L_\Lambda^-)_{\ell \ell}(\vec k, -\vec k) \\[3pt]
 & = B_+(\Lambda) + B_-(\Lambda) \,,
\end{align}
with the functions $B_\ell(\Lambda)$ defined by Eq.~\eqref{expr}. The differential equation for the $\vec p = \vec 0$ component
is therefore equivalent to the flow equation for the coupling constant~$g$ of a superconducting interaction (see Eq.~\eqref{tosolve}). Furthermore, the general solution of Eq.~\eqref{diffeq} with the 
initial condition
\begin{equation}
 g_{(\Lambda = \Lambda_0)}(\vec p) = g_0(\vec p)
\end{equation}
is given by
\begin{equation} \label{ana}
 g_\Lambda(\vec p) \h = \h g_0(\vec p) \, \bigg( 1 - g_0(\vec p) \h \int_{\Lambda_0}^{\Lambda} B_{\Lambda'}(\vec p) \h\hh \de \Lambda' \bigg)^{\!\!-1},
\end{equation}
which generalizes the result \eqref{expl_sol} derived above for a superconducting interaction. In principle, Eq. \eqref{ana} can be used to determine the flow of the interaction vertex from the attractive onsite interaction at the initial scale to the superconducting interaction at the critical scale.

Our final remark in this section regards the question of why it is sufficient (for our model) to choose all the representative momenta on the Fermi lines (of any band) instead of taking a two-dimensional grid of discrete momenta in the whole Brillouin zone: From the expressions \eqref{def_lmp}--\eqref{disc_2b}, one can see that the loop terms $(L_\Lambda^\mp)_{\ell_1 \ell_2}(\vec k_1, \vec k_2)$ \h are singular if both $\vec k_1$ and $\vec k_2$ lie on \linebreak the Fermi lines of the respective bands $\ell_1$ and $\ell_2$\hh, such that $e_{\ell_1}(\vec k_1) = e_{\ell_2}(\vec k_2) = 0$\,. This motivates a simplification of the momentum dependence of the interaction vertex by a projection to the Fermi lines. The quality of this approximation has been discussed for the different RG schemes in Refs.~\onlinecite{Sa98, HM, SH00, Metzner}. The fact that the projected vertex function is constant along paths transversal to the Fermi lines, hence may become large also away from the Fermi lines, usually leads to a slight overestimation of the coupling functions, hence a more conservative estimate of the stopping scale. The particle-particle flow considered here provides an example where the coupling function really is constant along certain lines in momentum space. In our analytical solution, Eq.~\eqref{form} with $g_\Lambda(\vec k_1 + \vec k_2)$ given by Eq.~\eqref{ana}, the flow coefficients are largest when the external momenta satisfy $\vec k_1 = -\vec k_2$, which does not restrict $\vec k_1$ and $\vec k_2$ to the Fermi lines. In fact, this analytical solution is even independent of the distance of $\vec k_1$ or $\vec k_2$ from the Fermi~lines.

\section{Mean-field theory} \label{sec_MF}

\subsection{Definitions}

By starting from an attractive, local interaction at the ultraviolet scale, we have thus far obtained the effective superconducting interaction at the stopping scale $\Lambda_{*}$ (which is close to the critical scale $\Lambda_{\rm c}$) given by Eq.~\eqref{eq_eff_coeff}. This can be written equivalently as
\begin{equation} \label{sc_int}
\begin{aligned}
 \hat V_{\Lambda_*} & = \frac 1 2 \h \int \! \db^2 \vec k \int \! \db^2 \vec k' \sum_{s_1,\ldots, s_4} V_{s_1 s_2 s_3 s_4}(\vec k, \vec k') \\[3pt]
 & \quad \, \times \hat a_{s_1}\dag(-\vec k) \, \hat a_{s_2}\dag(\vec k) \, \hat a_{s_3}(\vec k') \, \hat a_{s_4}(-\vec k') \,,
\end{aligned}
\end{equation}
where we have introduced an interaction kernel of only two momentum arguments (which in our case is momentum independent),
\begin{align}
 V_{s_1 s_2 s_3 s_4}(\vec k, \vec k') & = \frac{g }{2} \, ( \delta_{s_1 s_3} \delta_{s_2 s_4} - \delta_{s_1 s_4} \delta_{s_2 s_3} ) \\[5pt]
 & = -\frac{g }{2} \, [\ii \sigma_y]_{s_2 s_1} [\ii \sigma_y]_{s_3 s_4} \,. \label{singlet}
\end{align}
As in Ref.~\onlinecite{Reiss}, we use mean-field theory to approximately describe the electronic degrees of freedom below the energy scale $\Lambda_*$. Hence, we restrict all wave vectors to a shell around the Fermi lines given by
\begin{equation} \label{what}
 |e_n(\vec k) | \h \equiv \h | E_n(\vec k) - \mu | \h < \h \Lambda_{*}\,,
\end{equation}
where $n \in \{-, +\}$\h. Mean-field theory allows to calcu\-{}late---starting from a superconducting interaction of the form \eqref{sc_int}---the gap function and the order parameter. For this, we proceed analogous to Ref.~\onlinecite{Sig91} and generalize the results presented there to the case without spin SU(2) symmetry. The mean-field ansatz consists in replacing the quartic interaction \eqref{sc_int} by the quadratic {\itshape mean-field interaction},
\begin{align}
 \hat V^{\rm mf} & = \frac 1 2 \int \! \db^2 \vec k \int \! \db^2 \vec k' \sum_{s_1, \ldots, s_4} V_{s_1 s_2 s_3 s_4}(\vec k, \vec k') \label{eq_mf} \\[3pt] \nonumber
 & \quad \, \times \hat a\dag_{s_1}(-\vec k) \h \hat a\dag_{s_2}(\vec k) \, \big\langle \hat a_{s_3}(\vec k') \h \hat a_{s_4}(-\vec k') \big\rangle + \mathrm{H.\h a.} \,,
\end{align}

\vspace{3pt} \noindent
where ``$\mathrm{H.\h a.}$'' denotes the Hermitian adjoint. Consequently, the effective Hamiltonian at the critical scale, \begin{equation} \label{effHam}
 \hat H_{\Lambda_*} = \hat H^0 + \hat V_{\Lambda_*} \,,
\end{equation}
is replaced by the {\itshape mean-field Hamiltonian,}
\begin{equation} \label{mfHam}
\hat H^{\rm mf} = \hat H^0 + \hat V^{\rm mf} \,.
\end{equation}
The latter is quadratic and can in principle be solved exactly. However, the expectation values in Eq.~\eqref{eq_mf} have to be evaluated with respect to the mean-field Hamiltonian itself, i.e.,
\begin{equation} \label{eq_mf_1}
 \langle \hat A \rangle = \frac{1}{Z^{\rm mf}} \, \mathrm{Tr} \h \big( \e^{-\beta (\hat H^{\rm mf} - \mu \hat N)} \hat A \h \big)\,,
\end{equation}
with
\begin{equation} \label{eq_mf_2}
Z^{\rm mf} = \mathrm{Tr} \h \big( \e^{-\beta (\hat H^{\rm mf} - \mu \hat N)} \h \big) \,.
\end{equation}
Therefore, $\hat H^{\rm mf}$ has to be determined as a self-consistent solution of Eqs.~\eqref{eq_mf} and \eqref{mfHam}--\eqref{eq_mf_2}.

The expectation value
\begin{equation} \label{eq_order}
\Psi_{s s'}(\vec k) = \big \langle \hat a_s(\vec k) \h \hat a_{s'}(-\vec k) \big \rangle
\end{equation}
is called the superconducting {\itshape order parameter}, while
\begin{align} \label{eq_gap}
 & \Delta_{s s'}(\vec k) \\[2pt] \nonumber
 & = -\int \! \db^2 \vec k' \sum_{s_3, s_4} V_{s' s s_3 s_4}(\vec k, \vec k') \, \big \langle \hat a_{s_3}(\vec k') \h \hat a_{s_4}(-\vec k') \big \rangle
\end{align}
is called the {\itshape gap function} (or {\itshape pair potential}). The mean-field interaction can 
be written in terms of the gap function as
\begin{equation}
\begin{aligned}
 \hat V^{\rm mf} & = \frac 1 2 \int \! \db^2 \vec k \, \sum_{s_1, s_2} \Delta_{s_1 s_2}(\vec k) \, \hat a_{s_1}\dag(\vec k) \h \hat a_{s_2}\dag(-\vec k) \\
 & \quad \, + \mathrm{H.\h a.} \,,
\end{aligned}
\end{equation}
which is seen from Eq.~\eqref{eq_mf} by substituting $\vec k \to -\vec k$ and using that
\begin{equation}
 V_{s_1 s_2 s_3 s_4}(\vec k, \vec k') = -V_{s_2 s_1 s_3 s_4}(-\vec k, \vec k') \,.
\end{equation}
In the following, we will solve the mean-field theory for the spin-singlet interaction \eqref{singlet}, and thereby derive explicit expressions for both the gap function and the order parameter in our model. First, we obtain immediately from Eq.~\eqref{singlet} the spin structure and momentum dependence of the gap function,
\begin{equation}
\begin{aligned}
 & \Delta_{ss'}(\vec k) = \frac{g }{2}\, [\ii\sigma_y]_{ss'} \\[3pt]
 & \hspace{0.5cm} \times \int \! \db^2 \vec k' \sum_{s_3, \h s_4} [\ii\sigma_y]_{s_3 s_4} \, \big \langle \hat a_{s_3}(\vec k') \h \hat a_{s_4}(-\vec k') \big \rangle \,. 
\end{aligned} \smallskip
\end{equation}
In matrix notation, we can write this as
\begin{equation} \label{eq_gapform}
 \Delta(\vec k) = \ii \sigma_y \h \Delta_0 \,, 
\end{equation}
where we have defined the scalar {\itshape gap parameter}
\begin{equation} \label{eq_defd}
 \Delta_0 = \frac{g }{2} \h \int \! \db^2 \vec k \sum_{s_3, \h s_4} [\ii\sigma_y]_{s_3 s_4} \, \big \langle \hat a_{s_3}(\vec k) \h \hat a_{s_4}(-\vec k) \big \rangle \,.
\end{equation}
In order to determine this parameter, we first have to calculate the order parameter, which in turn depends on the gap function. Therefore, $\Delta_0$ must be determined self-consistently as a solution of the {\itshape gap equation} (see Sec.~\ref{sec_gapeq}). Up to this parameter, however, the form of the gap function is already fixed by Eq.~\eqref{eq_gapform}: it is independent of the Bloch momentum $\vec k$ and the chemical potential~$\mu$, and it has a {\itshape singlet} spin structure.

\bigskip
\subsection{Bogoliubov transformation}

Calculating the order parameter requires to diagonalize the quadratic mean-field Hamiltonian. First, we formally rewrite the mean-field Hamiltonian (with the contribution $-\mu \hat N$ from the particle-number operator) as follows:
\begin{widetext}
\begin{equation}
\begin{aligned}
 \hat H^{\rm mf} - \mu \hat N & = \frac 1 2 \int \! \db^2 \vec k \sum_{s_1, \h s_2} \label{eq_mfham} \big( \h \hat a\dag_{s_1}(\vec k), \, \hat a_{s_1}(-\vec k) \h \big) \\[3pt]
 & \quad \, \times \left( \! \begin{array}{cc} H^0_{s_1 s_2}(\vec k) - \mu \h \delta_{s_1 s_2} & \Delta_{s_1 s_2}(\vec k) \\[8pt] -\Delta_{s_1 s_2}^*(-\vec k) & -(H^0)^*_{s_1 s_2}(-\vec k) + \mu \h \delta_{s_1 s_2} \end{array} \hspace{-1pt} \right) \left( \! \begin{array}{c} \hat a_{s_2}(\vec k) \\[8pt] \hat a\dag_{s_2}(-\vec k) \end{array} \! \right) \,.
\end{aligned}
\end{equation}
\end{widetext}
{\linespread{1.1}\selectfont
Here, $H^0_{s_1 s_2}\mh(\vec k)$ 
is the free Hamiltonian matrix, expressed in terms of the functions $f(\vec k)$ and $\vec g(\vec k)$ by Eq.~\eqref{eq_Ham}, and $\Delta_{s_1 s_2}(\vec k)$ the gap function given by Eq.~\eqref{eq_gapform}. In Sec.~\ref{sec_Rashba}, we have diagonalized $H^0(\vec k)$ as \par}
\begin{equation}
 U\dag(\vec k) \h H^0(\vec k) \h U(\vec k) = E(\vec k) \,,
\end{equation}
with the unitary matrix $U(\vec k)$ given by Eqs.~\eqref{expU1}--\eqref{expU}, and the diagonal matrix of eigenvalues
\begin{equation}
 E(\vec k) = \left( \! \begin{array}{cc} E_-(\vec k) & 0 \\[5pt] 0 & E_+(\vec k) \end{array} \mh \right) ,
\end{equation}
with $E_{\mp}(\vec k) = f(\vec k) \mp |\vec g(\vec k)|$\h.
In the following, we will often denote the momentum dependencies by a subscript, e.g.~$H^0_{\vec k} \equiv H^0(\vec k)$, in order to lighten the notation. We proceed as in Ref.~\onlinecite{Sig91}, defining the $(4 \times 4)$ matrix
\begin{equation} \label{eq_todiag}
 \mathcal H \k = \left( \! \begin{array}{cc} H^0\k -\mu & \Delta\k \\[6pt] -\Delta^*\mk & -(H^{0}\mk)^* + \mu \end{array} \mh \right) , 
\end{equation}
which appears in the mean-field Hamiltonian \eqref{eq_mfham}. Note that the gap function is antisymmetric,
\begin{equation}
 \Delta_{s_1 s_2}(\vec k) = -\Delta_{s_2 s_1}(-\vec k) \,,
\end{equation}
and therefore $\mathcal H_{\vec k}$ is Hermitian. The diagonalization of the mean-field Hamiltonian is performed by means of a {\itshape Bogoliubov transformation,}
\begin{align}
 \hat a_{s}(\vec k) & = \sum_n \big( X_{sn}(\vec k) \, \hat b_n(\vec k) + Y_{sn}(\vec k) \, \hat b\dag_n(-\vec k) \big) \,, \\[5pt]
 \hat a\dag_{s}(-\vec k) & = \sum_n \big( Y_{sn}^*(-\vec k) \, \hat b_n(\vec k) + X_{sn}^*(-\vec k) \, \hat b\dag_n(-\vec k) \big) \,.
\end{align}
We seek $X\k \equiv X_{sn}(\vec k)$ and $Y\k \equiv Y_{sn}(\vec k)$ such that
the $(4 \times 4)$ matrix
\begin{equation} \label{eq_seek}
 \mathcal U\k = \left( \! \begin{array}{cc} \, X\k & Y\k \\[5pt] Y\mk^* & X\mk^* \, \end{array} \! \right) \smallskip
\end{equation}
has the following properties:
(i) it is unitary,
\begin{equation} \label{eq_unit}
 \mathcal U\k\dag \,\h \mathcal U\k = 1\,,
\end{equation}
and (ii) it diagonalizes $\mathcal H\k$, i.e.,
\begin{equation} \label{eq_diagonalize}
 \mathcal U\k\dag \, \mathcal H\k \,\h \mathcal U\k = \mathcal E\k \,,
\end{equation}
where $\mathcal E\k$ is the $(2 \times 2)$ diagonal matrix of eigenvalues, which turns out to be of the following form:
\begin{widetext}
\begin{equation}
 \mathcal E\k = \left( \!\! \begin{array}{cc} \varepsilon\k & 0 \\[5pt] 0 & -\varepsilon\mk \end{array} \! \right)
 \equiv \left( \begin{array}{cccc} \varepsilon_-(\vec k) & 0 & 0 & 0 \\[5pt] 0 & \varepsilon_+(\vec k) & 0 & 0 \\[5pt] 0 & 0 & -\varepsilon_-(-\vec k) & 0 \\[5pt] 0 & 0 & 0 & -\varepsilon_+(-\vec k) \, \end{array} \right). \label{eq_bigm}
\end{equation}
With this, the mean-field Hamiltonian \eqref{eq_mfham} is diagonalized as
\begin{equation}
 \hat H^{\rm mf} - \mu \hat N = \frac 1 2 \int \! \db^2 \vec k \, \sum_{n} \h \big( \h \hat b\dag_n(\vec k) \h , \, \hat b_n(-\vec k) \h \big) \left( \! \begin{array}{cc} \varepsilon_n(\vec k) & 0 \\[8pt] 0 & -\varepsilon_n(-\vec k) \, \end{array} \! \right) \left( \! \begin{array}{c} \hat b_n(\vec k) \\[8pt] \hat b\dag_n(-\vec k) \end{array} \! \right)\,.
\end{equation}
\end{widetext}
By substituting $\vec k \to -\vec k$, this is equivalent to
\begin{equation}
 \hat H^{\rm mf} - \mu \hat N = \int \! \db^2 \vec k \, \sum_{n} \varepsilon_{n}(\vec k) \, \hat b\dag_{n}(\vec k) \h \hat b_n(\vec k) \,.
\end{equation}
A lengthy calculation analogous to (Ref.~\onlinecite{Sig91}, Appendix A) yields the eigenvalues
\begin{equation} \label{eq_eigenvalues}
 \varepsilon_{\mp}(\vec k) = \sgn(e_{\mp}(\vec k)) \, \sqrt{e_{\mp}(\vec k)^2 + \Delta_0^2} \,,
\end{equation}
where $e_{\mp}(\vec k) = E_{\mp}(\vec k) - \mu$, and $\sgn(x) = x / |x|$ denotes the sign function. (The latter was introduced such that for $\Delta_0 \to 0$, the eigenvalues $\varepsilon_{\mp}(\vec k)$ approach the respective eigenenergies $e_{\mp}(\vec k)$ of the non-interacting system.) Furthermore, we obtain the following expressions for the $(2 \times 2)$ matrices of the Bogoliubov transformation:
\begin{align}
 X\k & = U\k \h (\varepsilon\k + e\k) \, \frac{1}{\sqrt{(\varepsilon\k + e\k)^2 + \Delta_0^2}} \,, \label{resu} \\[3pt]
 Y\k & = -\Delta\k \h U\mk^* \, \frac{1}{\sqrt{(\varepsilon\k + e\k)^2 + \Delta_0^2}} \,. \label{resv}
\end{align}
Here, $U_{\vec k} \equiv U(\vec k)$ is the unitary matrix diagonalizing the free Hamiltonian. Furthermore, $e\k \equiv E\k - \mu$ and $\varepsilon\k$ are diagonal matrices containing the eigenvalues of the free Hamiltonian and, respectively, the mean-field Hamiltonian. The above formulas \eqref{resu}--\eqref{resv} generalize the result (2.13) in Ref.~\onlinecite{Sig91} to the case without SU(2) symmetry. 
We remark that in deriving these results, we have only used the property
\begin{equation} \label{iddelta}
 \Delta\k\dag \h \Delta\k = \Delta_0^2
\end{equation}
of the singlet gap function, and the identity
\begin{equation}
 \Delta\k \h U\mk^* = \Delta_0 \h \e^{-\ii\varphi\k} \h U\k \h \sigma_z \,,
\end{equation}
which follows from Eqs.~\eqref{expU1}--\eqref{expU} by assuming time-reversal symmetry (such that $\vec g(-\vec k) = -\vec g(\vec k)$). Therefore, our results for the eigenvalues and eigenvectors of the mean-field Hamiltonian do not only apply to the concrete Rashba model, but to any time-reversal-symmetric Hamiltonian of the form \eqref{eq_Ham} with a singlet superconducting interaction.

Having diagonalized the mean-field Hamiltonian, it is no more difficult to calculate the order parameter \eqref{eq_order}: In terms of the new annihilation and creation operators $\hat b_\ell(\vec k)$ and $\hat b_\ell\dag(\vec k)$, we can write
\begin{align}
 & \Psi_{ss'}(\vec k) = \sum_{\ell, \ell'} \Big\langle \left( X_{s\ell}(\vec k) \, \hat b_{\ell}(\vec k) + Y_{s\ell}(\vec k) \, \hat b\dag_\ell(-\vec k) \right) \nonumber \\[0pt]
 & \times \left( X_{s'\ell'}(-\vec k) \, \hat b_{\ell'}(-\vec k) + Y_{s'\ell'}(-\vec k) \, \hat b\dag_{\ell'}(\vec k) \right) \Big\rangle \,. \label{eq_72}
\end{align}
Using the relation
\begin{equation}
 \big \langle \h \hat b\dag_\ell(\vec k) \, \hat b_{\ell'}(\vec k) \big \rangle = \delta_{\ell \ell'} \, n_\ell(\vec k)
\end{equation}
with
\begin{equation}
 n_\ell(\vec k) = \frac{1}{\phantom{a\dag}\hspace{-0.35cm}\e^{\beta \varepsilon_{\ell}(\vec k) } + 1} \,,
\end{equation}
we obtain from Eq.~\eqref{eq_72},
\begin{equation}
\begin{aligned}
 \Psi_{ss'}(\vec k) & = \sum_\ell X_{s\ell}(\vec k) \, Y_{s'\ell}(-\vec k) \, (1 - n_\ell(\vec k)) \\[2pt]
 & \quad \, + \sum_\ell Y_{s\ell}(\vec k) \, X_{s'\ell}(-\vec k) \, n_{\ell}(\vec k) \,.
\end{aligned} \smallskip
\end{equation}
In matrix form, this can be written compactly as
\begin{equation} \label{eqX_this}
 \Psi\k = X\k \, (1-n\k) \, Y\mk\T + Y\k \, n\k \, X\mk\T \,.
\end{equation}
Putting the matrices $X\k$ and $Y\k$ given by \eqref{resu}--\eqref{resv} into this formula yields after some algebra the concise expression
\begin{equation}
 \Psi_{\vec k} \h = \h U\k \, \frac{1 - 2 n\k}{2 \varepsilon\k} \, U\k^\dagger \, \Delta\k \h  \equiv \h  U\k \, \gamma(\varepsilon\k) \, U\k\dag \, \Delta\k \,,
\end{equation}
where we have defined the function
\begin{equation} \label{defgamma}
 \gamma(\varepsilon) = \frac{1}{2\varepsilon} \, \tanh \mh \left( \frac{\beta \varepsilon}{2} \right) \,.
\end{equation}
In the zero-temperature limit, this reduces to
\begin{equation} \label{redu}
 \lim_{\beta \to \infty} \gamma(\varepsilon) = \frac{1}{2|\varepsilon|} \,.
\end{equation}
An even more concrete expression for the order parameter can be obtained by writing
\begin{widetext}
\begin{equation}
 \gamma(\varepsilon\k) \equiv \left( \begin{array}{cc} \gamma(\varepsilon_-(\vec k)) & 0 \\[5pt] 0 & \gamma(\varepsilon_+(\vec k)) \end{array} \right)
 = \frac{\gamma(\varepsilon_-(\vec k)) + \gamma(\varepsilon_+(\vec k))}{2} \, \mathbbm 1 + \frac{\gamma(\varepsilon_-(\vec k)) - \gamma(\varepsilon_+(\vec k))}{2} \, \sigma_z \,,
\end{equation}
and using the unitarity of $U(\vec k)$ as well as the property
\begin{equation} \label{eq_prop2}
 U(\vec k) \h \sigma_z \h U\dag(\vec k) = -\frac{\vec g(\vec k)}{|\vec g(\vec k)|} \cdot \vec \sigma \h \equiv \h - \hat{\vec g}(\vec k) \cdot \vec \sigma \,,
\end{equation}
which follows from Eqs.~\eqref{expU1}--\eqref{expU}. We thereby arrive at the following formula:
\begin{equation} \label{eq_res}
 \Psi(\vec k) = \Delta_0 \, \frac{\gamma(\varepsilon_-(\vec k)) + \gamma(\varepsilon_+(\vec k))}{2} \, \ii\sigma_y 
 \, - \, \Delta_0 \, \frac{\gamma(\varepsilon_-(\vec k)) - \gamma(\varepsilon_+(\vec k))}{2} \, [\h\hat{\vec g}(\vec k) \cdot \vec \sigma] \, \ii\sigma_y \,.
\end{equation}
\end{widetext}
This is our result for the order parameter matrix $\Psi(\vec k) \equiv \Psi_{ss'}(\vec k)$.
In contrast to the gap function \eqref{eq_gapform}, the order parameter depends nontrivially on the Bloch momentum~$\vec k$ and on the chemical potential $\mu$, where the latter is implicitly contained in the eigenvalues
$\varepsilon_n(\vec k)$ given by 
Eq.~\eqref{eq_eigenvalues}. With its mixed singlet and triplet contributions, the order parameter indicates an unconventional superconducting phase.\cite{Sig91, Mineev}

\subsection{Singlet and triplet amplitudes}

\begin{figure}[t]
\vspace{-1.5cm}
\includegraphics[width=\columnwidth]{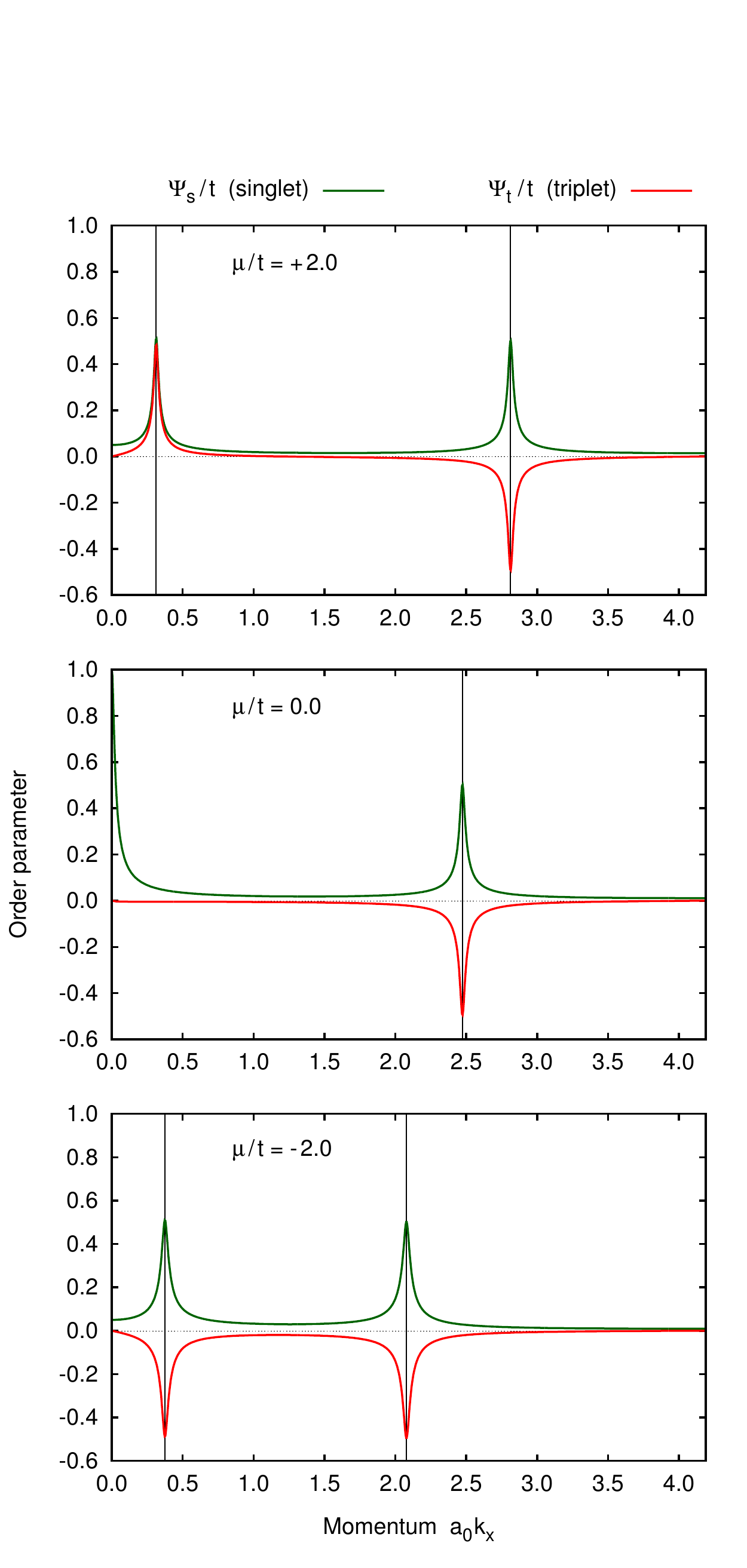}
\caption{Spin singlet and triplet amplitudes of the order parameter, for $\Delta_0/t = 0.1$. Vertical lines mark the positions of the two Fermi lines for the respective values of the chemical potential. \label{fig_order}}
\end{figure}

We define the {\itshape (spin-) singlet and triplet amplitudes} $\Psi_{\rm s}(\vec k)$ and $\Psi_{\rm t}(\vec k)$ of the order parameter by the equation
\begin{equation} \label{eq_sitri}
 \Psi(\vec k) = \Psi_{\rm s}(\vec k) \, \ii\sigma_y + \Psi_{\rm t}(\vec k) \, [ \h\hat{\vec g}(\vec k) \cdot \vec \sigma] \, \ii\sigma_y \,.
\end{equation}
(For a group-theoretical definition of these quantities, see Ref.~\onlinecite{Neu10}.) Our result \eqref{eq_res} implies that
\begin{align}
 \Psi_{\rm s}(\vec k) & = \Delta_0 \, \frac{\gamma(\varepsilon_+(\vec k)) + \gamma(\varepsilon_-(\vec k))}{2} \,, \label{eq_sitri_1} \\[6pt]
 \Psi_{\rm t}(\vec k) & = \Delta_0 \, \frac{\gamma(\varepsilon_+(\vec k)) - \gamma(\varepsilon_-(\vec k))}{2} \,. \label{eq_sitri_2}
\end{align}
In the zero-temperature limit, these formulas reduce to (cf.~Ref.~\onlinecite{Neu10}, Eq.~(2.18))
\begin{align}
 \Psi_{\rm s}(\vec k) & = \frac{\Delta_0}{4} \left( \frac{1}{|\varepsilon_+(\vec k)|} + \frac{1}{|\varepsilon_-(\vec k)|} \h \right), \label{eq_zero_1} \\[5pt]
 \Psi_{\rm t}(\vec k) & = \frac{\Delta_0}{4} \left( \frac{1}{|\varepsilon_+(\vec k)|} - \frac{1}{|\varepsilon_-(\vec k)|} \h \right). \label{eq_zero_2}
\end{align}
For small energies, i.e., in the vicinity of the band crossing, the dispersion of our tight-binding model is approximately described by the ideal Rashba model \eqref{eq_rashba}. Hence, near $\mu = 0$, the singlet and triplet amplitudes of the order parameter depend essentially only on the modulus $|\vec k|$. Figure~\ref{fig_order} shows these  amplitudes as a function of $k_x$ for three different values of the chemical potential $\mu$ (above, at, and below the band crossing), assuming a small value of the scalar gap parameter, $\Delta_0/t = 0.1$\h.

We can understand these results qualitatively as follows: First, we restrict ourselves to momenta $\vec k$ satisfying the condition \eqref{what}. For small enough $\Delta_0$, we can then estimate using Eq.~\eqref{eq_eigenvalues},
\begin{equation}
 \left\{ \begin{array}{ll} |\varepsilon_-(\vec k)| \ll |\varepsilon_+(\vec k)| \,, & \quad \textnormal{if } \, |e_-(\vec k)| \h < \h \Lambda_* \,, \\[5pt] |\varepsilon_+(\vec k)| \ll |\varepsilon_-(\vec k)| \,, & \quad \textnormal{if } \, |e_+(\vec k)| \h < \h \Lambda_* \,. \end{array} \right.
\end{equation}
From Eqs.~\eqref{eq_zero_1}--\eqref{eq_zero_2}, we therefore obtain
\begin{equation}
 \left\{ \begin{array}{ll} \Psi_{\rm s}(\vec k) \approx \displaystyle \frac{\Delta_0}{4 \h |\varepsilon_-(\vec k)|} \approx -\Psi_{\rm t}(\vec k) \,, & \ \textnormal{if } \, |e_-(\vec k)| \h < \h \Lambda_* \,, \\[15pt]
 \Psi_{\rm s}(\vec k) \approx \displaystyle \frac{\Delta_0}{4 \h |\varepsilon_+(\vec k)|} \approx \Psi_{\rm t}(\vec k) \,, & \ \textnormal{if } \, |e_+(\vec k)| \h < \h \Lambda_* \,. \end{array} \right.
\end{equation}
This means, if $\vec k$ is close to the Fermi line of the lower or upper band, then the singlet and triplet amplitudes are of equal magnitude and have the opposite or same sign, respectively. This is indeed clearly seen in Fig.~\ref{fig_order}. In particular, if the Fermi level is above the band crossing ($\mu > 0$), then there is one Fermi line for each band, and hence the ratio between $\Psi_{\rm s}$ and $\Psi_{\rm t}$ changes sign in the Brillouin zone as shown in the uppermost panel of Fig.~\ref{fig_order}.

\subsection{Gap equation and critical temperature} \label{sec_gapeq}

So far, we have calculated the gap function $\Delta_{ss'}(\vec k)$ and the order parameter $\Psi_{ss'}(\vec k)$ up to the scalar gap parameter $\Delta_0$. The latter was defined in Eq.~\eqref{eq_defd}, which can be written equivalently in terms of the order parameter and a trace over the spin indices as
\begin{equation}
 \Delta_0 = \frac{g }{2} \int \! \db^2 \vec k \,\h \mathrm{Tr} \h \big( \Psi(\vec k) \h [\ii \sigma_y]^{\mh\dagger} \hh \big) \,.
\end{equation}
Inserting our result for the order parameter, Eqs.~\eqref{eq_sitri}--\eqref{eq_sitri_2}, and using that
\begin{equation}
\mathrm{Tr} \h \big( \h \hat{\vec g}(\vec k) \cdot \vec\sigma \big) = 0 \,,
\end{equation}
we obtain immediately
\begin{align}
 \Delta_0 & = \frac{g }{2} \int \! \db^2 \vec k\,\h 2 \hh \Psi_{\rm s}(\vec k)  \\[3pt]
 & = \frac{g }{2} \, \Delta_0 \int \! \db^2 \vec k \, \big( \gamma(\varepsilon_+(\vec k)) + \gamma(\varepsilon_-(\vec k)) \hh \big) \,,
\end{align}
which is equivalent to the scalar gap equation
\begin{equation}
 1 = \frac{g }{2} \int \! \db^2 \vec k \,\sum_n \frac{1}{2 \h \varepsilon_n(\vec k)} \h \tanh \mh \left( \frac{\beta \varepsilon_n(\vec k)}{2} \right). \label{eq_gapeq}
\end{equation}
Note that this equation agrees with the standard form of the gap equation in the SU(2)-symmetric case (see e.g.~Ref.~\onlinecite{Vollhardt}). The right-hand side depends on the gap parameter $\Delta_0$ through the mean-field energies $\varepsilon_n(\vec k)$ (see Eq.~\eqref{eq_eigenvalues}). Equation \eqref{eq_gapeq} can be used to calculate $\Delta_0$ \linebreak

\pagebreak \noindent
depending on the inverse temperature $\beta$, the chemical potential $\mu$ and the coupling constant $g$.

Before coming to the solution of Eq.~\eqref{eq_gapeq} in the zero-temperature limit, we remark that the gap equation also allows to estimate the critical temperature $T_{\rm c}$\h, which is defined as the temperature where the gap vanishes.\cite{Vollhardt} In the limit $\Delta_0 \to 0$, we obtain from Eq.~\eqref{eq_gapeq} the {\itshape linearized gap equation}
\begin{equation} \label{lge}
\begin{aligned}
 1 = \frac{g }{2} \int \! \db^2 \vec k \, \sum_n \frac{1}{2 \h (E_n(\vec k) - \mu)} \h \tanh \! \left( \frac{E_n(\vec k) - \mu}{2 \h k_B T_{\rm c}} \right),
\end{aligned}
\end{equation}
where $E_n(\vec k)$ are the eigenvalues of the free Hamiltonian. In terms of the total density of states \eqref{total_dos}, we can write this equation as
\begin{equation}
 1 = \frac{g }{2} \int_{\mu-\Lambda_{*}}^{\mu+\Lambda_{*}} \! \de E \, \frac{D(E)}{2(E-\mu)} \, \tanh \! \left( \frac{E-\mu}{2 k_B T_{\rm c}} \right) , \label{lingap} 
\end{equation}
where we have re-introduced the stopping scale $\Lambda_{*}$. Now, provided that $\mu$ is away from the band minimum, one may approximate the density of states by its value at the chemical potential, $D(E) \approx D(\mu)$. 
The gap equation~\eqref{lingap} then yields the standard estimate for the critical temperature \footnote{See Eq.~(3.48) in  Ref.~\onlinecite{Vollhardt}; the factor $2$ in the exponent of our Eq.~\eqref{est_temp} comes from the fact that $D(\mu)$ refers to both bands of the model, and hence in the absence of the spin splitting would be twice the density of states defined in Ref.~\onlinecite{Vollhardt}.}
\begin{equation} \label{est_temp}
 k_B T_{\rm c} = C_0 \h \Lambda_{*} \exp \mh \left( -\frac{2}{g \hh D(\mu)} \right) ,
\end{equation}
with $C_0 \approx 1.134$. The parameter $\Lambda_*$, which in pure mean-field studies is usually introduced by hand as a ``cutoff energy'' (sometimes identified with the Debye frequency\cite{Vollhardt}), has in our approach a concrete meaning as the stopping scale of the RG flow.

\subsection{Solving the gap equation}

In this subsection, we will solve the scalar gap equation \eqref{eq_gapeq} in the zero-temperature limit. We will first provide analytical expressions for the asymptotics of the solution, and then present our numerical solution for $\Delta_0$ as a function of $\mu$ and $g $\h. For $\beta \to \infty$, Eq.~\eqref{eq_gapeq} reduces to
\begin{equation}
 1 = \frac{g }{4} \int \! \db^2 \vec k \,\sum_n \frac{1}{\sqrt{(E_n(\vec k) - \mu)^2 + \Delta_0^2}} \,.
\end{equation}
In terms of the density of states \eqref{total_dos}, we can write this equivalently as
\begin{equation} \label{gap}
 1 = \frac{g }{4} \int_{\mu - \Lambda_{*}}^{\mu + \Lambda_{*}} \de E \, \frac{D(E)}{\sqrt{(E-\mu)^2 + \Delta_0^2}} \,.
\end{equation}
We are most interested in the solution of the gap equation for $\mu < 0$, and in particular near the band minimum where the density of states diverges. For $E < 0$, the dispersion of the tight-binding model can be approximated by the ideal Rashba model \eqref{eq_rashba}, whose density of states is given by Eq.~\eqref{dos_explicit}. By putting this into Eq.~\eqref{gap}, we obtain
\begin{equation} \label{gap_2}
\begin{aligned}
 1 & = g \, \frac{\sqrt 3}{16 \pi} \h \frac{(a_0 k_{\rm R})^2}{E_{\rm R}} \int_{\mu - \Lambda_{*}}^{\mu + \Lambda_{*}} \de E \\[2pt]
 & \quad \, \times \frac{1}{\sqrt{1 + E / E_{\rm R}}} \, \frac{1}{\sqrt{(E-\mu)^2 + \Delta_0^2}} \,.
\end{aligned}
\end{equation}
Note that $\mu$ and $E$ are measured relative to the band crossing at $\vec k = \vec 0$, while the minimum of the lower band has the negative energy $E = -E_{\rm R}$. For simplicity, we now ignore the integration boundaries depending on $\Lambda_*$ and instead integrate over the whole interval \linebreak $-E_{\rm R} \leq E \leq 0$. Furthermore, as we are interested in the case where $\mu \approx -E_{\rm R}$, we define the dimensionless variables
\begin{equation}
 \bar \mu \equiv \frac{\mu + E_{\rm R}}{E_{\rm R}} \,, \qquad \bar E \equiv \frac{E + E_{\rm R}}{E_{\rm R}} \,, \qquad \bar \Delta \equiv \frac{\Delta_0}{E_{\rm R}} \,,
\end{equation}
as well as the dimensionless coupling constant
\begin{equation}
 \bar g \h \equiv \h g \, \frac{\sqrt 3}{16 \pi} \h \frac{(a_0 k_{\rm R})^2}{E_{\rm R}} \,.
\end{equation}
In terms of these new variables, we can write the gap equation \eqref{gap_2} more compactly as
\begin{equation} \label{gap_4}
 1 = \bar g \int_0^1 \! \de \bar E \, \frac{1}{\sqrt{\bar E\!\!\phantom{\raisebox{1pt}{$^2$}}}} \, \frac{1}{\sqrt{(\bar E-\bar \mu\hh)\raisebox{1pt}{$^{2}$} + \bar \Delta\raisebox{1pt}{\mh$^{2}$}}} \,.
\end{equation}
We now analyze the asymptotics of the solution of this equation, focusing on two particular cases: $1 \h \gg \h \bar \mu \h \gg \h \bar g^{\hh2}$ \h and \h$\bar \mu \h = \h 0$\h.

\bigskip \noindent
{\bfseries Case 1:} $1 \h \gg \h \bar \mu \h \gg \h \bar g^{\hh2}$\h{\bfseries.}
We substitute in Eq.~\eqref{gap_4}
\begin{equation}
 x = \bar E / \bar \mu \,, \qquad \delta = \bar \Delta / \bar \mu \,. 
\end{equation}
Then, we can write the gap equation as
\begin{equation} \label{gap_5}
 1 = \frac{\bar g}{\sqrt{\bar \mu\!\!\phantom{(}}} \, \int_0^{1/\bar\mu} \de x \, \frac{1}{\sqrt{x}} \, \frac{1}{\sqrt{(x-1)^2 + \delta^2}} \,.
\end{equation}
By our case assumptions, the prefactor on the right-hand side is small, i.e.,
\begin{equation}
 \frac{\bar g}{\sqrt{\bar \mu\!\!\phantom{(}}} \ll 1 \,.
\end{equation}
Therefore, the integral in Eq.~\eqref{gap_5} must be large, which in turn is only possible if $\delta$ is small. We split the integral into two parts (using that $\bar \mu < 1/2$):
\begin{equation}
 \int_0^{1/\bar \mu} \de x = \int_2^{1/\bar \mu} \de x + \int_0^2 \de x \,.
\end{equation}
The first integral can be estimated as follows, using that $\delta \ll 1$ and $\bar \mu \ll 1$:
\begin{equation}
\begin{aligned}
 & \int_2^{1/\bar \mu} \de x \, \frac{1}{\sqrt{x}} \, \frac{1}{\sqrt{(x-1)^2 + \delta^2}} \\[2pt]
 & \approx \h \int_2^{\infty} \de x \, \frac{1}{\sqrt{x}} \, \frac{1}{x - 1} \h = \h 2 \h \arsinh \h (1) \,, \label{c1}
\end{aligned}
\end{equation}
where we have obtained the last result using {\itshape Mathema\-{}tica}.\cite{Mathematica} In the second integral, we substitute $y = x-1$:
\begin{align}
 & \int_0^2 \de x \, \frac{1}{\sqrt{x}} \, \frac{1}{\sqrt{(x-1)^2 + \delta^2}} \nonumber \\[3pt]
 & = \int_{-1}^1 \de y \, \frac{1}{\sqrt{y^2 + \delta^2}} \, \frac{1}{\sqrt{1 + y}} \\[3pt]
 & = \int_{0}^1 \de y \, \frac{1}{\sqrt{y^2 + \delta^2}} \h \left( \frac{1}{\sqrt{1 + y}} + \frac{1}{\sqrt{1 - y}} \, \right).
\end{align}
This integral can in turn be split into two parts: one which is singular for $\delta \to 0$,
\begin{equation}
 \int_{0}^1 \de y \, \frac{2}{\sqrt{y^2 + \delta^2}} \h = \h 2 \h \arsinh \mh \left(\frac 1 \delta\right) \h \approx \h 2 \h \ln \mh \left( \frac 2 \delta \right) , \label{c2}
\end{equation}
and one which approaches a constant value,
\begin{equation}
\begin{aligned}
 & \int_0^1 \de y \, \frac{1}{y} \left( \frac{1}{\sqrt{1 + y}} + \frac{1}{\sqrt{1 - y}} - 2 \, \right) \\[6pt]
 & = -2 \h \arsinh\h(1) + 4 \h \ln \h(2) \,. \label{c3}
\end{aligned}
\end{equation}
This last integral has again been evaluated using {\itshape Mathema\-{}tica}.\cite{Mathematica} Adding Eqs.~\eqref{c1}, \eqref{c2}, and \eqref{c3} yields
\begin{equation}
\begin{aligned}
 & \int_0^{1/\bar \mu} \de x \, \frac{1}{\sqrt{x}} \, \frac{1}{\sqrt{(x-1)^2 + \delta^2}} \\[3pt]
 & \approx \h 2 \h \ln \mh \left(\frac 2 \delta\right) + 4 \h \ln\h(2) \h = \h 2 \h \ln \mh \left(\frac 8 \delta\right).
\end{aligned}
\end{equation}
The gap equation \eqref{gap_5} can thus be approximated as
\begin{equation}
 1 = \frac{2 \bar g}{\sqrt{\bar \mu\!\!\phantom{(}}} \, \ln \mh \left(\frac 8 \delta\right).
\end{equation}
From this, we obtain the estimate
\begin{equation}
 \delta = 8 \h \exp\mh\bigg( {- \frac{\sqrt{\bar \mu\!\!\phantom{(}}}{2 \bar g}} \h \bigg),
\end{equation}
which is equivalent to
\begin{equation} \label{result_1}
 \bar \Delta = 8 \h \bar \mu \h \exp\mh\bigg( {- \frac{\sqrt{\bar \mu\!\!\phantom{(}}}{2 \bar g}} \h \bigg).
\end{equation}
In particular, by our case assumption \h$\bar \mu \hh \gg \hh \bar g^2$\h, we can verify a posteriori the condition $\delta \ll 1$ which we have used in the derivation. Furthermore, by noting that the density of states, Eq.~\eqref{dos_explicit}, can be written for $\bar \mu < 1$ as
\begin{equation}
 D_{\rm R}(\bar \mu) \h = \h \frac{4 \h \bar g}{\raisebox{1pt}{$g $}} \, \frac{1}{\sqrt{\bar \mu\!\!\phantom{(}}} \,,
\end{equation}
we see that the result \eqref{result_1} is equivalent to
\begin{equation}
 \bar \Delta \h = \h 8 \h \bar \mu \, \exp \mh \left( -\frac{2}{g \hh D_{\rm R}(\bar \mu)} \right) ,
\end{equation}
or in terms of the original parameters,
\begin{equation}
 \Delta_0 = 8 \left(\mu + E_{\rm R} \right) \exp \mh \left( -\frac{2}{g \hh D_{\rm R}(\mu)} \right) .
\end{equation}
Note in particular the exponent, which coincides with the usual exponent in the SU(2) symmetric case.

\bigskip \noindent
{\bfseries Case 2:} $\bar \mu = 0$\h{\bfseries.} 
This case corresponds to the situation where the chemical potential is precisely at the band minimum. We then obtain from Eq.~\eqref{gap_4},
\begin{equation}
 1 = \bar g \, \int_0^1 \de \bar E \, \frac{1}{\sqrt{\bar E\!\!\phantom{\raisebox{1pt}{$^2$}}}} \, \frac{1}{\sqrt{\bar E\raisebox{1pt}{$^2$} + \bar \Delta\mh\raisebox{1pt}{$^2$} \phantom{(}\!\!}} \,.
\end{equation}
By substituting $x = \bar E/\bar \Delta$, this is equivalent to
\begin{equation} \label{from_here}
 1 = \frac{\bar g}{\sqrt{\bar \Delta\!\!\phantom{\raisebox{1pt}{$^2$}}}} \int_0^{1/\bar \Delta} \de x \, \frac{1}{\sqrt x} \h \frac{1}{\sqrt{x^2 + 1}} \,.
\end{equation}
This integral has two contributions,
\begin{equation}
 \int_0^{1/\bar \Delta} \de x = \int_0^\infty \de x - \int_{1/\bar \Delta}^{\infty} \de x \,.
\end{equation}
The first integral yields a constant,
\begin{equation}
 \int_0^{\infty} \de x \, \frac{1}{\sqrt x} \h \frac{1}{\sqrt{x^2 + 1}} = C \,, \label{e1}
\end{equation}
which can be evaluated using {\itshape Mathematica}\cite{Mathematica} as
\begin{equation}
 C \h = \h \frac{8}{\sqrt \pi} \, \Gamma(5/4)^2 \h \approx \h 3.71 \,.
\end{equation}
The second integral can be estimated as follows, assuming that $\bar \Delta \ll 1$:
\begin{equation}
 \int_{1/\bar \Delta}^{\infty} \h \de x\, \frac{1}{\sqrt x} \h \frac{1}{\sqrt{x^2 + 1}} \h \approx \h \int_{1/\bar \Delta}^{\infty} \, \frac{\de x}{x^{\nicefrac 3 2}} \h = \h 2 \h \sqrt{\bar \Delta} \,. \label{e2}
\end{equation}
Combining Eqs.~\eqref{e1} and \eqref{e2}, we obtain
\begin{equation}
 \int_{0}^{1/\bar \Delta} \h \de x \, \frac{1}{\sqrt x} \h \frac{1}{\sqrt{x^2 + 1}} \h \approx \h C - 2 \h \sqrt{\bar \Delta} \,.
\end{equation}
The gap equation \eqref{from_here} now reduces to
\begin{equation}
 1 = \frac{\bar g}{\sqrt{\bar \Delta\!\!\phantom{\raisebox{1pt}{$^2$}}}} \h \Big( C - 2 \sqrt{\bar \Delta\!\!\phantom{\raisebox{1pt}{$^2$}}} \, \Big) \approx \frac{\bar g}{\sqrt{\bar \Delta\!\!\phantom{\raisebox{1pt}{$^2$}}}} \, C \,,
\end{equation}
where the last estimate applies again for $\bar \Delta \ll 1$. Thus, we obtain
\begin{equation} \label{result_2}
 \bar \Delta = (\bar g \h C)^2 \,,
\end{equation}
or in terms of the original parameters,
\begin{equation}
 \Delta_0 = \frac{g ^2}{E_{\rm R}} \h \bigg( \frac{\sqrt 3 \h C}{16 \pi} \bigg)^{\!\!2} \h (a_0 k_{\rm R})^4 \,. \label{res2}
\end{equation}
In particular, the condition $\bar \Delta \ll 1$ is fulfilled for \mbox{$\bar g \ll 1$,} which means that the above calculation (just as the discussion in Case 1) is only valid for sufficiently small coupling parameters.

\begin{figure}[t]
\begin{center}
 \includegraphics[width=\columnwidth]{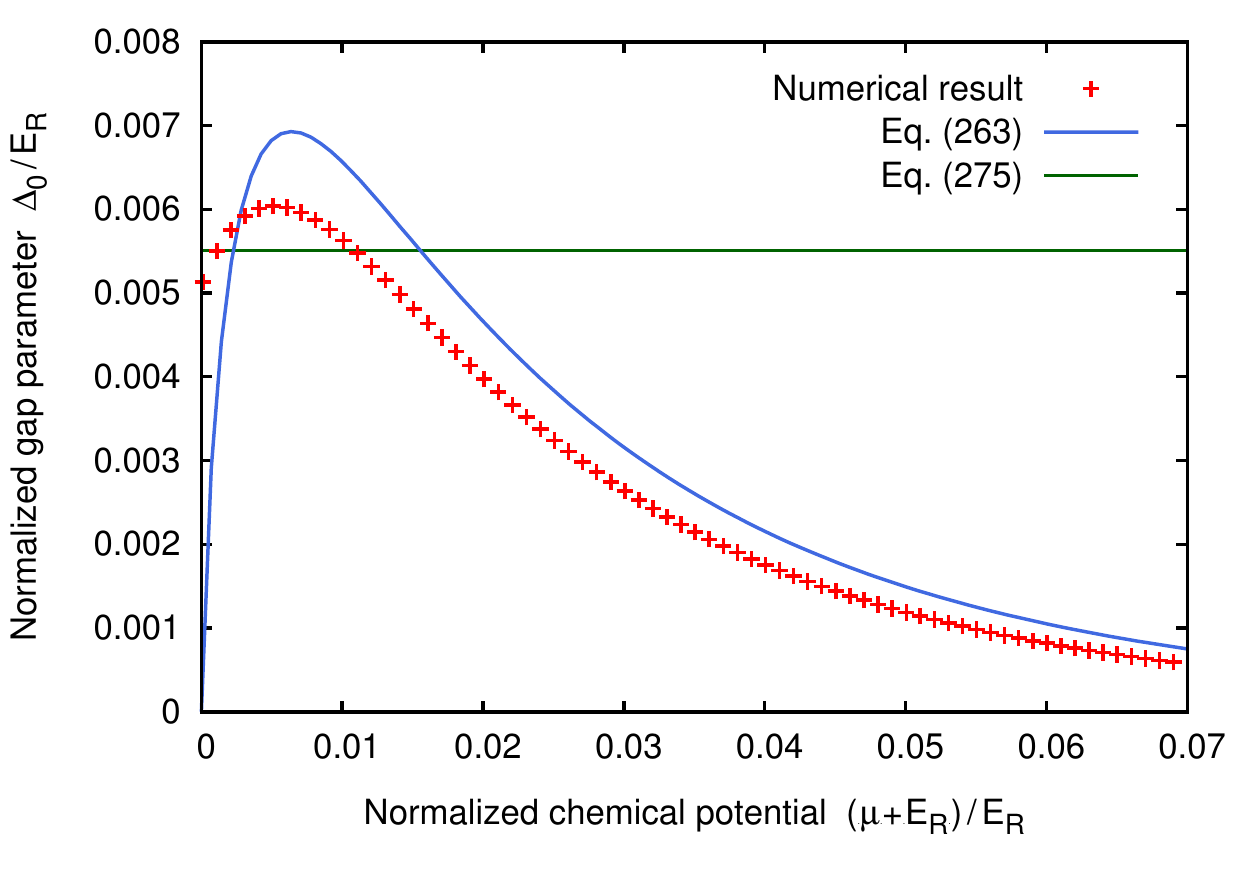}
 \caption{Numerical solution of the scalar gap equation \eqref{gap_4} for $\bar g = 0.02$, and comparison with the analytical results for the asymptotics of the solution, Eqs.~\eqref{result_1} and \eqref{result_2}. \label{fig_mu_dep}}

\bigskip
\bigskip \noindent
 \includegraphics[width=\columnwidth]{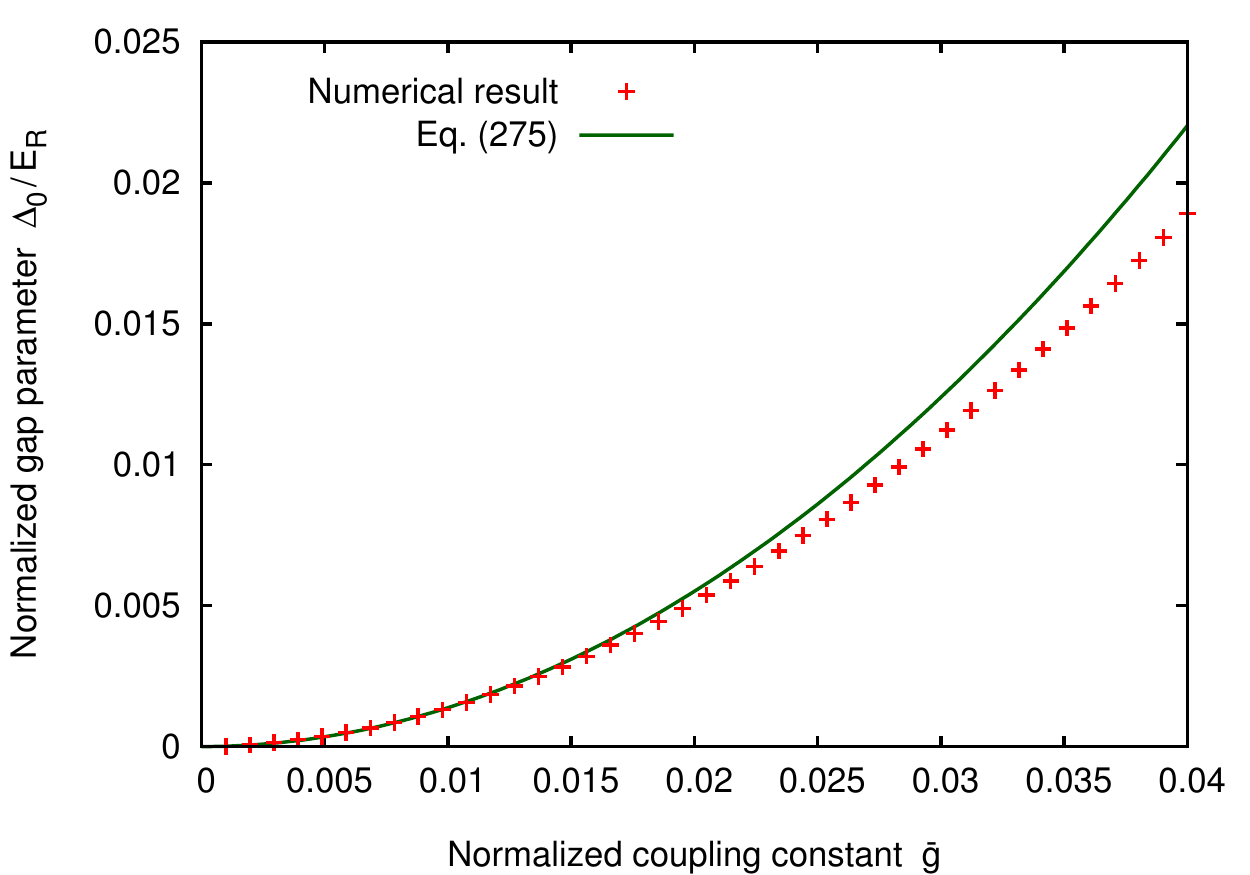}
 \caption{Numerical solution of the scalar gap equation \eqref{gap_4} for $\bar \mu = 0.001$, and comparison with the analytical result for $\bar \mu = 0$, Eq.~\eqref{result_2}. \label{fig_g_dep}}
\end{center}
\end{figure}

For the numerical solution of Eq.~\eqref{gap_4}, we have used the function $\texttt{fzero}$ from {\itshape GNU Octave}.\cite{octave} We have fixed the coupling parameter to a small value, $\bar g = 0.02$, and solved the implicit equation for the gap parameter $\bar\Delta$.
Figure~\ref{fig_mu_dep} shows the resulting dependence of $\bar \Delta$ on the chemical potential~$\bar \mu$. The characteristic features of the asymptotic solution are clearly reproduced in the numerical result: (i) the positive value of $\bar \Delta(\bar \mu = 0)$, (ii) the maximum of $\bar \Delta(\bar \mu)$ at small $\bar \mu$, and (iii) the exponential decay for large $\bar \mu$. Even quantitatively, there is a good agreement between the numerical data and the analytical results given by Eqs.~\eqref{result_1} and \eqref{result_2}.
Finally, we have fixed the chemical potential to a tiny value ($\bar \mu = 0.001$) and plotted the dependence of the gap parameter $\bar \Delta$ on the coupling constant $\bar g$. The result is shown in Fig.~\ref{fig_g_dep}. One clearly sees the quadratic dependence on $\bar g$, and the agreement with the analytical result \eqref{result_2}  becomes perfect for small coupling constants.

\bigskip
\section{Conclusion} \label{sec_conclusion}

We have implemented a functional RG flow without SU(2) spin symmetry and used it to analyze the Rashba model with an attractive, local interaction. This model has a Hamiltonian whose kinetic term is not SU(2) spin invariant, while the interaction is invariant, and it is one of the simplest two-band correlated fermion models. Our RG flow results in an effective interaction slightly above the critical scale which is SU(2) invariant and attractive between singlet Cooper pairs of fermions with momenta $\vec k$ and $-\vec k$\h. We have applied mean-field theory to this interaction in order to calculate the gap function and the order parameter. The gap function is a pure singlet function, but the order parameter (defined as the expectation value of a Cooper pair field) has a nontrivial decomposition into singlet and triplet parts.
While it is not surprising that an attractive interaction drives superconductivity, the symmetries of the gap function and the order parameter were not a priori obvious in the case without SU(2) symmetry. Besides these results, our analysis has also provided clarifications about more general theoretical issues, which we summarize and discuss further in the following.

The RG flow employed here does not require any a priori assumption on the type of symmetry breaking that may happen. Indeed, it is a standard level-two truncation of the fermionic RG equations in the symmetric phase.
The fact that a local (hence in momentum space, constant) bare interaction gives rise to a singlet-pairing effective interaction at low scales, and that this interaction is with very high accuracy given by $(\vec k, \h -\vec k)$ pairing, is thus a genuine result and not an input of our analysis.
Besides the truncation, the projection to the Fermi lines performed here (as in most earlier studies) is the main approximation used. This approximation has been justified by power counting for the asymptotics of the flow of one-band models at low scales,\cite{SH00, Metzner} but it is used more generally to make the RG equations amenable to numerics.
In comparison to earlier definitions of this projection for multiband models, a crucial point of our work is a refined Fermi surface projection that makes no assumption on the relative importance of the contributions from different bands, and which is therefore compatible with the transformation between spin and band indices at all scales. It is this property that leads to the SU(2) invariance of the effective interaction. Our projection gives a flow in which at all scales, not only degrees of freedom near to the Fermi surface {\em and in the conduction band}, but also ones near to the Fermi surface and in higher bands (which have a propagator that is non-singular) contribute in an essential way to the flow.

This goes against the intuition in general discussions that ``high-energy degrees of freedom, once integrated over, should no longer explicitly enter the RG equations at lower scales\pp , and one might even worry that some double counting was involved in our procedure. There is, however, no double counting problem here. Whether the above intuition is correct or not depends on the scheme and approximations that are used, and the projection simply has to match the properties of the scheme. In the one-line irreducible scheme that we use, the functional RG equations contain both a single-scale propagator $S_\Lambda$ and a full propagator $G^2_\Lambda$\h. In the former,
the energy must be close to the flowing scale, $|E| \approx \Lambda$\h, but the latter contains only a condition that the energy is at least as big as the RG scale, $|E| \ge \Lambda$\h. There is therefore no room for an additional assumption that all internal band indices save for the low-energy band are removed, and there is also no need for it since any such restriction is automatically enforced by the above-mentioned support properties of $S_\Lambda$ and $G_\Lambda$\h. Conversely, momentum and band index configurations of the external variables that include upper bands may be driven nontrivially in the flow. Specifically, for the Cooper pair interaction, the condition $\vec p_1 + \vec p_2 = \vec 0$ \h and momentum conservation allow that such configurations couple to the singular flow in the particle-particle channel of the conduction band, i.e.~to configurations $(\vec k, \h -\vec k)$ and band indices~$\ell$ such that $|e_{\ell} (\vec k) | \approx \Lambda$\h, no matter whether $e_{n_1}\mh(\vec p_1)$ is small. This is seen explicitly in our equations \eqref{first}--\eqref{third} for the subsystem of couplings $g_{++}$\h, $g_{--}$\h, $g_{+-}$\h, where $\ell=-$\h, and the flow of $g_{--}$\h, i.e.~the coupling of the low-energy degrees of freedom, drags along all the other couplings, so that they all remain equally large at low scales if they are equal at the initial scale (as is the case for a local interaction). More generally, we have shown that our numerical solution of the RG equations is consistent with an analytical resummation of the particle-particle ladder. The latter provides a general explanation for the SU(2) symmetry of the effective interaction.

We remark that in other schemes, the RG equations are arranged differently. In the Polchinski scheme,\cite{ZS} there is indeed only a single-scale propagator on the right-hand side of the flow equation. However, in that hierarchy of equations, the flow for the four-point function is driven by the six-point function. It is well known that the analog of the level-two truncation in this scheme is to truncate the RG equation for the six-point function to the tree term with two four-point functions (see e.g.~Ref.~\onlinecite{ZS}), but this introduces a scale integral over $\Lambda'$ from $\Lambda_0$ to $\Lambda$ in the RG equation for the four-point function, and hence leads to a similar situation as above. In the Wick-ordered scheme,\cite{Sa98, HM} all internal lines are indeed at or below the scale $\Lambda$, so that a simpler projection scheme may suffice. The special kinematics of the Wick-ordered scheme has also been discussed in the treatment of the constrained random phase approximation (cRPA) and downfolding by functional RG methods.\cite{Honerkamp}

Furthermore, our mean-field analysis has shown that one generally has to distinguish between the two different notions of a {\itshape gap function} and an {\itshape order parameter.} The reason why they are different lies in the Bogoliubov transformation: while the gap function can be deduced directly from the superconducting interaction, the order parameter requires the diagonalization of the mean-field Hamiltonian, Eq.~\eqref{eq_mfham}, and therefore depends on the concrete form of its eigenvectors and eigenenergies. Unless the free Hamiltonian is diagonal (such that \h$\vec g(\vec k) = \vec 0$ \h in Eq.~\eqref{eq_Ham}), the resulting order parameter will have a triplet contribution as given explicitly in Eq.~\eqref{eq_res}. This result for the order parameter is not restricted to the particular model, but applies more generally to any time-reversal symmetric Hamiltonian of the form \eqref{start}. Our formulas for the Bogoliubov transformation and the resulting order parameter therefore generalize the results of Ref.~\onlinecite{Sig91} to the case without SU(2) symmetry. Finally, by analytically and numerically solving the scalar gap equation for the Rashba model, we have shown that the gap size attains a constant value as the chemical potential approaches the minimum of the lower band. This constant value grows with the square of the strength of the superconducting interaction.

This work confirms the functional RG as a tool for investigating Fermi liquid instabilities in multiband and less symmetric systems, in particular in the absence of SU(2) symmetry. While we have applied it here to the case of the Rashba model with an attractive onsite interaction, our procedure can be generalized straightforwardly to investigate more complicated interactions such as spin-dependent nearest-neighbor or long-range interactions. Furthermore, it would be interesting to include a Zeeman term in the free Hamiltonian, which in Refs. \onlinecite{Fujimoto08, Sau10, Alicea10} is an ingredient for the predicted topological superconductivity. Combined with mean-field theory, the functional RG approach allows for an unambiguous prediction of the superconducting gap function and the order parameter. We expect this to be useful for an unbiased theoretical
description of the low-temperature properties of correlated electron materials with spin-orbit coupling.

\bigskip
\bigskip

\begin{acknowledgments}
This work was supported by the DFG Research Unit FOR 723. M.\,M.\,S.~is supported by the Grant No.~ERG-AdG-290623. We thank Ryotaro Arita, Mohammad S. Bahramy, Andreas Eberlein, Tilman Enss, Mario Fink, Lukas Janssen, Julian Lichtenstein, Titus Neupert, \linebreak Christian Platt, Daniel D.~Scherer, Ronald Starke, \linebreak Ronny Thomale and Kambis Veschgini for discussions.
\end{acknowledgments}

\appendix

\section{Tight-binding Rashba model} \label{app_Rashba}

In this first appendix, we explain our conventions for the tight-binding description of electronic states on the hexagonal Bravais lattice in two dimensions. We derive the Rashba spin splitting from symmetry conditions, and subsequently construct a minimal tight-binding model which displays Rashba spin splitting near the center of the Brillouin zone. Furthermore, we derive the corresponding symmetry conditions for a two-particle interaction in a second-quantized framework.

\subsection{Hexagonal Bravais lattice} \label{subsec_basis}

\begin{figure}[t]
\begin{center}
\includegraphics[width=0.9\columnwidth]{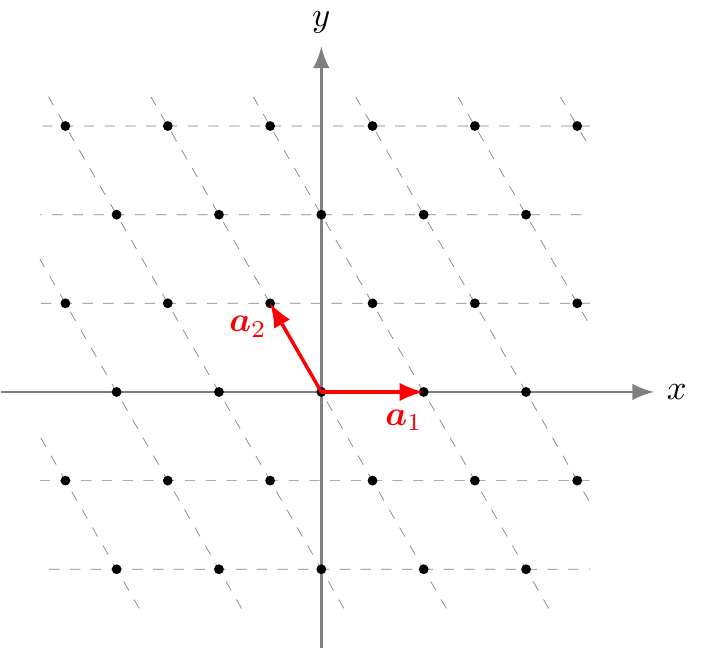}
\caption{Hexagonal Bravais lattice (direct lattice) with primitive vectors $\vec a_1$ and $\vec a_2$. \label{fig_direct_lattice}}

\bigskip
\includegraphics[width=0.8\columnwidth]{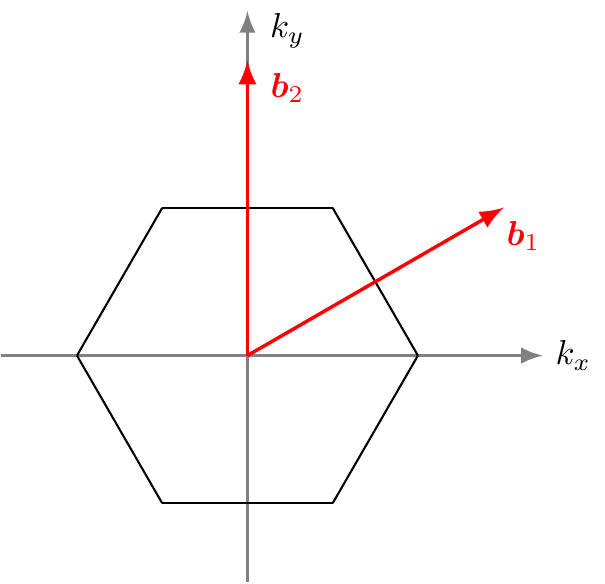}
\caption{First Brillouin zone (dual space) and primitive vectors $\vec b_1, \vec b_2$ of the reciprocal lattice. \label{fig_Brillouin}}
\end{center}
\end{figure}

All our formulas are valid in SI units.\cite{SI} The {\itshape hexagonal Bravais lattice} in two dimensions (Ref.~\onlinecite[Table 2.2]{Levy}) is defined by
\begin{equation}
 \Gamma = \big\{\vec R = r_1 \vec a_1 + r_2 \vec a_2 \h ; \, r_1, r_2 \in \mathbb Z \big\} \,,
\end{equation}
in terms of the primitive vectors
\begin{align}
 \vec a_1 & = a_0 \, (1, \h 0)^{\rm T} \,, \\[2pt]
 \vec a_2 & = a_0 \, \bigg({-\frac 1 2} \h, \, \frac{\sqrt 3}{2} \h \bigg)^{\!\!\rm T},
\end{align}
where $a_0$ is the lattice constant. This Bravais lattice is also referred to as the {\itshape direct lattice} (or {\itshape direct space}). By contrast, the {\itshape reciprocal lattice} $\Gamma^*$ consists of all vectors~$\vec K$ with the property
\begin{equation}
 \e^{\ii\vec K \cdot \vec R} = 1 \quad \forall \, \vec R \in \Gamma \,.
\end{equation}
It is given explicitly by
\begin{equation}
 \Gamma^* = \big\{\vec K = r_1 \vec b_1 + r_2 \vec b_2 \h ; \, r_1, r_2 \in \mathbb Z \big\} \,,
\end{equation}
where the primitive vectors of the reciprocal lattice are defined by the condition ($i, j \in \{1, 2\}$)
\begin{equation}
 \vec a_i \cdot \vec b_j = 2\pi \, \delta_{ij} \,,
\end{equation}
and given explicitly by
\begin{align}
 \vec b_1 & = \frac{2\pi}{a_0} \, \bigg(1 \h, \, \frac{1}{\sqrt 3} \h \bigg)^{\!\!\rm T} , \\[3pt]
 \vec b_2 & = \frac{2\pi}{a_0} \, \bigg(0 \h, \, \frac{2}{\sqrt 3} \h \bigg)^{\!\!\rm T} .
\end{align}
The {\itshape (first) Brillouin zone} $\mathcal B \subseteq \mathbb R^2$ (also referred to as {\itshape dual space} or {\itshape Bloch momentum space}) is defined as the set of all points $\vec k$ which lie closer to the origin $\vec K = \vec 0$ than to any other reciprocal lattice vector. Defining for $\vec K \in \Gamma^*$ the set
\begin{equation}
 \mathcal B_{\vec K} = \{ \vec k + \vec K \,; \, \vec k \in \mathcal B \} \,,
\end{equation}
which is the Brillouin zone $\mathcal B$ shifted by the reciprocal lattice vector $\vec K$, we have for all $\vec K, \vec K' \in \Gamma^*$ the two properties
\begin{equation}
 \mathcal B_{\vec K} \cap \mathcal B_{\vec K'} = \emptyset \quad \textnormal{if} \ \vec K \not = \vec K' \,,
\end{equation}
and
\begin{equation}
 \bigcup_{\vec K \in \Gamma^*} \mathcal B_{\vec K} = \mathbb R^2 \,.
\end{equation}
The hexagonal Bravais lattice $\Gamma$ and the corresponding Brillouin zone $\mathcal B$ are shown in Figs.~\ref{fig_direct_lattice} and \ref{fig_Brillouin}, respectively.

We consider a simple two-band model described by a non-interacting Hamiltonian $\hat H$ acting on wave functions defined on the Bravais lattice. \footnote{For the purposes of this article, we may take such a model Hamiltonian $\hat H$ defined on the Bravais lattice as our starting point. On a more fundamental level, such a description should be derived from a fundamental Hamiltonian for the electrons in the periodic potential of the nuclei, which are assumed to be fixed at the sites of the Bravais lattice. This fundamental Hamiltonian acts on electronic wave functions defined on the continuous three-dimensional space~$\mathbb R^3$ (i.e., {\itshape position space}). For any lattice vector $\vec R$, the Hamiltonian matrix $H(\vec R)$ then corresponds to matrix elements of the fundamental Hamiltonian between two Wannier functions centered at lattice sites with distance vector $\vec R$. In a \mbox{tight-binding} description, one neglects these matrix elements if the distance $|\vec R|$ exceeds a few 
lattice sites (see Ref.~\onlinecite[\mbox{Chap.~1}]{Hubbard})}
In this model, any state vector $\ket{\psi}$ can be characterized by its wave function in direct space or in dual space,
\begin{align}
 \psi(\vec R, s) & = \langle \vec R, s \h | \h \psi \rangle \,, \label{wave_1} \\[3pt]
 \psi(\vec k, s) & = \langle \vec k, s \h | \h \psi \rangle \,. \label{wave_2}
\end{align}
Here, $\vec R \in \Gamma$ is a direct lattice vector, $s \in \{\uparrow, \downarrow\}$ a spin index and $\vec k \in \mathcal B$ a dual vector (or Bloch momentum). In the following (as in the main text) we also call $\vec k$ a ``momentum'' (or ``wave vector''), although strictly speaking {\itshape momentum space} refers to the continuous three-dimensional space~$\mathbb R^3$. In particular, momentum space is related by Fourier transformation to {\itshape position space} (see footnote 93), just as the dual space $\mathcal B$ is related to the direct space~$\Gamma$. The wave functions \eqref{wave_1}--\eqref{wave_2} are interrelated by Fourier transformation as follows:
\begin{align}
 \psi(\vec k, s) & = \sum_{\vec R} \psi(\vec R, s) \, \e^{-\ii \vec k \cdot \vec R} \,, \label{ft_1} \\[3pt]
 \psi(\vec R, s) & = \frac{1}{|\mathcal B|} \int_{\mathcal B} \de^2 \vec k \, \psi(\vec k, s) \, \e^ {\ii \vec k \cdot \vec R} \,, \label{ft_2}
\end{align}
where
\begin{equation}
 |\mathcal B| = \bigg( \frac{2\pi}{a_0} \bigg)^{\!\!2} \frac{2}{\sqrt 3}
\end{equation}
 is the area of the Brillouin zone. In particular, the basis vectors in direct space $\ket{\vec R, s}$ and the basis vectors in dual space $\ket{\vec k, s}$ are given by their wave functions in direct space as
\begin{align}
 \langle \vec R', s' \h | \h \vec R, s \rangle & = \delta_{\vec R, \hh \vec R'} \, \delta_{s, \hh s'} \,, \label{coin_1} \\[5pt]
 \langle \vec R', s' \h | \h \vec k, s \rangle & = \e^{\ii \vec k \cdot \vec R'} \h \delta_{s, \hh s'} \,.
\end{align}
The basis vectors in direct space are orthonormal,
\begin{equation}
 \langle \vec R, s \h | \h \vec R', s' \rangle = \delta_{\vec R, \vec R'} \, \delta_{s, s'} \,, \label{coin_2}
\end{equation}
and complete,
\begin{equation}
 \sum_{\vec R} \sum_s \ket{\vec R, s} \bra{\vec R, s} = 1 \,.
\end{equation}
Note that Eq.~\eqref{coin_2} formally coincides with Eq.~\eqref{coin_1}, although the interpretation is different.
Similarly, the basis vectors in dual space are orthonormal and complete in the sense that
\begin{equation}
 \langle \vec k, s \h | \h \vec k', s' \rangle = |\mathcal B| \, \delta^2(\vec k - \vec k') \, \delta_{s, s'} \,,
\end{equation}
and
\begin{equation}
 \frac{1}{|\mathcal B|} \int_{\mathcal B} \de^2 \vec k \, \sum_s \ket{\vec k, s} \bra{\vec k, s} = 1 \,.
\end{equation}
These properties can be derived using the identities
\begin{align}
 \delta_{\vec R, \vec R'} & = \frac{1}{|\mathcal B|} \int_{\mathcal B} \de^2 \vec k \,\h \e^{\ii \vec k \cdot (\vec R - \vec R') } \,, \\[7pt]
 |\mathcal B| \, \delta^2(\vec k - \vec k') & = \sum_{\vec R} \e^{\ii(\vec k - \vec k') \cdot \vec R } \,.
\end{align}
Note that $\delta_{\vec R, \vec R'}$ denotes the Kronecker delta,
\begin{equation}
 \delta_{\vec R, \vec R'} = \left\{ \begin{array}{ll} 1\,, & \textnormal{if} \ \vec R = \vec R' \,, \\[5pt]
 0 \,, & \textnormal{otherwise} \,, \end{array} \right.
\end{equation}
whereas $\delta^2(\vec k - \vec k')$ is the Dirac delta distribution which \linebreak is normalized such that
\begin{equation}
 \psi(\vec k) = \int_{\mathcal B} \de^2 \vec k' \, \delta^2(\vec k - \vec k') \, \psi(\vec k') \,.
\end{equation}
In particular, since $\de^2 \vec k$ has the unit of an inverse area, the Dirac delta distribution in dual space has the unit of an area. The abbreviation \eqref{nmeasure} will be useful in the following to simplify expressions.

Next, we define the Hamiltonian matrices in direct and in dual space. For this purpose, let the translation operators $T_{\vec R'}$, $\vec R' \in \Gamma$, be defined as linear operators acting on direct-space wave functions as
\begin{equation}
 (\hat T_{\vec R'} \psi)(\vec R, s) = \psi(\vec R - \vec R', s) \,.
\end{equation} 
In particular, they act on the basis vectors in direct and respectively dual space as
\begin{align}
 \hat T_{\vec R'} \ket{\vec R, s} & = \ket{\vec R + \vec R', s} \,, \\[5pt]
 \hat T_{\vec R'} \ket{\vec k, s} & = \e^{-\ii \vec k \cdot \vec R'} \ket{\vec k, s} \,.
\end{align}
We assume that the Hamiltonian is invariant under all lattice translations, i.e.,
\begin{equation}
 [\hat T_{\vec R}, \hat H ] = 0 \quad \forall \, \vec R \in \Gamma \,.
\end{equation}
For the matrix elements of the Hamiltonian in direct space, this implies
\begin{align}
 \langle \vec R, s \h | \h \hat H \h | \h \vec R', s' \rangle & = \langle \vec R - \vec R', s \h | \h \hat H \h | \h \vec 0, s' \rangle \\[5pt]
 & \equiv H_{ss'}(\vec R - \vec R') \,, \label{ham_direct}
\end{align}
where $H_{ss'}(\vec R - \vec R')$ is called the {\itshape Hamiltonian matrix in direct space.} On the other hand, in dual space we obtain
\begin{equation}
 \langle \vec k, s \h | \h \hat H \h | \h \vec k', s' \rangle = |\mathcal B| \, \delta^2(\vec k - \vec k') \, H_{ss'}(\vec k) \,,
\end{equation}
where the {\itshape Hamiltonian matrix in dual space} $H_{ss'}(\vec k)$ is related to the Hamiltonian matrix in direct space by
\begin{align}
 H_{ss'}(\vec k) & = \sum_{\vec R} H_{ss'}(\vec R - \vec R') \, \e^{-\ii\vec k\cdot (\vec R - \vec R')} \,, \label{ham_ft_1} \\[5pt]
 H_{ss'}(\vec R - \vec R') & = \int \! \db^2 \vec k \,\h H_{ss'}(\vec k) \, \e^{\ii\vec k \cdot (\vec R - \vec R')} \,. \label{ham_ft_2}
\end{align}
These formulas are analogous to the Fourier transformation of wave functions, Eqs.~\eqref{ft_1}--\eqref{ft_2}.

Any complex $(2 \times 2)$ matrix can be expanded in terms of the identity matrix $\mathbbm 1$ and the Pauli matrices $\vec \sigma = (\sigma_x, \sigma_y, \sigma_z)$. In particular, we can write the Hamiltonian matrix in dual space, $H(\vec k) \equiv H_{ss'}(\vec k)$, as follows:
\begin{equation} \label{exp}
 H(\vec k) = f(\vec k) \h \mathbbm 1 + \vec g(\vec k) \cdot \vec \sigma \,.
\end{equation}
The Hermiticity of the Hamiltonian operator,
\begin{equation}
 \hat H = \hat H^\dagger \,,
\end{equation}
implies that the Hamiltonian matrix in dual space is also Hermitian, i.e.,
\begin{equation}
 H(\vec k) = H^\dagger(\vec k)\,.
\end{equation}
As a consequence, the functions $f(\vec k)$, $g_x(\vec k)$, $g_y(\vec k)$, $g_z(\vec k)$ in Eq.~\eqref{exp} are real valued. In components, the representation \eqref{exp} of the Hamiltonian matrix reads as
\begin{equation} 
 H \equiv \left( \begin{array}{cc} H_{\uparrow \uparrow} & H_{\uparrow \downarrow} \\[5pt] H_{\downarrow \uparrow} & H_{\downarrow \downarrow} \end{array} \right) = \left( \begin{array}{cc} f + g_z & g_x - \ii g_y \\[5pt] g_x + \ii g_y & f - g_z \end{array} \right) \label{mat}
\end{equation}
(where we have suppressed the $\vec k$ dependencies). The Hamiltonian matrix in direct space, $H(\vec R) \equiv H_{ss'}(\vec R)$, can be expanded analogously as
\begin{equation}
 H(\vec R) = f(\vec R) \h \mathbbm 1 + \vec g(\vec R) \cdot \vec \sigma \,.
\end{equation}
However, Hermiticity implies that 
\begin{equation}
 H(\vec R) = H^\dagger(-\vec R) \,,
\end{equation}
hence $f(\vec R)$ and $\vec g(\vec R)$ are not necessarily real (see Tables \ref{tab_direct} and \ref{tab_dual}). The functions $f(\vec R), \, \vec g(\vec R)$ are related to $f(\vec k), \, \vec g(\vec k)$ by Fourier transformation analogous to Eqs.~\eqref{ham_ft_1}--\eqref{ham_ft_2}.

\subsection{Symmetry conditions} \label{sec_sym}

\begin{table*}[ht]
\renewcommand\arraystretch{2.2}
\begin{tabular}{p{0.2\textwidth} p{0.18\textwidth} p{0.29\textwidth} p{0.29\textwidth}}
\hline\hline
& Possible symmetry & \multicolumn{2}{l}{Transformation of basis functions} \\[-0.45cm]
& of the Hamiltonian
& \multicolumn{2}{l}{under the symmetry operation} \\[0.15cm]
\hline
Hermiticity & $\hat H = \hat H\dag$ & --- & --- \\
Time-reversal & $[\hat \varTheta, \hat H] = 0$ & $\hat \varTheta \h \ket{\vec R, s} = \sum_{s'} \ket{\vec R, s' \hh } \, [-\ii\sigma_y]_{s' \mh s}$ & $\hat \varTheta \h \ket{\vec k, s} = \sum_{s'} \ket{{-\vec k}, s' \hh } \, [-\ii\sigma_y]_{s' \mh s}$ \\
(Spatial inversion) & $[\hat P, \hat H] = 0$ & $\hat P \h \ket{\vec R, s} = \ket{{-\vec R}, s}$ & $\hat P \h \ket{\vec k, s} = \ket{{-\vec k}, s}$ \\
Three-fold rotation & $[\hat C_3, \hat H] = 0$ & $\hat C_3 \h \ket{\vec R, s} = \ket{C_3 \vec R, s} \, \e^{-\ii\frac{\pi}{3}s}$ & $\hat C_3 \h \ket{\vec k, s} = \ket{C_3 \h \vec k, s} \, \e^{-\ii\frac{\pi}{3}s}$ \\
Mirror reflection & $[\hat M_x, \hat H] = 0$ & $\hat M_x \h \ket{\vec R, s} = \sum_{s'} \ket{M_x \vec R, s' \hh} \, [\sigma_x]_{s' \mh s}$ & $\hat M_x \h \ket{\vec k, s} = \sum_{s'} \ket{M_x \hh \vec k, s' \hh } \, [\sigma_x]_{s' \mh s}$ \\[0.25cm]
\hline\hline
\end{tabular}
\caption{Possible symmetries of the Hamiltonian and their action on the basis functions. \label{tab_symmetries}}
\end{table*}
\begin{table*}
\renewcommand\arraystretch{2.2}
\begin{tabular}{p{0.2\textwidth} p{0.3\textwidth} p{0.23\textwidth} p{0.23\textwidth}}
\\[-0.6cm]
\hline \hline
& Hamiltonian matrix & \multicolumn{2}{l}{Pauli matrix representation} \\[-0.35cm]
& $H_{ss'}(\vec R) = \langle \vec R, s \h | \h \hat H \h | \h \vec 0, s'\rangle$
& \multicolumn{2}{l}{$H(\vec R) = f(\vec R) + \vec g(\vec R) \cdot \vec \sigma$} \\[0.15cm]
\hline
Hermiticity & $H(\vec R) = H\dag(-\vec R)$ & $f(\vec R) = f^*(-\vec R)$ & $\vec g(\vec R) = \vec g^*(-\vec R)$ \\
Time-reversal & $H(\vec R) = [\ii\sigma_y]\dag \h H^*(\vec R) \, \ii \sigma_y$ & $f(\vec R) = f^*(\vec R)$ & $\vec g(\vec R) = -\vec g^*(\vec R)$ \\
(Spatial inversion) & $H(\vec R) = H(-\vec R)$ & $f(\vec R) = f(-\vec R)$ & $\vec g(\vec R) = \vec g(-\vec R)$ \\
Three-fold rotation & $H(\vec R) = \e^{\ii\frac{\pi}{3}\sigma_z} H(C_3 \vec R) \, \e^{-\ii\frac{\pi}{3}\sigma_z}$ & $f(\vec R) = f(C_3 \vec R)$ & $(C_3 \h \vec g)(\vec R) = \vec g(C_3 \vec R)$ \\
Mirror reflection & $H(\vec R) = \sigma_x \h H(M_x \vec R) \, \sigma_x$ & $f(\vec R) = f(M_x \vec R)$ & $(M_x \h \vec g)(\vec R) = -\vec g(M_x \vec R)$ \\[0.25cm]
\hline\hline
\end{tabular}
\caption{Possible symmetries of the Hamiltonian matrix in direct space. \label{tab_direct}}

\bigskip
\begin{tabular}{p{0.2\textwidth} p{0.3\textwidth} p{0.23\textwidth} p{0.23\textwidth}}
\\[-0.6cm]
\hline \hline
& Hamiltonian matrix & \multicolumn{2}{l}{Pauli matrix representation} \\[-0.35cm]
& $H_{ss'}(\vec k) = \langle \vec k, s \h |\h \hat H \h | \h \vec k, s'\rangle$
& \multicolumn{2}{l}{$H(\vec k) = f(\vec k) + \vec g(\vec k) \cdot \vec \sigma$} \\[0.15cm]
\hline
Hermiticity & $H(\vec k) = H\dag(\vec k)$ & $f(\vec k) = f^*(\vec k)$ & $\vec g(\vec k) = \vec g^*(\vec k)$ \\
Time-reversal & $H(\vec k) = [\ii\sigma_y]\dag \h H^*(-\vec k) \, \ii \sigma_y$ & $f(\vec k) = f^*(-\vec k)$ & $\vec g(\vec k) = -\vec g^*(-\vec k)$ \\
(Spatial inversion) & $H(\vec k) = H(-\vec k)$ & $f(\vec k) = f(-\vec k)$ & $\vec g(\vec k) = \vec g(-\vec k)$ \\
Three-fold rotation & $H(\vec k) = \e^{\ii\frac{\pi}{3}\sigma_z} H(C_3 \h \vec k) \, \e^{-\ii\frac{\pi}{3}\sigma_z}$ & $f(\vec k) = f(C_3 \h \vec k)$ & $(C_3 \h \vec g)(\vec k) = \vec g(C_3 \h \vec k)$ \\
Mirror reflection & $H(\vec k) = \sigma_x \h H(M_x \hh \vec k) \, \sigma_x$ & $f(\vec k) = f(M_x \h \vec k)$ & $(M_x \h \vec g)(\vec k) = -\vec g(M_x \h \vec k)$ \\[0.25cm]
\hline\hline
\end{tabular}
\caption{Possible symmetries of the Hamiltonian matrix in dual space. \label{tab_dual}}
\end{table*}

Besides the lattice translations, we consider the following symmetry operations which are defined by their action on direct-space wave functions (see~Ref.~\onlinecite{Messiah}):
\begin{itemize}
 \item[(i)] {\itshape Time-reversal:}
\begin{equation}
 (\hat \varTheta \psi)(\vec R, s) = \sum_{s'} \left[-\ii\sigma_y\right]_{ss'} \psi^*(\vec R, s') \,,
\end{equation}
where ``${}^*$'' denotes the complex conjugation.
 \item[(ii)] {\itshape Spatial inversion:}
\begin{equation}
 (\hat P \psi)(\vec R, s) = \psi(-\vec R, s) \,,
\end{equation}
which leaves the spin invariant.
 \item[(iii)] {\itshape Three-fold rotation:}
\begin{equation} \label{eq_def_C3}
 (\hat C_3 \psi)(\vec R, s) = \sum_{s'} \left[\e^{-\ii \frac{\pi}{3} \sigma_z} \right]_{ss'} \psi(C_3^{-1} \vec R, s') \,, \hspace{-0.4cm}
\end{equation}
where the matrix $C_3$ acting on $\vec R \in \Gamma$ is given by
\begin{equation} \label{def_C32}
 C_3 = \left( \begin{array}{cc} \cos \frac{2\pi}{3} & -\sin \frac{2\pi}{3} \\[5pt] \sin \frac{2\pi}{3} & \cos \frac{2\pi}{3} \end{array} \right) = \left( \begin{array}{cc} -\frac 1 2 & -\frac{\sqrt{3}}{2} \\[5pt] \frac{\sqrt{3}}{2} & -\frac 1 2 \end{array} \right). \hspace{-0.6cm}
\end{equation}
If we identify the spin indices as \,$\uparrow \ \equiv +1$\,, $\downarrow \ \equiv -1$\,, such that formally,
\begin{equation}
[\sigma_z]_{ss'} = s \, \delta_{ss'} \,,
\end{equation}
we can write Eq.~\eqref{eq_def_C3} more compactly as
\begin{equation}
 (\hat C_3 \h \psi)(\vec R, s) = \e^{-\ii\frac{\pi}{3} s} \, \psi(C_3^{-1} \vec R, s) \,.
\end{equation}
\item[(iv)] {\itshape Mirror reflection:}
\begin{equation}
 (\hat M_x \h \psi)(\vec R, s) = \sum_{s'} \left[\sigma_x\right]_{ss'} \psi(M_x \vec R, s') \,,
\end{equation}
where $M_x$ acting on $\vec R \in \Gamma$ is defined as
\begin{equation} \label{def_Mx2}
 M_x = \bigg( \begin{array}{cc} \! -1 & 0 \\[3pt] \! 0 & 1 \end{array} \bigg).
\end{equation}
This matrix is self inverse, i.e., $M_x^{-1} = M_x$\h.
\end{itemize}

\vspace{3pt} \noindent
The above symmetry operations defined on the two-dimen\-{}sional Bravais lattice $\Gamma$ are derived from their respective counterparts on the three-dimensional hexagonal lattice, $\Gamma \times \mathbb Z$, by the restriction to a single plane (i.e., to the lattice points $(R_x, R_y, R_z)$ with $R_z = 0$). In particular, (iii) and (iv) correspond to the symmetries $C_3$ (rotation by $2\pi/3$ around the $z$ axis) and $\sigma_{\rm v}$ (reflection in the vertical $yz$ plane) of the three-dimensional point-group $C_{3\mathrm v}$, which describes the Rashba semiconductor BiTeI.\cite{Ishizaka, Bahramy}
The transformation properties of the basis functions (in direct and in dual space) under the symmetry operations (i)--(iv) are shown in Table~\ref{tab_symmetries}.

We now assume that the Hamiltonian $\hat H$ is Hermitian and invariant under the symmetries (i), (iii), and (iv).  We shall, however, {\itshape not} assume that the Hamiltonian is invariant under spatial inversion (ii). In fact, the crystal structure of BiTeI lacks inversion symmetry, and this is one main reason---besides the large atomic spin-orbit coupling of the bismuth atoms---for the bulk Rashba spin splitting in this material (see Refs.~\onlinecite{Ishizaka}, \onlinecite{Bahramy}, and the following discussion). The possible symmetries of the Hamiltonian {\itshape operator} (see Table \ref{tab_symmetries}) can be expressed equivalently in terms of the Hamiltonian {\itshape matrix} in direct or in dual space, or in terms of the functions $f$ and $\vec g$ introduced in Appendix~\ref{subsec_basis}. These symmetry conditions are shown in Table \ref{tab_direct} (direct space) and in Table \ref{tab_dual} (dual 
space), where we have included for the sake of completeness also the inversion symmetry.

In the following, we show that the symmetries (i), (iii) and (iv) already imply that near $\vec k = \vec 0$, the model is described by the Rashba Hamiltonian, Eq.~\eqref{eq_rashba}: First, {\itshape Hermiticity} implies that $f(\vec k)$ and $\vec g(\vec k)$ are real functions. By {\itshape time-reversal symmetry,} they satisfy (see Table \ref{tab_dual})
\begin{align}
 f(\vec k) = f(-\vec k) \,, \label{odd_f} \\[5pt]
 \vec g(\vec k) = -\vec g(-\vec k) \,. \label{odd_g}
\end{align}
(Note that inversion symmetry (ii) would imply $\vec g(\vec k) = \vec g(-\vec k)$, which combined with Eq.~\eqref{odd_g} yields $\vec g(\vec k) \equiv \vec 0$. Hence, requiring time-reversal sym\-{}metry and inversion symmetry at the same time necessarily leads to a spin-degenerate band, which is indeed well known, see e.g. Ref.~\onlinecite[Sec.~16.4]{Dresselhaus}) The above equations \eqref{odd_f}--\eqref{odd_g} imply that $f$ is an even function, while $g_x, g_y, g_z$ are odd functions in the Bloch momentum $\vec k$. We can therefore expand them to quadratic order in $\vec k$ as
\begin{align}
 f(\vec k) & = f(\vec 0) + \sum_{\alpha, \h \beta} F_{\alpha\beta} \h k_\alpha \h k_\beta \,, \label{zw_1} \\[3pt]
 g_i(\vec k) & = \sum_\alpha G_{i \alpha} \h k_\alpha \,, \label{zw_2}
\end{align}
where $i \in \{x, y, z\}$, $\alpha, \beta \in \{x, y\}$, and $F_{\alpha \beta}, \,G_{i \alpha}$ are real constants. Without loss of generality (by a constant energy shift), we may set \mbox{$f(\vec 0) = 0$}. In matrix notation, Eqs.~\eqref{zw_1}--\eqref{zw_2} can then be written as
\begin{equation}
 f(\vec k) \h = \h \vec k^{\rm T} F \h \vec k \h = \h (k_x, k_y) \left(\begin{array}{cc} F_{xx} & F_{xy} \\[5pt] F_{yx} & F_{yy} \end{array} \right) \left( \begin{array}{c} k_x \\[5pt] k_y \end{array} \right) \label{zw_3}
\end{equation}
and, respectively,
\begin{equation}
 \vec g(\vec k) \h = \h G \h \vec k \h =\h \left(\begin{array}{cc} G_{xx} & G_{xy} \\[5pt] G_{yx} & G_{yy} \\[5pt] G_{zx} & G_{zy} \end{array} \right) \left( \begin{array}{c} k_x \\[5pt] k_y \end{array} \right) . \label{zw_4}
\end{equation}
We go on to study the consequences of the {\itshape three-fold rotation symmetry}. By Table \ref{tab_dual}, this implies
\begin{align}
 f(\vec k) & = f(C_3 \h \vec k) \,, \label{zw_5} \\[3pt]
 (\bar C_3 \h \vec g)(\vec k) & = \vec g\left(C_3 \h \vec k\right) \,. \label{zw_6}
\end{align}
In the last equation, we have explicitly distinguished between the matrix $C_3$ acting on $\vec k \in \mathcal B$ (which is the $(2 \times 2)$ matrix given by Eq.~\eqref{def_C32}), and the $(3 \times 3)$ matrix $\bar C_3$ acting 
on $\vec g \in \mathbb R^3$, i.e.,
\begin{equation}
 \bar C_3 = \left( \begin{array}{ccc} -\frac 1 2 & -\frac{\sqrt{3}}{2} & 0 \, \\[5pt] \frac{\sqrt{3}}{2} & -\frac 1 2 & 0 \, \\[5pt] 0 & 0 & 1 \, \end{array} \right).
\end{equation}
In terms of the matrices $F$ and $G$, Eqs.~\eqref{zw_5}--\eqref{zw_6} are equivalent to
\begin{align}
 F & = C_3^{\rm T} \h F \h C_3 \,, \\[5pt]
 G & = \bar C_3^{\rm T} \h G \h C_3 \,.
\end{align}
These conditions imply
\begin{align}
 F_{xx} & = F_{yy} \,, \\[3pt]
 F_{xy} & = -F_{yx} \,,
\end{align}
the analogous equations for $G$, and furthermore,
\begin{equation}
 G_{zx} = G_{zy} = 0 \,.
\end{equation}
Hence, the matrices $F$ and $G$ take the form
\begin{equation}
F = \left( \mh \begin{array}{cc} F_{xx} & F_{xy} \\[5pt] -F_{xy} & F_{xx} \end{array} \right), \quad
G = \left( \! \begin{array}{cc} G_{xx} & G_{xy} \\[5pt] -G_{xy} & G_{xx} \\[5pt] 0 & 0 \end{array} \right).
\end{equation}
Finally, we consider the {\itshape mirror reflection symmetry}. From Table \ref{tab_dual}, we obtain the conditions
\begin{align}
f(\vec k) & = f(M_x \h \vec k) \,, \label{zw_7} \\[3pt]
(\bar M_x \h \vec g)(\vec k) & = - \vec g(M_x \h \vec k) \,, \label{zw_8}
\end{align}
where the matrix $\bar M_x$ acting on $\vec g \in \mathbb R^3$ is defined as
\begin{equation} \label{def_Mx}
 \bar M_x = \left( \begin{array}{ccc} -1 & \, 0 \, & \, 0 \, \\[5pt] 0 & \, 1 \, & \, 0 \, \\[5pt] 0 & \, 0 \, & \, 1 \, \end{array} \right).
\end{equation}
In terms of the matrices $F$ and $G$, Eqs.~\eqref{zw_7}--\eqref{zw_8} are equivalent to
\begin{align}
 F & = M_x \h F \h M_x \,, \\[5pt]
 G & = -\bar M_x \h G \h M_x \,.
\end{align}
These conditions imply
\begin{equation}
 F_{xy} = F_{yx} = 0
\end{equation}
and, respectively,
\begin{equation}
 G_{xx} = G_{yy} = 0 \,.
\end{equation}
We are therefore left with only two parameters $F_{xx}$ and $G_{xy}$, in terms of which the matrices $F$ and $G$ are completely determined as
\begin{equation}
F = \left( \begin{array}{cc} F_{xx} & 0 \\[5pt] 0 & F_{xx} \end{array} \right), \quad 
G = \left( \! \begin{array}{cc} 0 & G_{xy} \\[5pt] \! -G_{xy} & 0 \\[5pt] \! 0 & 0 \end{array} \right).
\end{equation}
By inserting these matrices into Eqs.~\eqref{zw_3}--\eqref{zw_4} and using Eq.~\eqref{exp}, we obtain the following Hamiltonian matrix up to quadratic order in $\vec k$:
\begin{align}
 H(\vec k) & = (\vec k^{\rm T} F \h \vec k) \, \mathbbm 1 + (G \h \vec k) \cdot \vec \sigma \\[5pt]
 & = F_{xx} \h (k_x^2 + k_y^2) \, \mathbbm 1 - G_{xy} \, (k_x \sigma_y - k_y \sigma_x ) \,.
\end{align}
This coincides with the Rashba Hamiltonian \eqref{eq_rashba} if we identify the parameters
\begin{align}
 F_{xx} \equiv \frac{E_{\rm R}}{\raisebox{-1pt}{$k_{\rm R}^2$}} \,, \quad \ G_{xy} \equiv -\frac{2 E_{\rm R}}{k_{\rm R}} \,.
\end{align}
Thus, we have shown that near \h$\vec k = \vec 0$, the Rashba Hamiltonian \eqref{eq_rashba} can be derived from symmetry considerations only, by assuming time-reversal symmetry as well as three-fold rotation and mirror reflection symmetries.

\subsection{Minimal tight-binding model} \label{sec_min}

We now construct a minimal tight-binding model on the hexagonal Bravais lattice which is invariant under time-reversal, three-fold rotation, and mirror reflection symmetries (but not under inversion symmetry). By the
argument of the preceding subsection, this model will necessarily be described near $\vec k = \vec 0$ by the Rashba Hamiltonian.
The minimal model will be defined in terms of the six nearest neighbor vectors (see Fig.~\ref{fig_direct_lattice})
\begin{align}
 \vec R_1 & = \vec a_1 \,, \label{nn1} \\[4pt]
 \vec R_2 & = \vec a_1 + \vec a_2 \,, \\[4pt]
 \vec R_3 & = \vec a_2 \,, \\[4pt]
 \vec R_4 & = -\vec a_1 \,, \\[4pt]
 \vec R_5 & = -\vec a_1 - \vec a_2 \,, \\[4pt]
 \vec R_6 & = -\vec a_2 \,. \label{nn6}
\end{align}
We assume that the Hamiltonian matrix in direct space (see Eq.~\eqref{ham_direct}),
\begin{equation} \label{zw_10}
 H_{ss'}(\vec R) = \langle \vec R, s \h | \h \hat H \h | \h \vec 0, s' \rangle \,,
\end{equation}
vanishes unless $\vec R = \vec R_j$ for some $j \in \{1, \ldots, 6 \}$ (this means, there is only ``electron hopping'' between nearest-neighbor sites). Consequently, the Hamiltonian matrix in direct space can be written as
\begin{equation} \label{zw_11}
 H_{ss'}(\vec R) = \sum_{j = 1}^{6} H_{ss'}(\vec R_j) \, \delta_{\vec R, \vec R_j} \,.
\end{equation}
In dual space, this implies by Fourier transformation (see Eq.~\eqref{ham_ft_1})
\begin{equation} \label{zw_11b}
H_{ss'}(\vec k) = \sum_{j = 1}^{6} H_{ss'}(\vec R_j) \, \e^{-\ii \vec k \h \cdot \vec R_j} \,.
\end{equation}
The expansions \eqref{zw_11}--\eqref{zw_11b} hold analogously for the functions $f$ and $\vec g$ defined in Sec.~\ref{subsec_basis}.
We now employ the constraints imposed on these functions by the Hermiticity as well as the time-reversal, three-fold rotation, and mirror reflection symmetries. The 
function $f(\vec R)$ has to be real and satisfy (see Table \ref{tab_direct})
\begin{align}
 f(\vec R) & = f(-\vec R) \,, \label{eq_symm_f_1} \\[5pt]
 f(\vec R) & = f(C_3 \h \vec R) \,, \\[5pt]
 f(\vec R) & = f(M_x \vec R) \label{eq_symm_f_3} \,.
\end{align}
Hence, if $f$ is restricted to nearest-neighbor vectors, it is completely determined by only one real parameter,
\begin{equation}
 f(\vec R_1) \equiv -t \in \mathbb R \,.
\end{equation}
By Eqs.~\eqref{eq_symm_f_1}--\eqref{eq_symm_f_3}, we then have
\begin{equation}
 f(\vec R_i) = -t \quad \forall i \in \{1, \ldots, 6\} \,.
\end{equation}
Therefore, $f$ is given in direct space by
\begin{equation}
 f(\vec R) = \sum_{i = 1}^6 f(\vec R_i) \, \delta_{\vec R, \vec R_i} = -t \, \sum_{i = 1}^6 \delta_{\vec R, \vec R_i} \,, \end{equation}
and in dual space by
\begin{align}
 & f(\vec k) = -t \, \sum_{i = 1}^6 \e^{-\ii \vec k \h \cdot \vec R_i} \\[3pt]
 & = -2t \, \big\{ \cos(\vec k \cdot \vec R_1) + \cos(\vec k \cdot \vec R_3) + \cos(\vec k \cdot \vec R_5) \big\} \,. \label{zw_9}
\end{align}
In terms of the dimensionless quantity $\vec \kappa = a_0 \hh \vec k$,
this can be written equivalently as
\begin{equation} \label{eq_minimal_model_f}
\begin{aligned}
 & f(\vec \kappa) =  \\[2pt]
 & -2 t \, \bigg\{ \cos(\kappa_x) + 2 \cos \! \bigg( \frac 1 2 \h \kappa_x \bigg) \cos \! \bigg( \frac{\sqrt 3}{2} \h \kappa_y \bigg) \bigg\} \,,
\end{aligned}
\end{equation}
where we have used Eqs.~\eqref{nn1}--\eqref{nn6}. Similarly, the function $\vec g(\vec R)$ has to be purely imaginary and satisfy
\begin{align}
 \vec g(\vec R) & = -\vec g(-\vec R) \,, \label{eq_symm_g_1} \\[5pt]
 (\bar C_3 \h \vec g)(\vec R) & = \vec g(C_3 \h \vec R) \,, \\[5pt]
 (\bar M_x \h \vec g)(\vec R) & = -\vec g(M_x \h \vec R) \,. \label{eq_symm_g_3}
\end{align}
In particular, using that $M_x \vec R_1 = -\vec R_1$, we find
\begin{align}
 (\bar M_x \h \vec g)(\vec R_1) & = -\vec g(M_x \vec R_1) \\[5pt]
 & = -\vec g(-\vec R_1) \\[5pt]
 & = \vec g(\vec R_1) \,,
\end{align}
and consequently, by Eq.~\eqref{def_Mx},
\begin{equation}
 g_x(\vec R_1) = 0 \,.
\end{equation}
By restricting $\vec g(\vec R)$ to nearest-neighbor vectors, it is therefore completely determined by two real parameters $\alpha, \gamma \in \mathbb R$, which we define as
\begin{equation}
 g_y(\vec R_1) \equiv \ii \alpha \,, \qquad g_z(\vec R_1) \equiv \ii \gamma \,.
\end{equation}
Thus, $\vec g$ is given in direct space by
\begin{align}
 \vec g(\vec R) & = \sum_{i = 1}^6 \vec g(\vec R_i) \, \delta_{\vec R, \vec R_i} \\[5pt]
 & = \vec g(\vec R_1) \, \delta^-_{\vec R, \hh \vec R_1} + (\bar C_3 \h \vec g)(\vec R_1) \, \delta^-_{\vec R, \h C_3 \vec R_1} \nonumber \\[5pt]
 & \quad \, + (\bar C_3^{\rm T} \vec g)(\vec R_1) \, \delta^-_{\vec R, \h C_3^{\rm T} \mh \vec R_1} \,,
\end{align}
where we have abbreviated
\begin{equation}
 \delta^-_{\vec R, \hh  \vec R'} = \delta_{\vec R, \hh  \vec R'} - \delta_{\vec R, \hh  -\vec R'} \,.
\end{equation}
In components, this is equivalent to
\begin{widetext}
\begin{align}
\left( \mh \begin{array}{c} g_x(\vec R) \\[3pt] g_y(\vec R) \end{array} \mh \right) & =
 \ii\alpha \left\{
 \left( \mh \begin{array}{c} 0 \\[3pt] 1 \end{array} \mh \right) \delta^-_{\vec R, \hh  \vec R_1} + 
 \left( \! \begin{array}{c} \! -\nicefrac {\sqrt 3} 2 \\[3pt] \! - \nicefrac 1 2 \end{array} \! \right) \delta^-_{\vec R, \hh  \vec R_3} + 
 \left( \! \begin{array}{c} \nicefrac {\sqrt 3} 2 \\[3pt] - \nicefrac 1 2 \end{array} \mh \right) \delta^-_{\vec R, \hh  \vec R_5} \right\} \\[5pt]
g_{z}(\vec R) & =
 \ii \gamma \, \Big\{ \delta^-_{\vec R, \hh  \vec R_1} + \delta^-_{\vec R, \hh  \vec R_3} + \delta^-_{\vec R, \hh  \vec R_5} \Big\} \,,
\end{align}
or by Fourier transformation,
\begin{align}
\left( \mh \begin{array}{c} g_x(\vec k) \\[3pt] g_y(\vec k) \end{array} \mh \right) & =
 2 \alpha \, \left\{
 \left( \mh \begin{array}{c} 0 \\[3pt] 1 \end{array} \mh \right) \sin(\vec k \cdot \vec R_1) + 
 \left( \! \begin{array}{c} \! -\nicefrac {\sqrt 3} 2 \\[3pt] \! - \nicefrac 1 2 \end{array} \! \right) \sin(\vec k \cdot \vec R_3) + 
 \left( \! \begin{array}{c} \nicefrac {\sqrt 3} 2 \\[3pt] - \nicefrac 1 2 \end{array} \mh \right) \sin(\vec k \cdot \vec R_5) \right\} \\[5pt]
g_{z}(\vec R) & =
 2 \gamma \, \Big\{ \sin(\vec k \cdot \vec R_1) + \sin(\vec k \cdot \vec R_3) + \sin(\vec k \cdot \vec R_5) \Big\} \,.
\end{align}
With $\vec \kappa$ defined in Eq.~\eqref{dimensionless_kappa}, we obtain the following explicit expressions:
\begin{align}
 g_x(\vec \kappa) & = {-2 \alpha} \, \sqrt 3 \, \cos \! \bigg( \frac 1 2 \h \kappa_x \bigg) \sin\! \bigg( \frac{\sqrt 3}{2} \h \kappa_y \bigg) \,, \label{eq_minimal_model_gx} \\[4pt]
 g_y(\vec \kappa) & = 2 \alpha \, \bigg\{ \sin(\kappa_x) + \sin \! \bigg( \frac 1 2 \h \kappa_x \bigg) \cos \! \bigg( \frac{\sqrt 3}{2} \h \kappa_y \bigg) \bigg\} \,, \label{eq_minimal_model_gy} \\[5pt]
 g_z(\vec \kappa) & = 2 \gamma \, \bigg\{ \sin(\kappa_x) - 2\sin \! \bigg( \frac 1 2 \h \kappa_x \bigg) \cos \! \bigg( \frac{\sqrt 3}{2} \h \kappa_y \bigg) \bigg\} \,. \label{eq_minimal_model_gz}
\end{align}

\vspace{0.5cm}
\end{widetext}

\noindent
In summary, the {\itshape minimal tight-binding model} is defined \linebreak in dual space by Eq.~\eqref{exp}, where the functions $f(\vec k)$ and $\vec g(\vec k)$ are expressed in terms of the real ``hopping'' parameters $t$, $\alpha$, and $\gamma$ by Eqs.~\eqref{eq_minimal_model_f} and \eqref{eq_minimal_model_gx}--\eqref{eq_minimal_model_gz}.

For small wave vectors, more precisely for
\begin{equation}
 |\vec \kappa| = a_0 |\vec k| \ll 1 \,,
\end{equation}
we can expand the functions $f(\vec k)$ and $\vec g(\vec k)$ around $\vec k = \vec 0$. By approximating
\begin{equation}
 \sin x \approx x \,, \qquad
 \cos x \approx 1 - \frac{x^2}{2} \,,
\end{equation}
we obtain from Eqs.~\eqref{eq_minimal_model_f} and \eqref{eq_minimal_model_gx}--\eqref{eq_minimal_model_gz} the following expressions, which are valid to second order in the momentum:
\begin{equation}
 f(\vec \kappa) = -6 t + \frac{3 t}{2} \h (\kappa_x^2 + \kappa_y^2) \,, \label{eq_expand_f}
\end{equation}
and
\begin{align}
 g_x(\vec \kappa) & = -3 \alpha \h \kappa_y \,, \\[6pt]
 g_y(\vec \kappa) & = 3 \alpha \h \kappa_x \,, \\[6pt]
 g_z(\vec \kappa) & = 0 \,.
\end{align}
Note that the parameter $\gamma$ does not appear at all in this second-order expansion. For the Hamiltonian matrix, we thus obtain 
\begin{equation}
H(\vec k) = \frac{3 t}{2} \h a_0^2 \h (k_x^2 + k_y^2) \h \mathbbm 1 + 3 \alpha \h a_0 \h ( k_x \sigma_y - k_y \sigma_x ) \,,
\end{equation}
where we have neglected the constant energy shift $(-6 t)$ in Eq.~\eqref{eq_expand_f}. As expected, the above expression coincides again with the Rashba Hamiltonian \eqref{eq_rashba} if we identify the parameters
\begin{equation}
 E_{\rm R} \equiv \frac{3 \hh \alpha^2}{2 \hh t} \,, \qquad k_{\rm R} \equiv \frac{\alpha}{a_0 \hh t} \,.
\end{equation}
Note that in the main text, we have set the model para\-{}meter~$\gamma$ to zero such that $g_z(\vec k)$ vanishes identically (not only to second order in the momentum). The formulas \eqref{eq_minimal_model_f} and \eqref{eq_minimal_model_gx}--\eqref{eq_minimal_model_gz} then coincide with Eqs.~\eqref{eq_f}--\eqref{eq_gz} in the main text.

\subsection{Second quantization} \label{subsec_sq}

In the previous subsection, we have defined the Hamiltonian $\hat H$ of a minimal tight-binding model by specifying its matrix elements in direct space,
\begin{equation}
 H_{ss'}(\vec R - \vec R') = \langle \vec R, s \h | \h \hat H \h | \h \vec R', s' \rangle \,.
\end{equation}
The (first-quantized) Hamiltonian can thus be written as
\begin{equation}
 \hat H = \sum_{\vec R, \h \vec R'} \h \sum_{s, \h s'} H_{ss'}(\vec R - \vec R') \, |\vec R, s \rangle \langle \vec R', s'| \,.
\end{equation}
By Fourier transformation, we obtain the corresponding representation in dual space,
\begin{equation}
 \hat H = \int \! \db^2 \vec k \, \sum_{s, \h s'} H_{ss'}(\vec k) \, |\vec k, s \rangle \langle \vec k, s'| \,.
\end{equation}
Under second quantization, the Hamiltonian turns into (see e.g.~Ref.~\onlinecite{Giuliani})
\begin{equation}
 \mathcal Q(\hat H) = \sum_{\vec R, \h \vec R'} \h \sum_{s, \h s'} H_{ss'}(\vec R - \vec R') \, \hat a_s\dag(\vec R) \h \hat a_{s'}(\vec R') \,,
\end{equation}
or equivalently,
\begin{equation} \label{second}
 \mathcal Q(\hat H) = \int \! \db^2 \vec k \, \sum_{s, \h s'} H_{ss'}(\vec k) \, \hat a_s\dag(\vec k) \h \hat a_{s'}(\vec k) \,.
\end{equation}
The second-quantized Hamiltonian $\mathcal Q(\hat H)$ is defined as a Hermitian operator by its action on fermionic $N$-particle states,
\begin{equation}
\begin{aligned}
 & \mathcal Q(\hat H) \, \ket{\psi_1} \wedge \ldots \wedge \ket{\psi_N} \\[2pt]
 & = (\hat H \ket{\psi_1}) \wedge \ldots \wedge \ket{\psi_N} + \ldots \\[2pt]
 & \quad \, + \ket{\psi_1} \wedge \ldots \wedge (\hat H \ket{\psi_N}) \,.
\end{aligned}
\end{equation}
The operators
\begin{align}
 \hat a_s(\vec R) & \equiv \hat a(\ket{\vec R, s}) \,, \label{ann_direct} \\[5pt]
 \hat a_s\dag(\vec R) & \equiv \hat a\dag(\ket{\vec R, s}) \label{cre_direct}
\end{align}
annihilate and create, respectively, a basis vector $\ket{\vec R, s}$ with direct lattice vector $\vec R$ and spin $s$. Similarly, the operators
\begin{align}
 \hat a_s(\vec k) & \equiv \hat a(\ket{\vec k, s}) \,, \label{ann_dual} \\[3pt]
 \hat a\dag_s(\vec k) & \equiv \hat a\dag(\ket{\vec k, s}) \label{cre_dual}
\end{align}
annihilate and create, respectively, a basis vector $\ket{\vec k, s}$ with Bloch momentum $\vec k$ and spin $s$. Note that for any single-particle state $\ket{\varphi}$, the annihilation operator $\hat a(\ket{\varphi})$ is defined by its action on fermionic $N$-particle states as 
\begin{equation}
\begin{aligned}
 & \hat a(\ket{\varphi}) \, \ket{\psi_1} \wedge \ldots \wedge \ket{\psi_N} \\[1pt]
 & = \frac{1}{\sqrt N} \sum_{i = 1}^N (-1)^{i-1} \h \langle \varphi \h | \h \psi_i \rangle \\[1pt]
 & \quad \, \times \ket{\psi_1} \wedge \ldots \wedge \widehat{\ket{\psi_i}} \wedge \ldots \wedge \ket{\psi_N} \,,
\end{aligned}
\end{equation}
where the notation in the last line means that the vector $\ket{\psi_i}$ is omitted from the antisymmetrized product. The corresponding creation operator $\hat a\dag(\ket{\varphi})$ is defined by
\begin{equation}
\begin{aligned}
 & \hat a\dag(\ket{\varphi}) \, \ket{\psi_1} \wedge \ldots \wedge \ket{\psi_N} \\[3pt]
 & = \sqrt{N+1} \ \ket{\varphi} \wedge \ket{\psi_1} \wedge \ldots \wedge \ket{\psi_N} \,.
\end{aligned}
\end{equation}
In particular, the maps
\begin{align}
 \ket{\varphi} & \mapsto \hat a(\ket{\varphi}) \,, \\[3pt]
 \ket{\varphi} & \mapsto \hat a\dag(\ket{\varphi}) \,,
\end{align}
are antilinear and linear respectively in the sense that
\begin{align}
 \hat a(\lambda_1 \ket{\varphi_1} + \lambda_2 \ket{\varphi_2}) & = \lambda_1^* \h \hat a(\ket{\varphi_1}) + \lambda_2^* \h \hat a(\ket{\varphi_2}) \,, \label{antilin} \\[5pt]
 \hat a\dag(\lambda_1 \ket{\varphi_1} + \lambda_2 \ket{\varphi_2}) & = \lambda_1 \h \hat a\dag(\ket{\varphi_1}) + \lambda_2 \h \hat a(\ket{\varphi_2}) \,. \label{lin}
\end{align}
As the notation suggests, the creator $\hat a_s\dag(\ket{\varphi})$ is the Hermitian adjoint of the annihilator $\hat a_s(\ket{\varphi})$. The operators \eqref{ann_direct}--\eqref{cre_direct} in direct space are related to their counterparts \eqref{ann_dual}--\eqref{cre_dual} in dual space by
\begin{align}
 \hat a_s(\vec R) & = \int \! \db^2 \vec k \,\h \hat a_s(\vec k) \, \e^{\ii\vec k \cdot \vec R} \,, \\[1pt]
 \hat a\dag_s(\vec R) & = \int \! \db^2 \vec k \,\h \hat a\dag_s(\vec k) \, \e^{-\ii\vec k \cdot \vec R} \,,
\end{align}
which can be shown using the relation
\begin{equation}
 \ket{\vec R, s} = \frac{1}{|\mathcal B|} \int_{\mathcal B} \de^2 \vec k \, |\vec k, s \rangle \, \e^{-\ii\vec k \cdot \vec R}
\end{equation}
and the properties \eqref{antilin}--\eqref{lin}.

\begin{table*}[ht]
\renewcommand\arraystretch{2.2}
\begin{tabular}{p{0.2\textwidth} p{0.38\textwidth} p{0.38\textwidth}}
\hline\hline
& Annihilation operator & Creation operator \\[-0.35cm]
& $\hat a_s(\vec k) = \hat a(\ket{\vec k, s})$ & $\hat a\dag_s(\vec k) = \hat a\dag(\ket{\vec k, s})$ \\[0.15cm]
\hline
Time-reversal & $\hat \varTheta \, \hat a_s(\vec k) \, \hat \varTheta^{-1} = \sum_{s'} \hat a_{s'}(-\vec k) \, [-\ii\sigma_y]_{s' \mh s}$ & $\hat \varTheta \, \hat a\dag_s(\vec k) \, \hat \varTheta^{-1} = \sum_{s'} \hat a\dag_{s'}(-\vec k) \, [-\ii\sigma_y]_{s' \mh s}$ \\
(Spatial inversion) & $\hat P \, \hat a_s(\vec k) \, \hat P^{-1} = \hat a_{s}(-\vec k)$ & $\hat P \, \hat a\dag_s(\vec k) \, \hat P^{-1} = \hat a\dag_{s}(-\vec k)$ \\
Three-fold rotation & $\hat C_3 \, \hat a_s(\vec k) \, \hat C_3^{-1} = \hat a_{s}(C_3 \hh \vec k) \, \e^{\ii\frac{\pi}{3} s}$ & $\hat C_3 \, \hat a\dag_s(\vec k) \, \hat C_3^{-1} = \hat a\dag_{s}(C_3 \hh \vec k) \, \e^{-\ii\frac{\pi}{3} s}$ \\
Mirror reflection & $\hat M_x \, \hat a_s(\vec k) \, \hat M_x^{-1} = \sum_{s'} \hat a_{s'}(M_x \hh \vec k) \, [\sigma_x]_{s' \mh s}$ & $\hat M_x \, \hat a\dag_s(\vec k) \, \hat M_x^{-1} = \sum_{s'} \hat a\dag_{s'}(M_x \hh \vec k) \, [\sigma_x]_{s' \mh s}$ \\[0.25cm]
\hline\hline
\end{tabular}
\caption{Transformation laws of the annihilation and creation operators in dual space. \label{tab_snd}}

\bigskip
\bigskip
\renewcommand\arraystretch{2.2}
\begin{tabular}{p{0.2\textwidth} p{0.16\textwidth} p{0.60\textwidth}}
\hline\hline
& Interaction & Interaction kernel \\[-0.4cm]
& operator $\hat V$
& in dual space (see Eq.~\eqref{def_coeff}) \\[0.15cm]
\hline
Hermiticity & $\hat V = \hat V\dag$ & $V_{s_1 s_2 s_3 s_4}(\vec k_1, \vec k_2, \vec k_3) = V_{s_4 s_3 s_2 s_1}^*(\vec k_1 + \vec k_2 - \vec k_3, \vec k_3, \vec k_2)$ \\[5pt]
Time-reversal & $[\hat \varTheta, \hat V] = 0$ & $V_{s_1 s_2 s_3 s_4}(\vec k_1, \vec k_2, \vec k_3)$ \\[-3pt]
& & \hspace{0.2cm} $= \, \sum_{t_1, \ldots, t_4} [\ii\sigma_y]\dag_{s_1 t_1} \h [\ii\sigma_y]\dag_{s_2 t_2} \, V_{t_1 t_2 t_3 t_4}^*(-\vec k_1, -\vec k_2, -\vec k_3) \, [\ii\sigma_y]_{t_3 s_3} \h [\ii\sigma_y]_{t_4 s_4}$ \\[5pt]
(Spatial inversion) & $[\hat P, \hat V] = 0$ & $V_{s_1 s_2 s_3 s_4}(\vec k_1, \vec k_2, \vec k_3) = V_{s_1 s_2 s_3 s_4}(-\vec k_1, -\vec k_2, -\vec k_3)$ \\[5pt]
Three-fold rotation & $[\hat C_3, \hat V] = 0$ & $V_{s_1 s_2 s_3 s_4}(\vec k_1, \vec k_2, \vec k_3)$ \\[-3pt]
& & \hspace{0.2cm} $= \, \e^{\ii \frac{\pi}{3} s_1} \, \e^{\ii\frac{\pi}{3} s_2} \, V_{s_1 s_2 s_3 s_4}(C_3 \h \vec k_1, C_3 \h \vec k_2, C_3 \h \vec k_3) \, \e^{-\ii\frac{\pi}{3} s_3} \, \e^{-\ii\frac{\pi}{3} s_4}$ \\[5pt]
Mirror reflection & $[\hat M_x, \hat V] = 0$ & $V_{s_1 s_2 s_3 s_4}(\vec k_1, \vec k_2, \vec k_3)$ \\[-3pt]
& & \hspace{0.2cm} $= \, \sum_{t_1, \ldots, t_4} [\sigma_x]_{s_1 t_1} \h [\sigma_x]_{s_2 t_2} \, V_{t_1 t_2 t_3 t_4}(M_x \vec k_1, M_x \vec k_2, M_x \vec k_3) \, [\sigma_x]_{t_3 s_3} \h [\sigma_x]_{t_4 s_4}$ \\[0.3cm]

\hline\hline
\end{tabular}
\caption{Possible symmetries of the two-particle interaction. \label{tab_int}}
\end{table*}

Under symmetries, the annihilation and creation operators transform as
\begin{equation} \label{zw_12}
 \mathcal Q(\hat X) \, \hat a^{(\dagger)}(\ket{\vec k, s}) \, \mathcal Q(\hat X)^{-1} = \hat a^{(\dagger)}(\hat X \h \ket{\vec k, s}) \,,
\end{equation}
where $\hat X = \hat \varTheta, \, \hat P, \, \hat C_3$, or $\hat M_x$\h. In this equation, $\mathcal Q(\hat X)$ denotes the second-quantized symmetry operator, which is defined as an (anti)unitary operator by its action on fermionic $N$-particle states,
\begin{equation}
 \mathcal Q(\hat X) \, \ket{\psi_1} \wedge \ldots \wedge \ket{\psi_N} = \hat X \ket{\psi_1} \wedge \ldots \wedge \hat X \ket{\psi_N} \,.
\end{equation}
Its inverse is given by
\begin{equation}
 \mathcal Q(\hat X)^{-1} = \mathcal Q(\hat X^{-1}) \,.
\end{equation}
Using Eq.~\eqref{zw_12} and the transformation laws for the basis vectors $\ket{\vec k, s}$ shown in Table \ref{tab_symmetries}, one can deduce the explicit transformation laws for the operators $\hat a_s(\vec k)$ and $\hat a\dag_s(\vec k)$ as shown in Table \ref{tab_snd}. Using these transformation laws and the symmetries of the Hamiltonian matrix in dual space (Table \ref{tab_dual}), one recovers again the symmetries of the second-quantized Hamiltonian \eqref{second}, i.e.,
\begin{equation}
 \mathcal Q(\hat X) \h \mathcal Q(\hat H) \h \mathcal Q(\hat X)^{-1} = \mathcal Q(\hat H)
\end{equation}
for $\hat X = \hat \varTheta, \, \hat C_3$ or $\hat M_x$. In the following (as in the main text), we will suppress the operation of second quantization in the notation and simply write, for example, $\hat X \equiv \mathcal Q(\hat X)$. The distinction between first- and second-quantized operators should, however, always be clear from the context.

In addition to the quadratic Hamiltonian \eqref{second}, we consider a quartic (two-particle) interaction of the form
\begin{equation} \label{int}
\begin{aligned}
 \hat V & = \frac 1 2 \sum_{\vec R_1, \ldots \vec R_4} \sum_{s_1, \ldots s_4} V_{s_1 \ldots s_4}(\vec R_1, \ldots, \vec R_4) \\[5pt]
 & \quad \, \times \hat a\dag_{s_1}(\vec R_1) \h \hat a\dag_{s_2}(\vec R_2) \h \hat a_{s_4}(\vec R_4) \h \hat a_{s_3}(\vec R_3)
\end{aligned}
\end{equation}
(note the order of the annihilation operators). We assume that $\hat V$ is translation invariant just as the free Hamiltonian (see Sec.~\ref{subsec_basis}), i.e.,
\begin{equation}
 [\hat T_{\vec R}, \h \hat V] = 0 \quad \forall \, \vec R \in \Gamma \,.
\end{equation}
This implies that
\begin{equation}
\begin{aligned}
 & V_{s_1 \ldots s_4}(\vec R_1, \vec R_2, \vec R_3, \vec R_4) \\[5pt]
 & = V_{s_1 \ldots s_4}(\vec R_1 - \vec R_4, \h \vec R_2 - \vec R_4, \h \vec R_3 - \vec R_4, \h \vec 0) \,,
\end{aligned}
\end{equation}
or in dual space,
\begin{equation} \label{Vk}
\begin{aligned}
 & V_{s_1 \ldots s_4}(\vec k_1, \vec k_2, \vec k_3, \vec k_4) = V_{s_1 \ldots s_4}(\vec k_1, \vec k_2, \vec k_3) \\[5pt]
 & \times |\mathcal B| \, \sum_{\vec K} \delta^2(\vec K + \vec k_1 + \vec k_2 - \vec k_3 - \vec k_4) \,.
\end{aligned}
\end{equation}
Note that in the last line, we formally sum over all reciprocal lattice vectors $\vec K$. The condition that all momenta $\vec k_1, \ldots, \vec k_4$ lie in the first Brillouin zone, however, fixes precisely one vector $\vec K$ which contributes to the sum.
Hence, the translation-invariant interaction term \eqref{int} can be written in dual space as
\begin{equation} \label{def_coeff}
\begin{aligned}
 & \hat V = \frac 1 2 \, \int \! \db^2 \vec k_1 \int \! \db^2 \vec k_2 \int \! \db^2 \vec k_3 \int \! \db^2 \vec k_4 \\[5pt]
 & \times \, \sum_{\vec K} \h |\mathcal B| \, \delta^2(\vec K + \vec k_1 + \vec k_2 - \vec k_3,  \h\vec k_4) \\[2pt]
 & \times \sum_{s_1, \ldots, s_4} V_{s_1 \ldots s_4}(\vec k_1, \vec k_2, \vec k_3) \\[3pt]
 & \times \h \hat a^\dagger_{s_1}(\vec k_1) \h \hat a^\dagger_{s_2}(\vec k_2) \h \hat a_{s_4}(\vec k_4) \h \hat a_{s_3}(\vec k_3) \,. 
\end{aligned}
\end{equation}
This formula corresponds to Eq.~\eqref{def_coeff_m} in the main text.

Without loss of generality, we may assume that the interaction kernel is antisymmetric with respect to its first two and its last two arguments, i.e.,
\begin{align}
 & V_{s_1 s_2 s_3 s_4}(\vec k_1, \vec k_2, \vec k_3) = -V_{s_2 s_1 s_3 s_4}(\vec k_2, \vec k_1, \vec k_3) \\[6pt]
 & = -V_{s_1 s_2 s_4 s_3}(\vec k_1, \vec k_2, \vec k_1 + \vec k_2 - \vec k_3) \,.
\end{align}
Furthermore, we assume that $\hat V$ is Hermitian and invariant under time-reversal, three-fold rotation, and mirror reflection symmetries. These conditions on the operator $\hat V$ translate again---using the transformation properties of the creation and annihilation operators shown in Table~\ref{tab_snd}---into conditions on the interaction kernel, which are shown in Table \ref{tab_int}. (For the sake of completeness, \linebreak

\pagebreak \noindent
the table contains also the spatial inversion, although we do not 
assume that $\hat V$ is invariant under this symmetry.) These 
symmetries apply to the initial interaction, Eq.~\eqref{eq_onsite}, as well as to the scale-dependent effective interaction given by Eq.~\eqref{fund_rel} in the main text.

\section{Statistical field theory} \label{app_not}

In this second appendix, we fix our conventions for the temperature Green functions, in terms of which the renormalization group equations have been formulated in the main text. For the non-expert in statistical field theory, we also provide a brief summary of the most important definitions and properties of temperature Green functions.

\subsection{Temperature Green functions} \label{sec_temperature_green_functions}

For defining the temperature Green functions of our second-quantized tight-binding model, let
\begin{equation} \label{b1}
 \hat K = \hat H - \mu \hat N \,,
\end{equation}
where $\hat H = \hat H^0 + \hat V$ denotes the full Hamiltonian of the interacting many-fermion system (given by Eqs.~\eqref{second} and \eqref{def_coeff}; in Appendix \ref{app_Rashba}, $\hat H^0$ was denoted by $\hat H$ for simplicity). Furthermore, $\mu$ denotes the chemical potential and $\hat N$ the particle-number operator, which is  given in terms of the annihilation and creation operators of Sec.~\ref{subsec_sq} by
\begin{equation}
 \hat N = \int \! \db^2 \vec k \, \sum_s \hat a_s\dag(\vec k) \h \hat a_s(\vec k) \,.
\end{equation}
In terms of these operators, the grand canonical partition function can be written as
\begin{equation} \label{partition_function}
 Z  = \Tr\big(\e^{-\beta \hat K}\big) \,,
\end{equation}
where $\beta = 1/(k_{\rm B} T)$ is the inverse temperature. Furthermore, for any operator $\hat O$ and for $\tau \in \mathbb R$ we define
\begin{equation}
 \hat O(\tau) = \e^{\hat K \tau/ \hbar} \, \hat O \, \e^{-\hat K \tau/\hbar} \,,
\end{equation}
which is {\itshape analogous} to the time evolution in the Heisenberg picture if we identify $\tau = \ii t$. Therefore, $\tau$ is called ``imaginary time.'' For $n \geq 1$ and for $\tau_1,\ldots, \tau_{2n} \in [0, \beta]$, the {\itshape $2n$-point} {\itshape temperature Green function} (or {\itshape imaginary-time Green function}) $G^{2n}$ is defined as
\begin{equation} \label{def_GF}
\begin{aligned}
 & G^{2n}_{s_1 \ldots s_{2n}}(\vec R_1, \tau_1; \ldots; \vec R_{2n}, \tau_{2n}) \\[5pt]
 & = \frac{1}{Z} \, \Tr\big( \e^{-\beta \hat K} \h \mathcal T \big[ \hat a_{s_1}(\vec R_1, \tau_1) \ldots \hat a_{s_n}(\vec R_n, \tau_n) \\[5pt]
& \quad \, \times \hat a\dag_{s_{2n}}(\vec R_{2n}, \tau_{2n}) \ldots \hat a\dag_{s_{n+1}}(\vec R_{n+1}, \tau_{n+1}) \big] \big) \,.
\end{aligned}
\end{equation}
Here, the time-ordering operator $\mathcal T$ is defined with respect to the imaginary-time arguments. Thus, for any $m$ operators $\hat O_1(\tau_1), \ldots, \hat O_m(\tau_m)$, the time-ordered product is given by
\begin{equation}
\begin{aligned}
 & \mathcal T \big[ \hat O_1(\tau_1) \ldots \hat O_m(\tau_m) \big] = \sum_{\pi \in S_m} \mathrm{sgn}(\pi) \\[1pt]
 & \times \varTheta\big(\tau_{\pi(1)} - \tau_{\pi(2)} \big) \ldots \varTheta\big(\tau_{\pi(m-1)} - \tau_{\pi(m)} \big) \\[5pt]
 & \times \hat O_{\pi(1)}(\tau_{\pi(1)}) \ldots \hat O_{\pi(m)}(\tau_{\pi(m)}) \,.
\end{aligned}
\end{equation}
In this formula, $S_m$ denotes the symmetric group of $m$ elements, i.e., the group of all permutations of the set $\{1, \ldots, m\}$, and $\mathrm{sgn}(\pi)$ denotes the character of the permutation $\pi$. Furthermore, $\varTheta$ is the Heaviside step function defined by
\begin{equation}
 \varTheta(\tau_1 -\tau_2) = \left\{ \begin{array}{ll} 1 \,, & \textnormal{if} \ \, \tau_1 > \tau_2 \,, \\[5pt]
 0 \,, & \textnormal{if} \ \, \tau_1 < \tau_2 \,. \end{array} \right.
\end{equation}
Note that the Green functions defined by Eq.~\eqref{def_GF} are dimensionless.
In the following, we list the most important properties of the temperature Green functions (for details, see the standard references~\onlinecite{Fetter, ZinnJustin, Negele, Salmhofer}).

\bigskip
(i) {\itshape Antiperiodicity.} The temperature Green functions satisfy antiperiodic boundary conditions in the following sense: for each $i \in \{1, \ldots, 2n\}$, the values of $G^{2n}$ at $\tau_i = 0$
and $\tau_i = \hbar\beta$ are related by
\begin{equation}
\begin{aligned}
 & G^{2n}_{s_1 \ldots s_{2n}}(\vec R_1, \tau_1 ; \ldots; \vec R_i, 0 ; \ldots \vec R_{2n}, \tau_{2n}) \\[5pt]
 & = -G^{2n}_{s_1 \ldots s_{2n}}(\vec R_1, \tau_1 ; \ldots; \vec R_i, \hbar\beta ; \ldots \vec R_{2n}, \tau_{2n})\,.
\end{aligned}
\end{equation}
Therefore, the temperature Green functions---which were originally defined only for $\tau_i \in [0, \hbar\beta]$ by Eq.~\eqref{def_GF}---can be continued to antiperiodic functions of $\tau_i \in \mathbb R$.

\bigskip
(ii) {\itshape Fourier transformation.} The Fourier transforms of the temperature Green functions are defined by
\begin{align}
 & G^{2n}_{s_1 \ldots s_{2n}}(\vec k_1, \omega_1; \ldots; \vec k_{2n}, \omega_{2n}) \label{Fourier_Tr_1} \\[3pt] \nonumber
 & = \sum_{\vec R_1} \ldots \sum_{\vec R_{2n}} \,\h \frac{1}{(\hbar\beta)^{2n}} \int_{0}^{\hbar\beta} \! \de \tau_1 \ldots \int_0^{\hbar\beta} \! \de \tau_{2n} \\[2pt] \nonumber
 & \quad \, \times G^{2n}_{s_1 \ldots s_{2n}}(\vec R_1, \tau_1; \ldots; \vec R_{2n}, \tau_{2n}) \\[8pt] \nonumber
 & \quad \, \times \e^{-\ii\vec k_1 \cdot \vec R_1 + \ii\omega_1 \tau_1} \ldots \e^{-\ii\vec k_n \cdot \vec R_n + \ii\omega_n \tau_n} \\[5pt] \nonumber
 & \quad \, \times \e^{\ii\vec k_{n+1} \cdot \vec R_{n+1} - \ii\omega_{n+1} \tau_{n+1}} \ldots \e^{\ii\vec k_{2n} \cdot \vec R_{2n} - \ii\omega_{2n} \tau_{2n}} \,,
\end{align}
or conversely,
\begin{align}
 & G^{2n}_{s_1 \ldots s_{2n}}(\vec R_1, \tau_1; \ldots; \vec R_{2n}, \tau_{2n}) \label{Fourier_Tr_2} \\[3pt] \nonumber
 & = \frac{1}{|\mathcal B|^{2n}} \int_{\mathcal B} \de^2 \vec k_1 \ldots \int_{\mathcal B} \de^2 \vec k_{2n} \,\h \sum_{\omega_1} \ldots \sum_{\omega_{2n}} \\[2pt] \nonumber
 & \quad \, \times G^{2n}_{s_1 \ldots s_{2n}}(\vec k_1, \omega_1; \ldots; \vec k_{2n}, \omega_{2n}) \\[8pt] \nonumber
 & \quad \, \times \e^{\ii\vec k_1 \cdot \vec R_1 - \ii\omega_1 \tau_1} \ldots \e^{\ii\vec k_n \cdot \vec R_n - \ii\omega_n \tau_n} \\[5pt] \nonumber
 & \quad \, \times \e^{-\ii\vec k_{n+1} \cdot \vec R_{n+1} + \ii\omega_{n+1} \tau_{n+1}} \ldots \e^{-\ii\vec k_{2n} \cdot \vec R_{2n} + \ii\omega_{2n} \tau_{2n}} \,.
\end{align}
Here, $\omega_1, \ldots, \omega_{2n} \in \mathbbm M = \{ (2n+1) \h \pi / (\hbar\beta) \, ; \, n \in \mathbb N \h \}$ are fermionic Matsubara frequencies due to the antiperiodicity (property (i)) of the temperature Green functions. By our conventions, the Fourier transform of any function has the same dimension as the function itself. In particular, the temperature Green functions in the direct space/time domain and in the dual space/frequency domain are all dimensionless.

\bigskip
(iii) {\itshape Translation invariance.} The temperature Green functions are always translation invariant with respect to the imaginary-time arguments. If the Hamiltonian $\hat H$ is invariant with respect to spatial translations, then the same also applies to the Green functions. In particular, the two-point Green function can then be written as
\begin{equation} \label{eq_two}
 G^{2}_{s_1 s_2}(\vec R_1, \tau_1; \vec R_2, \tau_2) = G^2_{s_1 s_2}(\vec R_1 - \vec R_2, \tau_1 - \tau_2) \,,
\end{equation}
or equivalently, by Fourier transformation,
\begin{equation}
\begin{aligned}
 & G^2_{s_1 s_2}(\vec k_1, \omega_1; \vec k_2, \omega_2) \\[5pt]
 & = G^2_{s_1 s_2}(\vec k_1, \omega_1) \,\h |\mathcal B| \h \delta^2(\vec k_1 - \vec k_2) \, \delta_{\omega_1, \h \omega_2} \,.
\end{aligned}
\end{equation}
Similarly, the four-point Green function can be written in the Fourier domain as
\begin{align}
 & G^4_{s_1 \ldots s_4}(\vec k_1, \omega_1; \ldots; \vec k_4, \omega_4) \\[5pt] \nonumber
 & = G^4_{s_1 \ldots s_4}(\vec k_1, \omega_1; \vec k_2, \omega_2; \vec k_3, \omega_3) \\[6pt] \nonumber
 & \quad \, \times |\mathcal B| \h \sum_{\vec K} \delta^2(\vec K + \vec k_1 + \vec k_2 - \vec k_3 - \vec k_4) \, \delta_{\omega_1 + \omega_2, \h \omega_3 + \omega_4} \,,
\end{align}
where the reciprocal lattice vector $\vec K$ is fixed by the condition that all momenta $\vec k_1, \ldots, \vec k_4$ lie in the first Brillouin zone.

\bigskip
(iv) {\itshape Symmetries.} The Green functions defined by Eq.~\eqref{def_GF} inherit the symmetries of the Hamiltonian \eqref{b1}. In particular, the two-point Green function
\begin{equation}
 G^{2}_{s_1 s_2}(\vec k, \tau)
\end{equation}
(defined as the Fourier transform of Eq.~\eqref{eq_two} with respect to the spatial variables) transforms under symmetries in the same way as the free Hamiltonian matrix $H_{s_1 s_2}(\vec k)$ (see Table \ref{tab_dual}). Moreover, the four-point Green function in the Bloch momentum/time domain
\begin{equation}
 G^{4}_{s_1 \ldots s_4}(\vec k_1, \tau_1; \vec k_2, \tau_2; \vec k_3, \tau_3)
\end{equation}
transforms under symmetries in the same way as the interaction kernel $V_{s_1 \ldots s_4}(\vec k_1, \vec k_2, \vec k_3)$ (see Table \ref{tab_int}). In the frequency domain, time reversal also changes the sign of the Matsubara frequencies, i.e., time-reversal symmetry is expressed by
\begin{equation}
\begin{aligned}
 & G_{s_1 s_2}^2(\vec k, \omega) \\[4pt]
 & = \sum_{t_1, \h t_2} [\ii\sigma_y]^\dagger_{s_1 t_1} \h (G^2)^*_{t_1 t_2}(-\vec k, -\omega) \, [\ii\sigma_y]_{t_2 s_2} \,,
\end{aligned}
\end{equation}
and similar conditions on the higher $2n$-point Green functions. All other symmetry operations act only on the spatial variables and hence leave the frequency arguments unchanged.

\bigskip
(v) {\itshape Feynman graph expansion.} The temperature Green functions have a perturbative expansion in terms of the interaction kernel $V$ and the {\itshape free two-point Green function} (also called the {\itshape non-interacting Green function} or {\itshape covariance}), which is defined as
\begin{equation} \label{eq_formal_prod_equiv}
\begin{aligned}
 & C_{s_1 s_2}(\vec R_1, \tau_1; \vec R_2, \tau_2) \\[3pt]
 & = \frac{1}{Z^0} \h \Tr\big( \e^{-\beta\hat K^0} \h \mathcal T\big[ \hat a_{s_1}(\vec R_1, \tau_1) \h \hat a\dag_{s_2}(\vec R_2, \tau_2) \big] \big) \,,
\end{aligned}
\end{equation}
with $\hat K^0 = \hat H^0 - \mu \hat N$ and
\begin{equation}
 Z^0 = \Tr\big( \e^{-\beta \hat K^0} \big) \,.
\end{equation}
Concretely, the $k$th-order contribution to the $2n$-point Green function is given by the sum of all {\itshape bubble-free} Feynman graphs with $k$ interaction vertices and $2n$ external legs. Here, bubble-free means that every interaction vertex is connected (through a series of free Green function lines and interaction vertices) to {\itshape at least one} external point.

\subsection{Grassmann field integral} \label{subsec_Grassmann}

The {\itshape Grassmann field integral} (also called {\itshape functional integral,} or {\itshape path integral}) is useful for organizing the Feynman graph expansion of the fermionic temperature Green functions in an efficient way. We consider the Grassmann algebra $\mathscr A$ generated by the elements (called {\itshape Grassmann fields})
\begin{equation}
 \psi(x) \,, \ \bar\psi(x) \,,
\end{equation}
where $x = (\vec R, s, \tau)$ is a multi-index containing the lattice site $\vec R$, the spin variable $s$, and the imaginary-time variable $\tau$. We impose antiperiodic boundary conditions,
\begin{align}
 \psi(\vec R, s, \hbar\beta) & \equiv -\psi(\vec R, s, 0) \,, \label{bound_1} \\[3pt]
 \bar\psi(\vec R, s, \hbar\beta) & \equiv -\bar\psi(\vec R, s, 0) \,. \label{bound_2}
\end{align}
Furthermore, we consider the Grassmann algebra $\mathscr S$ of the {\itshape source fields}
\begin{equation}
 \eta(x) \,, \ \bar\eta(x) \,,
\end{equation}
which satisfy the same boundary conditions \eqref{bound_1}--\eqref{bound_2}, and which also anticommute with the fields $\psi(x), \bar\psi(x)$. The {\itshape Green function generator} $\mathcal Z \equiv \mathcal Z[\bar \eta, \eta]$ is an element of $\mathscr S$ defined by
\begin{equation} \label{eq_gen_func}
 \mathcal Z = \frac{1}{\mathcal N} \int \! \de \bar \psi \h \de \psi \,\h \e^{-\langle \bar \psi, \h Q \h \psi \rangle} \, \e^{-\beta V[\bar \psi, \psi] + \langle \bar \eta, \psi \rangle + \langle \eta, \bar \psi \rangle} \,,
\end{equation}
with the normalization constant
\begin{equation}
 \mathcal N = \int \! \de \bar \psi \h \de \psi \,\h \e^{-\langle \bar \psi, \h Q \h \psi \rangle} \,.
\end{equation}
In these equations, $Q = C^{-1}$ is the inverse of the free two-point Green function \eqref{eq_formal_prod_equiv}. Furthermore,
\begin{equation}
 \langle \bar \psi, \h Q \h \psi \rangle = \int \! \de x_1 \int \! \de x_2 \, \bar\psi(x_1) \, Q(x_1, x_2) \h \psi(x_2) \,,
\end{equation}
where the integration over multi-indices is defined as in Eq.~\eqref{int_multi_index}.
Moreover, we have defined
\begin{align}
 & V[\bar\psi, \psi] = \label{eq_toexplain}  \\[3pt]
 & \frac 1 2 \h \frac{1}{\hbar \beta} \int_0^{\beta} \de \tau \! \sum_{\vec R_1, \ldots, \vec R_4} \sum_{s_1, \ldots, s_4} V_{s_1\ldots s_4}(\vec R_1, \ldots,\vec R_4) \nonumber \\[4pt]
 & \times \bar\psi(\vec R_1, s_1, \tau) \h \bar\psi(\vec R_2, s_2, \tau) \h \psi (\vec R_4, s_4, \tau) \h \psi(\vec R_3, s_3, \tau) \,. \nonumber
\end{align}
In terms of the interaction kernel \eqref{is_int_ker},
this can be written equivalently as
\begin{equation}
\begin{aligned}
 & V[\bar \psi, \psi] = \frac 1 2 \h \int \! \de x_1 \int \! \de x_2 \int \! \de x_3 \int \! \de x_4 \\[5pt]
 & \times V(x_1, x_2, x_3, x_4) \, \bar \psi(x_1) \h \bar \psi(x_2) \h \psi(x_4) \h \psi(x_3) \,.
\end{aligned}
\end{equation}
Now, the partition function \eqref{partition_function} coincides with the field-independent term of the Green function generator \eqref{eq_gen_func},
\begin{equation}
 Z = \mathcal Z[0, 0] \,,
\end{equation}
and the temperature Green functions $G^{2n}$ can be represented as Grassmann derivatives of $\mathcal Z[\bar \eta, \eta]$ with respect to the source fields: for $n \geq 1$,
\begin{align}
 & G^{2n}(x_1, \ldots, x_{2n}) = \frac{1}{\mathcal Z[0, 0]} \, \bigg( \frac{\delta}{\delta \bar \eta(x_1)} \ldots \frac{\delta}{\delta \bar \eta(x_n)} \nonumber \\[5pt]
 & \times \left. \frac{\delta}{\delta \eta(x_{2n})} \ldots \frac{\delta}{\delta \eta(x_{n+1})} \bigg) \mathcal Z[\bar \eta, \eta] \, \right|_{\eta \h = \h \bar \eta \h = \h 0} \,. \label{eq_repr_toshow}
\end{align}
This important formula provides the connection between the {\itshape operator formalism} and the {\itshape path integral formalism} of statistical field theory.

\subsection{Connected Green functions}

The {\itshape connected Green function generator} $\mathcal W \equiv \mathcal W[\bar \eta, \eta]$ is defined as another element of the Grassmann algebra $\mathscr S$ of the source fields. It is the natural logarithm of the Green function generator $\mathcal Z$ divided by the field-independent term,
\begin{equation}
\mathcal W[\bar \eta, \eta] = \ln \frac{\mathcal Z[\bar \eta, \eta]}{\mathcal Z[0, 0]} = \ln \mathcal Z[\bar \eta, \eta] - \ln \mathcal Z[0, 0] \,.
\end{equation}
For $n \geq 1$, the {\itshape connected temperature Green functions} $G^{2n}_{\rm c}$ are defined as the Grassmann derivatives of $\mathcal W$:
\begin{equation} \label{eq_connGF}
\begin{aligned}
 & G_{\rm c}^{2n}(x_1, \ldots, x_{2n}) = \bigg( \frac{\delta}{\delta \bar \eta(x_1)} \ldots \frac{\delta}{\delta \bar \eta(x_n)} \\[3pt]
 & \times \left. \frac{\delta}{\delta \eta(x_{2n})} \ldots \frac{\delta}{\delta \eta(x_{n+1})} \bigg) \h \mathcal W[\bar \eta, \eta] \, \right|_{\eta \h = \h \bar \eta \h = \h 0} \,.
\end{aligned}
\end{equation}
The connected Green functions have the same properties (i)--(iv) as the ordinary Green functions. They also have a Feynman graph expansion in terms of the interaction kernel and the free two-point Green function: the $k$th-order contribution to the connected $2n$-point Green function is given by the sum of all {\itshape connected} Feynman graphs with $k$ interaction vertices and $2n$ external legs. Here, connected means that every interaction vertex is connected (through a series of free Green function lines and interaction vertices) to {\itshape every} external point. The connected Green functions are related to the ordinary Green functions by the following equations: for $n = 1$,
\begin{equation} \label{n1a}
 G^2(x_1, x_2) = G_{\rm c}^2(x_1, x_2) \,,
\end{equation}
for $n = 2$,
\begin{align}
 & G^4(x_1, \ldots, x_4) = G_{\rm c}^4(x_1, \ldots, x_4) \\[5pt] \nonumber
 & + G_{\rm c}^2(x_1, x_3) \h G_{\rm c}^2(x_2, x_4) - G_{\rm c}^2(x_1, x_4) \h G_{\rm c}^2(x_2, x_3) \,,
\end{align}
and similar equations for the higher $2n$-point Green functions (see e.g.~Ref.~\onlinecite{ZinnJustin}).

\smallskip
\subsection{One-line irreducible Green functions} \label{subsec_oli}

\begin{table*}[t]
\normalsize
\begin{tabular}{c}
\hline\hline \\[0.1cm]
\hspace{0.5cm} \includegraphics{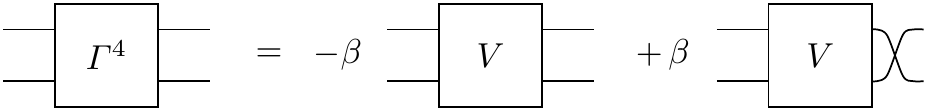} \hspace{0.5cm} \\[0.5cm]
\hline\hline
\end{tabular}

\vspace{5pt}
\caption{First-order Feynman graphs contributing to the one-line irreducible four-point Green function (representation by universal Feynman graphs, see Ref.~\onlinecite{Ronald}). Here, our convention is that $\varGamma^4(x_1, x_2, x_3, x_4)$ is represented by a square with four legs carrying the arguments $x_1$ (upper left), $x_2$ (lower left), $x_3$ (upper right), and $x_4$ (lower right). If two legs are crossed, the corresponding arguments are interchanged. Note that if the interaction kernel $V(x_1, x_2, x_3, x_4)$ is already antisymmetric in its last two arguments, then both terms on the right-hand side are equal.} \label{tab_feyn}

\end{table*}

The {\itshape one-line irreducible Green function generator} $\Gamma$ is defined as the {\itshape Legendre transform} of the connected Green function generator $\mathcal W$, i.e.,
\begin{equation}
 \Gamma = \mathcal W + \langle \bar \varphi, \eta \rangle + \langle \varphi, \bar \eta\rangle \,,
\end{equation}
 where the {\itshape new source fields} $\bar \varphi(x), \, \varphi(x)$ are defined by
\begin{equation}
 \bar \varphi(x) = \frac{\delta \mathcal W}{\delta \eta(x)} \,, \quad \varphi(x) = \frac{\delta \mathcal W}{\delta \bar \eta(x)} \,.
\end{equation}
We assume that these new source fields again generate the Grassmann algebra $\mathscr S$. For $n \geq 1$, the {\itshape one-line irreducible temperature Green functions} $\varGamma^{2n}$ are defined by
\begin{equation} \label{def_oli}
\begin{aligned}
 & \varGamma^{2n}(x_1, \ldots, x_{2n}) = \bigg( \frac{\delta}{\delta \bar \varphi(x_1)} \ldots \frac{\delta}{\delta \bar \varphi(x_n)} \\[3pt]
 & \times \left. \frac{\delta}{\delta \varphi(x_{2n})} \ldots \frac{\delta}{\delta \varphi(x_{n+1})} \bigg) \, \Gamma[\bar \varphi, \varphi] \, \right|_{\varphi \h = \h \bar \varphi \h = \h 0} \,.
\end{aligned}
\end{equation}
The one-line irreducible Green functions have again the same properties (i)--(iv) as the ordinary Green functions. In perturbation theory, $\varGamma^2$ is represented as
\begin{equation} \label{def_self_energy}
 \varGamma^2 = C^{-1} - \varSigma \,,
\end{equation}
where $C$ is the free two-point Green function and $\varSigma$ the {\itshape self-energy}. The $k$th-order contribution to the self-energy is given by the sum of all {\itshape amputated, one-line irreducible} Feynman graphs with $k$ interaction vertices and two external legs. Here, amputated means that all external Green function lines (i.e., the free two-point Green functions with external arguments) are removed; one-line irreducible means that by removing any single internal Green function line, the remaining Feynman graph is still connected. Similarly, for $n \geq 2$, \h$\varGamma^{2n}$ equals $(-1)^n$ times the sum of all amputated, one-line irreducible Feynman graphs with $2n$ external legs. In particular, the two first-order Feynman graphs contributing to $\varGamma^4$ are shown in Table \ref{tab_feyn}. The one-line irreducible Green functions are related to the connected 
Green functions by the following equations: for $n = 1$, 
\begin{equation} \label{n1b}
 \delta(x_1, x_2) = \int \! \de y_1 \,\h G_{\rm c}^2(x_1, y_1) \, \varGamma^2(y_1, x_2) \,,
\end{equation}
for $n = 2$,
\begin{align}
 & G_{\rm c}^4(x_1, \ldots, x_4) = \int \! \de y_1 \ldots \int \! \de y_4 \, G_{\rm c}^2(x_1, y_1) \\[2pt] \nonumber
 & \times G_{\rm c}^2(x_2, y_2) \h \varGamma^4(y_1, y_2, y_3, y_4) \, G_{\rm c}^2(y_3, x_3) \h G_{\rm c}^2(y_4, x_4) \,, 
\end{align}
and similar equations for $n \geq 3$ (see Refs.~\onlinecite{Kambis, Negele}).
Of particular interest is the one-line irreducible four-point Green function $\varGamma^4$, which is commonly interpreted as an ``effective two-body interaction between two particles propagating in a many-particle medium'' (Ref.~\onlinecite[p.~118]{Negele}; see also Ref.~\onlinecite[Chap.~6]{Nozieres}).

\begin{widetext}

\section{Projected RG equations} \label{app_proj}

In this last appendix, we derive the RG equations \eqref{disc_1}--\eqref{eq_21} for the projected interaction vertex, and we show how the mean-field interaction \eqref{to_derive} is deduced from our numerical result for the projected vertex.

\subsection{Derivation} \label{sec_num}

For deriving the RG equations \eqref{disc_1}--\eqref{eq_21}, we put our refined projection ansatz \eqref{ansatz} into Eqs.~\eqref{rg_eq}--\eqref{eq_phiphd}. Note that in \eqref{eq_phipp}--\eqref{eq_phiphd}, we have used the antisymmetry of the interaction vertex in its last two arguments to move the momentum $\vec k_4$ into the last argument of $V_\Lambda$, which is then fixed by momentum conservation, 
$\vec k_4 = \vec k_1 + \vec k_2 - \vec k_3$. This way of writing the RG equations will prove useful in the following, because it avoids ambiguities in the projection of $\vec k_4$ on a representative momentum. First, consider the loop terms given by Eq.~\eqref{def_lmp}. We can write them equivalently~as
\begin{equation} \label{eq_16}
 (L_{\Lambda}^{\mp})_{\ell_1 \ell_2}(\vec k_1, \vec k_2)
 = \Big( \dot \chi_\Lambda(e_{\ell_1}\mh(\vec k_1) ) \, \chi_\Lambda(e_{\ell_2}\mh(\vec k_2)) + \chi_\Lambda(e_{\ell_1}\mh(\vec k_1) ) \, \dot \chi_\Lambda(e_{\ell_2}\mh(\vec k_2) ) \Big) \, F^{\mp}_{\ell_1 \ell_2}(\vec k_1, \vec k_2) \,,
\end{equation}
with the functions $F^{\mp}_{\ell_1 \ell_2}(\vec k_1, \vec k_2)$ given by Eqs.~\eqref{disc_2a}--\eqref{disc_2b}.
Using the symmetry of $F^{\mp}$ under the exchange of its two arguments, we can also rewrite Eq.~\eqref{eq_16} as
\begin{equation}
 (L^\mp_{\Lambda})_{\ell_1 \ell_2}(\vec k_1, \vec k_2) = \dot \chi_\Lambda(e_{\ell_1}\mh(\vec k_1)) \, \chi_\Lambda(e_{\ell_2}\mh(\vec k_2)) \, F^\mp_{\ell_1 \ell_2}(\vec k_1, \vec k_2)
 \, + \, (\vec k_1, \ell_1) \h \leftrightarrow \h (\vec k_2, \ell_2) \,.
\end{equation}
We now evaluate Eq.~\eqref{rg_eq} at the representative momenta,
\begin{equation}
\begin{aligned}
 (\dot V_\Lambda)_{n_1 n_2 n_3 n_4}(\vec \pi_{i_1}, \vec \pi_{i_2}, \vec \pi_{i_3}) & = \big[ \h \varPhi_\Lambda^{\rm pp} + \h \varPhi_\Lambda^{\rm ph, c} + \h \varPhi_\Lambda^{\rm ph, d} \, \big]_{n_1 n_2 n_3 n_4} (\vec \pi_{i_1}, \vec \pi_{i_2}, \vec \pi_{i_3}) \,,
\end{aligned}
\end{equation}
and put the ansatz \eqref{ansatz} into the right-hand side of this equation. We consider separately the particle-particle, crossed particle-hole, and direct particle-hole terms given, respectively, by Eqs.~\eqref{eq_phipp}--\eqref{eq_phiphd}. 

\bigskip \noindent
{\itshape (i) Particle-particle term.} From Eq.~\eqref{eq_phipp}, we obtain
\begin{equation}
\begin{aligned}
 & (\varPhi_\Lambda^{\rm pp})_{n_1 n_2 n_3 n_4}(i_1, i_2, i_3) \equiv (\varPhi_\Lambda^{\rm pp})_{n_1 n_2 n_3 n_4}(\vec \pi_{i_1}, \vec \pi_{i_2}, \vec \pi_{i_3}) \\[6pt] 
 & = {- \sum_{\ell_1, \h \ell_2}} \h \frac{1}{|\mathcal B|} \int_{\mathcal B} \de^2 \vec k_1 \int_{\mathcal B} \de^2 \vec k_2 \, \sum_{\vec K} \delta^2(\vec K + \vec \pi_{i_1} + \vec \pi_{i_2} - \vec k_1, \vec k_2) \\[5pt] 
 & \quad \, \times \Big[ \h \dot \chi_\Lambda(e_{\ell_1}\mh(\vec k_1)) \,\h \chi_\Lambda(e_{\ell_2}\mh(\vec k_2)) \, F^-_{\ell_1 \ell_2}(\vec k_1, \vec k_2) \, + \, (\vec k_1, \ell_1) \h \leftrightarrow \h (\vec k_2, \ell_2) \, \Big] \\[5pt] 
 & \quad \, \times \sum_{j_1 = 1}^N \, \sum_{j_2 = 1}^N \mathbbm 1(\vec k_1 \in \mathcal B_{j_1}) \ \mathbbm 1(\vec k_2 \in \mathcal B_{j_2}) \ (V_\Lambda)_{n_1 n_2 \ell_1 \ell_2}(i_1, i_2, j_1) \ (V_\Lambda)_{\ell_1 \ell_2 n_3 n_4}(j_1, j_2, i_3) \,.
\end{aligned}
\end{equation}
In the second term in square brackets, we {\itshape relabel} the integration and summation variables
\begin{equation}
 \vec k_1 \leftrightarrow \vec k_2 \,, \quad \ell_1 \leftrightarrow \ell_2 \,, \quad j_1 \leftrightarrow j_2 \,.
\end{equation}
This leads to the equivalent expression
\begin{equation}
\begin{aligned}
 & (\varPhi_\Lambda^{\rm pp})_{n_1 n_2 n_3 n_4}(i_1, i_2, i_3) \\[5pt] 
 & = {-\sum_{\ell_1, \h \ell_2}} \h \frac{1}{|\mathcal B|} \int_{\mathcal B} \de^2 \vec k_1 \int_{\mathcal B} \de^2 \vec k_2 \, \sum_{\vec K} \delta^2(\vec K + \vec \pi_{i_1} + \vec \pi_{i_2} - \vec k_1, \vec k_2) \\[-2pt] 
 & \quad \, \times \dot \chi_\Lambda(e_{\ell_1}\mh(\vec k_1)) \ \chi_\Lambda(e_{\ell_2}\mh(\vec k_2)) \ F^-_{\ell_1 \ell_2}(\vec k_1, \vec k_2) \ \sum_{j_1 = 1}^N \, \sum_{j_2 = 1}^N \mathbbm 1(\vec k_1 \in \mathcal B_{j_1}) \, \mathbbm 1(\vec k_2 \in \mathcal B_{j_2}) \\[3pt] 
 & \quad \, \times \Big[ \h (V_\Lambda)_{n_1 n_2 \ell_1 \ell_2}(i_1, i_2, j_1) \ (V_\Lambda)_{\ell_1 \ell_2 n_3 n_4}(j_1, j_2, i_3) \, + \, (j_1, \ell_1) \h \leftrightarrow \h (j_2, \ell_2) \, \Big] \,.
\end{aligned}
\end{equation}
Here we have used that
\begin{equation}
 \delta^2(\vec K + \vec \pi_{i_1} + \vec \pi_{i_2} - \vec k_2, \h \vec k_1) = \delta^2(\vec K + \vec \pi_{i_1} + \vec \pi_{i_2} - \vec k_1, \h \vec k_2) \,.
\end{equation}
By performing the integral over $\vec k_2$ and changing the order of the summations and integrations, we obtain
\begin{align}
 & (\varPhi_\Lambda^{\rm pp})_{n_1 n_2 n_3 n_4}(i_1, i_2, i_3) \label{eq_11} \\[5pt] \nonumber
 & = {-\sum_{\ell_1, \h \ell_2}} \,\h \sum_{j_1 = 1}^N \,\h \sum_{j_2 = 1}^N \,\h \sum_{\vec K} \frac{1}{|\mathcal B|} \int_{\mathcal B} \de^2 \vec k_1 \, \mathbbm 1(\vec k_1 \in \mathcal B_{j_1}) \, \mathbbm 1(\vec K + \vec \pi_{i_1} + \vec \pi_{i_2} - \vec k_1 \in \mathcal B_{j_2}) \\[6pt] \nonumber
 & \quad \, \times \dot \chi_\Lambda(e_{\ell_1}\mh(\vec k_1)) \ \chi_\Lambda(e_{\ell_2}\mh(\vec K + \vec \pi_{i_1} + \vec \pi_{i_2} - \vec k_1)) \ F^-_{\ell_1 \ell_2}(\vec k_1, \vec K + \vec \pi_{i_1} + \vec \pi_{i_2} - \vec k_1) \\[8pt] \nonumber
 & \quad \, \times \Big[ \h (V_\Lambda)_{n_1 n_2 \ell_1 \ell_2}(i_1, i_2, j_1) \, (V_\Lambda)_{\ell_1 \ell_2 n_3 n_4}(j_1, j_2, i_3) \, + \, (j_1, \ell_1) \h \leftrightarrow \h (j_2, \ell_2) \, \Big] \,.
\end{align}
We now employ the following approximation: for $\vec k_1 \in \mathcal B_{j_1}$,
\begin{equation}
\mathbbm 1(\vec K + \vec \pi_{i_1} + \vec \pi_{i_2} - \vec k_1 \in \mathcal B_{j_2}) \h \approx \h
\mathbbm 1(\vec K + \vec \pi_{i_1} + \vec \pi_{i_2} - \vec \pi_{j_1} \in \mathcal B_{j_2}) \,.
\end{equation}
Then, Eq.~\eqref{eq_11} can be further simplified to
\begin{equation}
\begin{aligned}
 & (\varPhi_\Lambda^{\rm pp})_{n_1 n_2 n_3 n_4}(i_1, i_2, i_3) \\[5pt] 
 & = {-\sum_{\ell_1, \h \ell_2}} \ \sum_{j_1 = 1}^N \,\h \sum_{j_2 = 1}^N \,\h \sum_{\vec K} \mathbbm 1(\vec K + \vec \pi_{i_1} + \vec \pi_{i_2} - \vec \pi_{j_1} \in \mathcal B_{j_2}) \ (L^-_\Lambda)_{\ell_1 \ell_2}(i_1, i_2, j_1) \\[5pt]
 & \quad \, \times \Big[ \h (V_\Lambda)_{n_1 n_2 \ell_1 \ell_2}(i_1, i_2, j_1) \ (V_\Lambda)_{\ell_1 \ell_2 n_3 n_4}(j_1, j_2, i_3) \, + \, (j_1, \ell_1) \h \leftrightarrow \h (j_2, \ell_2) \, \Big] \,,
\end{aligned}
\end{equation}
where we have defined
\begin{equation}
\begin{aligned}
 & (L^-_\Lambda)_{\ell_1 \ell_2}(i_1, i_2, j_1) \\[5pt]
 & = \frac{1}{|\mathcal B|} \int_{\mathcal B_{j_1}} \!\! \de^2 \vec k_1 \,\h \dot \chi_\Lambda(e_{\ell_1}\mh(\vec k_1)) \,\h \chi_\Lambda(e_{\ell_2}\mh(\vec K + \vec \pi_{i_1} + \vec \pi_{i_2} - \vec k_1)) \, F^-_{\ell_1 \ell_2}(\vec k_1, \vec K + \vec \pi_{i_1} + \vec \pi_{i_2} - \vec k_1) \,.
\end{aligned}
\end{equation}
These equations are identical to Eqs.~\eqref{eq_18} and \eqref{eq_22} in the main text.

\bigskip \noindent
{\itshape (ii) Particle-hole terms.} Next, we obtain from Eq.~\eqref{eq_phiphc},
\begin{align}
 & (\varPhi_\Lambda^{\rm ph, c})_{n_1 n_2 n_3 n_4}(i_1, i_2, i_3) \equiv (\varPhi_\Lambda^{\rm ph, c})_{n_1 n_2 n_3 n_4}(\vec \pi_{i_1}, \vec \pi_{i_2}, \vec \pi_{i_3}) \\[6pt] \nonumber
 & = {-2\sum_{\ell_1, \h \ell_2}} \frac{1}{|\mathcal B|} \int_{\mathcal B} \de^2 \vec k_1 \int_{\mathcal B} \de^2 \vec k_2 \, \sum_{\vec K} \delta^2(\vec K + \vec \pi_{i_1} - \vec \pi_{i_3} + \vec k_1, \vec k_2) \\[5pt] \nonumber
 & \quad \, \times \Big[ \h \dot \chi_\Lambda(e_{\ell_1}\mh(\vec k_1)) \,\h \chi_\Lambda(e_{\ell_2}\mh(\vec k_2)) \, F^+_{\ell_1 \ell_2}(\vec k_1, \vec k_2) \, + \, (\vec k_1, \ell_1) \h \leftrightarrow \h (\vec k_2, \ell_2) \, \Big] \\[5pt] \nonumber
 & \quad \, \times \sum_{j_1 = 1}^N \, \sum_{j_2 = 1}^N \mathbbm 1(\vec k_1 \in \mathcal B_{j_1}) \ \mathbbm 1(\vec k_2 \in \mathcal B_{j_2}) \ (V_\Lambda)_{\ell_1 n_1 \ell_2 n_3}(j_1, i_1, j_2) \ (V_\Lambda)_{n_2 \ell_2 \ell_1 n_4}(i_2, j_2, j_1) \,.
\end{align}
In the second term in square brackets, we relabel again
\begin{equation}
 \vec k_1 \leftrightarrow \vec k_2 \,, \quad \ell_1 \leftrightarrow \ell_2 \,, \quad j_1 \leftrightarrow j_2 \,,
\end{equation}
and use that
\begin{equation}
 \sum_{\vec K} \delta^2(\vec K + \vec \pi_{i_1} - \vec \pi_{i_3} + \vec k_2, \h \vec k_1) = \sum_{\vec K} \delta^2(\vec K + \vec \pi_{i_3} - \vec \pi_{i_1} + \vec k_1, \h \vec k_2) \,.
\end{equation}
Performing the same steps as for the particle-particle term, we arrive at Eq.~\eqref{eq_20}. Finally, Eq.~\eqref{eq_phiphd} obviously implies Eq.~\eqref{eq_21}, and this completes the derivation.

\subsection{Mean-field interaction} \label{app_sc_int}

We now show explicitly how the interaction \eqref{to_derive} is deduced from our numerical result \eqref{eq_eff} for the projected interaction vertex. By putting Eq.~\eqref{eq_eff} into the projection ansatz \eqref{ansatz}, we obtain the following interaction operator (analogous to Eq.~\eqref{def_coeff_m} for the initial interaction),
\begin{equation} \label{app_zw_1}
\begin{aligned}
 \hat V & =-\frac 1 2 \, \int \! \db^2 \vec k_1 \int \! \db^2 \vec k_2 \int \! \db^2 \vec k_3 \sum_{s_1, \ldots, s_4}  \, \sum_{i_1 = 1}^N \h \sum_{i_2 = 1}^N \h \sum_{i_3 = 1}^N \h (-S) \, ( \delta_{s_1 s_3} \delta_{s_2 s_4} - \delta_{s_1 s_4} \delta_{s_2 s_3} ) \, \mathbbm 1(\vec \pi_{i_1} \mh = \mh -\vec \pi_{i_2}) \\[5pt]
 & \quad \, \times \mathbbm 1(\vec k_1 \in \mathcal B_{i_1}) \, \mathbbm 1(\vec k_2 \in \mathcal B_{i_2}) \, \mathbbm 1(\vec k_3 \in \mathcal B_{i_3}) \,\h \hat a\dag_{s_1}(\vec k_1) \h \hat a\dag_{s_2}(\vec k_2) \h \hat a_{s_3}(\vec k_3) \h \hat a_{s_4}(\vec k_4) \,.
\end{aligned}
\end{equation}
We now evaluate the sums over patch indices and the integral over $\vec k_1$ for each fixed $\vec k_2$ and $\vec k_3$\hh: First, the indices $i_2$ and $i_3$ are uniquely determined by the characteristic functions
\begin{equation}
 \mathbbm 1(\vec k_2 \in \mathcal B_{i_2}) \, \mathbbm 1(\vec k_3 \in \mathcal B_{i_3}) \,.
\end{equation}
The index $i_1$ is subsequently fixed by the function
\begin{equation}
 \mathbbm 1(\vec \pi_{i_1} \mh = \mh -\vec \pi_{i_2}) \,.
\end{equation}
Finally, the integral over $\vec k_1$ is restricted by the function
\begin{equation} \label{char}
 \mathbbm 1(\vec k_1 \in \mathcal B_{i_1})
\end{equation}
to the patch $\mathcal B_{i_1}$. We can therefore approximate
\begin{equation}
 \hat a\dag_{s_1}(\vec k_1) \approx \hat a\dag_{s_1}(\vec \pi_{i_1}) = \hat a\dag_{s_1}(-\vec \pi_{i_2}) \approx \hat a\dag_{s_1}(-\vec k_2) \,,
\end{equation}
and similarly,
\begin{equation}
 \hat a\dag_{s_4}(\vec k_4) \approx \hat a\dag_{s_4}(-\vec k_3) \,. 
\end{equation}
The integrand then depends on $\vec k_1$ only through the characteristic function \eqref{char}, and we can evaluate the integral~as
\begin{equation}
 \int \! \db^2 \vec k_1 \, \mathbbm 1(\vec k_1 \in \mathcal B_{i_1}) \h \equiv \h \frac{1}{|\mathcal B|} \int \! \db^2 \vec k_1 \, \mathbbm 1(\vec k_1 \in \mathcal B_{i_1}) \h = \h \frac{|\mathcal B_{i_1}|}{|\mathcal B|} \,.
\end{equation}
Here, $|\mathcal B|$ is the area of the Brillouin zone, whereas $|\mathcal B_i|$ is the area of the patch $\mathcal B_i$. We assume that all patches are of comparable size, such that
\begin{equation}
 \frac{|\mathcal B_{i_1}|}{|\mathcal B|} \approx \frac 1 N \,,
\end{equation}
with $N$ the number of patches. Thus we obtain from Eq.~\eqref{app_zw_1},
\begin{equation}
\begin{aligned}
 \hat V & =-\frac 1 2 \, \frac 1 N \int \! \db^2 \vec k_2 \int \! \db^2 \vec k_3 \sum_{s_1, \ldots, s_4} (-S) \, ( \delta_{s_1 s_3} \delta_{s_2 s_4} - \delta_{s_1 s_4} \delta_{s_2 s_3} ) \, \hat a\dag_{s_1}(-\vec k_2) \h \hat a\dag_{s_2}(\vec k_2) \h \hat a_{s_3}(\vec k_3) \h \hat a_{s_4}(-\vec k_3) \,,
\end{aligned}
\end{equation}
which coincides with Eq.~\eqref{to_derive} in the main text.
\pagebreak
\end{widetext}

\newpage

\bibliography{rashba_bib}

\end{document}